\documentclass{bmcart}

\usepackage[utf8]{inputenc} 

\def\includegraphics{}

\startlocaldefs
\endlocaldefs

\usepackage{array}
\usepackage{caption}
\usepackage{graphicx}
\usepackage{siunitx}
\usepackage{colortbl}
\usepackage{multirow}
\usepackage{hhline}
\usepackage{calc}
\usepackage{tabularx}
\usepackage{threeparttable}
\usepackage{xspace}
\usepackage{colortbl}
\PassOptionsToPackage{hyphens}{url}
\usepackage{hyperref}
\usepackage{longtable}
\usepackage{booktabs}
\usepackage{nameref}
\usepackage{amsmath}

\hypersetup{
	colorlinks,%
	citecolor=black,%
	filecolor=cyan,%
	linkcolor=black,%
	urlcolor=black
}

\PassOptionsToPackage{numbers/authoryear}{natbib}

\usepackage{soul}
\usepackage[colorinlistoftodos,prependcaption,textsize=tiny]{todonotes}
\usepackage[normalem]{ulem}

\AtBeginDocument{
  \let\namerefOld\nameref
  \renewcommand{\nameref}[1]{{\bf  \color{red}{\namerefOld{#1}}}}
}

\def \check#1{{}}

\newcommand{\yrv}[1]{{\textcolor{black}{#1}}}

\newcommand{\yrepj}[1]{{\textcolor{black}{#1}}}
\newcommand{\swepj}[1]{{\textcolor{black}{#1}}}
\newcommand{\yrrr}[1]{{\textcolor{black}{#1}}}
\newcommand{\swvv}[1]{{\textcolor{black}{#1}}}
\newcommand{\wam}[1]{{\textcolor{black}{#1}}}

\newcommand{\method}{Section~\ref{sec:method}\xspace}
\newcommand{\smapp}{{\textcolor{blue}{\it Supplemental Material}}\xspace}
\newcommand{\smappsec}[1]{{\textcolor{blue}{{\it Supplemental Material}: \ref{#1}}\xspace}}

\newcommand{\olle}{{\tt\small \color{black} OLLE}\xspace}
\newcommand{\olles}{{\tt\small \color{black} OLLE}'s\xspace}

\newcommand{\bigword}{{\color{black} LoFF words}\xspace}
\newcommand{\Bigword}{{\color{black} LoFF words}\xspace}
\newcommand{\bigw}{{\color{black} LoFF word}\xspace}

\usepackage{xr-hyper}

\makeatletter
\newcommand*{\addFileDependency}[1]{
  \typeout{(#1)}
  \@addtofilelist{#1}
  \IfFileExists{#1}{}{\typeout{No file #1.}}
}
\makeatother

\begin{document}

\begin{frontmatter}

\begin{fmbox}
\dochead{Research}


\newcommand{\fulltitle}{Mapping Language Literacy At Scale: A Case Study on Facebook}
\title{\fulltitle}


\author[
   addressref={aff1},                   
   corref={aff1},                       
   email={yurulin@pitt.edu}   
]{\inits{YRL}\fnm{Yu-Ru} \snm{Lin}}
\author[
   addressref={aff2},
   email={xx@fb.com}
]{\inits{SW}\fnm{Shaomei} \snm{Wu}}
\author[
   addressref={aff3},
   email={xx@fb.com}
]{\inits{WM}\fnm{Winter} \snm{Mason}}


\address[id=aff1]{
  \orgname{School of Computing and Information, University of Pittsburgh}, 
  \city{Pittsburgh, PA},                              
  \cny{USA}                                    
}
\address[id=aff2]{%
  \orgname{AImpower.org},
  \city{Mountain View, CA},
  \cny{USA}
}
\address[id=aff3]{%
  \orgname{Meta},
  \city{Menlo Park, CA},
  \cny{USA}
}



\end{fmbox}

\begin{abstractbox}

\begin{abstract} 

\swepj{Literacy is one of the most fundamental skills for people to access and navigate today's digital environment.} 
\swepj{This work systematically studies the language literacy skills of online populations for more than 160 countries and regions across the world, including many low-resourced countries where official literacy data are particularly sparse.} Leveraging public data on Facebook, we develop a population-level literacy estimate for the online population that is based on aggregated and de-identified public posts written by adult Facebook users globally, significantly improving both the coverage and resolution of existing literacy tracking data. 
We found that, on Facebook, women collectively show higher language literacy than men in many countries, but substantial gaps remain in Africa and Asia. Further, our analysis reveals a considerable regional gap within a country that is associated with multiple socio-technical inequalities, suggesting an ``inequality paradox'' -- where the online language skill disparity interacts with offline socioeconomic inequalities in complex ways. These findings have implications for global women's empowerment and socioeconomic inequalities. 
\end{abstract}


\begin{keyword}
\kwd{global literacy}
\kwd{global inequalities}
\kwd{social media demography}
\kwd{information accessibility}
\kwd{cross-language measurement}
\end{keyword}

\end{abstractbox}

%

\end{frontmatter}



\section{Introduction}\label{sec:intro}

Literacy, the ability to comprehend and produce textual information, is known as the foundation for many important personal and social functions. For individuals, the lack of literacy skills is associated with reduced access to education~\cite{NCES:2002,kutner2007literacy,schutz2008education}, employment~\cite{NCES:2002,NCES:2007,kutner2007literacy,ferrer2006effect,bonikowska2008literacy}, social benefits~\cite{schwerdt2018literacy}, as well as poorer health outcomes~\cite{dewalt2005health,OECD:2013} and lower civic engagement \cite{NCES:2002,NCES:2007,OECD:2013,gerger2008}. Collectively, literacy is considered a prerequisite for democracy and socioeconomic development \cite{bonikowska2008literacy,gerger2008}. 

\swepj{Despite a substantial
increase in global literacy rates over recent decades, there were still 750 million adults – two-thirds of whom were women – remaining illiterate in 2016~\cite{unesco2017}. The rise of digital communication technology has brought new challenges to those with limited literacy skills: as more and more public, professional, and social communications shift to the digital, text-mediated environment, a lack of literacy skills can not only exclude people from the information and resources available online but also expose them to greater (mis)informational vulnerability and harms~\cite{mundial2016education,bach2018poverty}. 
With most existing literacy programs and research focusing on school children and educational settings, we see a significant gap in our understanding of \emph{literacy practices and challenges in the digital environment}. 
In this study, we take a data-driven approach, leveraging the data available on Facebook -- the most widely adopted social media platform with a third of the world's population using it regularly~\cite{Meta_2022Q3_report} - to obtain a representative and up-to-date sample of literacy activities (e.g. reading and writing textual content) by the global online population.}

\swepj{
This study systemically examines the \textit{language literacy skills of online populations} (henceforth called  \textit{online language literacy}) for more than 160 countries and regions around the globe. We introduce a new population-level measure called {\it online language literacy estimate} ({\it \olle}) that is based on aggregated and de-identified written content posted publicly on Facebook. Thanks to the reach of Facebook to hundreds of millions of active users from low-resourced regions such as Africa, Latin American, and South East Asia~\cite{Meta_2022Q3_earnings}, our measure is able to estimate and track population-level language literacy at an unprecedented level of coverage, resolution, and timeliness comparing to traditional literacy assessment methods~\cite{rammstedt2016introduction}, while achieving an overall strong correlation with available official data. With \olle calculated for different gender, country, and regional groups across the world, we capture the disparities in online literacy across broad geographical areas and explore gender and regional literacy gaps under a diverse set of societal contexts. Our results not only quantify the association between online language literacy gaps and offline inequality metrics, but also uncover the complex interaction between literacy, Internet adoption, and civic participation for women. In summary, the main contributions of our work are:}
\swepj{
\begin{itemize}
    \item We develop a global online language literacy estimate (\olle) using Facebook data from over 160 countries in 12 languages. 
    \item We evaluate our measure with existing offline population literacy benchmarks, showing a strong correlation and broader coverage than current official data.
    \item We demonstrate how the online language literacy measure can be used to track gender and regional literacy gaps and unpack the complex societal context around literacy and literacy development.
\end{itemize}
}
\swepj{
The rest of the paper is structured as follows: Sec.~\ref{sec:related-work} offers a literature review of related work to contextualize our study. Sec.~\ref{sec:method} describes our methodology and the dataset used for developing the online language literacy estimate (\olle). Sec.~\ref{sec:results} validates the resulting \olle's with existing literacy assessment data and presents an overview of online language literacy skills across the world. We also share a few applications of \olle in studying and understanding regional and gender inequalities globally. Sec.~\ref{sec:discussion} discusses the implications and limitations of this study, and concludes our work. 
}


\section{Background and Related Work}\label{sec:related-work}
\subsection{Population Literacy Assessment}
Recognizing the importance of literacy in reducing poverty and expanding lifelong opportunities, the United Nations has included \textit{literacy} as part of its Sustainable Development Goals (Goal Target 4.6)~\cite{mundial2016education,SDG17}. However, tracking population-level literacy development for different demographics globally remains challenging, with most existing datasets incomplete, dated, or costly to obtain~\cite{50Yearso58:online}.

Worldwide, the United Nations Educational, Scientific and Cultural Organization (UNESCO) has been tracking country-level literacy rates for major demographics such as youth, adults, men, and women~\cite{unesco-literacy-data}. However, their data is based on self-declaration of reading and writing skills, often collected by asking the head of the household to answer questions like: ``\textit{Can you (and others in your home) read and write a simple sentence?}'' As a result, the data may overstate actual skills and not capture any notion of functional literacy~\cite{50Yearso58:online}. Even after adding a simple test of reading skills in the data collection process, the results only group people into three big categories -- illiterate, functional literate, and literate -- and cannot measure literacy on a continuum. Despite issues in the UNESCO data, they are still a major reference point, especially for developing countries and regions where the government infrastructures for census and population surveys are scarce. 

In the developed world, many countries have invested significant efforts to develop and implement modern literacy assessments that capture population literacy skills beyond a simple {\it literacy-illiteracy dichotomy}. In the US, the National Adult Literacy Survey (NALS) has been funded by the federal government in 1992 and 2003~\cite{NCES:2002,NCES:2007}. Internationally, there have been coordinated efforts to assess adult literacy skills through programs such as the Program for International Assessment of Adult Competencies (PIAAC), involving 39 countries and regions since its inception in 2012~\cite{PIAAC}. While these assessments provide more granular and contextualized literacy skill measures, they are expensive to administrate and hard to scale: both NALS and PIAAC were conducted once per decade, requiring 8 to 10 months to conduct the surveys and interviews, and a few years to compile the results~\cite{NCES:2002,NCES:2007,PIAAC}.

As a result, the Global Alliance to Monitor Learning has recently made the call to develop ``efficient'', and ``light''  methodologies to gather nuanced, standardized data that allows for cross-national tracking and comparison~\cite{50Yearso58:online}. This work directly responded to this call, by proposing a data-driven method that leverages social media data to estimate the literacy skills of diverse geographic and demographic populations in a cost-effective way with unprecedented coverage. Although our data were collected from only one platform -- Facebook, its high penetration in many parts of the world allows our method to capture the literacy skills of the entire population, especially populations with high Internet adoption.

\subsection{Digital Literacy}
Although closely related to \textit{digital literacy} -- the ability to operate and communicate through digital technology~\cite{hargittai2009update}, language literacy is composed of fundamental language skills such as reading, writing, and numeracy,  that are often a prerequisite for digital literacy~\cite{bach2018poverty,dimaggio2001digital,mckinsey2014}. In fact, research has shown that the lack of language literacy skills is a top barrier to Internet access and technology adoption~\cite{bach2018poverty,dimaggio2001digital,mckinsey2014}. 

As human society enter an increasingly technological and informational-rich age, modern literacy assessment programs such as PIAAC also include the assessment of ``problem-solving skills in technology-rich environment'', showing several demographic differences and similarities in literacy and digital literacy proficiency in the developed countries~\cite{OECD:2013}. For example, while the gender gap in favor of men was observed with digital literacy skills, there is a very small or non-existent gender difference in literacy skills. Similarly, the age gap in favor of young people was more observed with digital literacy than with literacy~\cite{OECD:2013}. The results from PIAAC also showed a significant interaction effect between gender, age, and socio-economic backgrounds on literacy and digital literacy~\cite{OECD:2013}, inspiring us to explore similar trends for the broader global population covered by this research. 

With over 27,000 new Internet users every hour and many of them from traditionally low-resourced regions~\cite{Theriseo30:online}, this work measures and characterizes the language literacy skills of the population that is already online - as captured on Facebook, laying the foundation for future research on more contextualized literacies such as digital literacy and information literacy.

\subsection{Literacy and Social Media}

Most studies of literacy in the social media context focus on youth and their literacy practice. For example, many studies documented the young social media users' practice of ``remixing'' - creating, uploading, selecting, copy-pasting, combining, and co-producing content in their profiles and timelines, noting a new literacy practice that is more collaborative, dynamic, and multi-model than traditional, print literacy~\cite{Erstad-2007, Alvermann-2008, Greenhow-2009, Perkel-2010, Greenhow-2012,Davies-2012}. As a result, the task of ``reading'' has changed significantly, becoming more technically simple yet socially complex while ~\cite{Kress-2003,Erstad-2007}.

Numerous research examined the relationship between social media and literacy development, especially, for socially disadvantaged groups and English Language Learners (ELLs). Most of these works supported the benefits of social media in literacy development. For example, \cite{Clark-2009} found that reading and writing blogs enhanced the confidence in writing for young people in the UK; \cite{Sabaruddin-2019} documented the use of Facebook is positively correlated with improved English skills for college students in Indonesia; and \cite{Black-2008}) argued that social media technologies can support ELLs to develop valuable print literacy, based on longitudinal ethnographic studies of adolescent ELLs literate and social activities around online fandom communities. However, some research also suggested that social media use can negatively impact the reading culture and academic performance of students~\cite{Kojo-2018}.

As misinformation on social media became a public concern~\cite{Vosoughi-2018-science,Edelson-2021}, some recent work highlighted the importance of literacy skills in the social media age. For example, data from the PISA 2018 reading assessment showed that less than 10\% of the 15-year-old students in OECD countries had the reading proficiency level to distinguish facts from opinions, which could significantly impact their abilities to assess the quality and credibility of information spread on social media~\cite{OECD-2021-pisa}. While an increasing amount of attention has been devoted to developing the digital and media literacy of online populations, this work underlines the prevalence of language literacy challenges and calls for future research in understanding the scale and impact of misinformation on low literacy populations.

\color{black}

\section{Methods}\label{sec:method}

\subsection{Facebook dataset}\label{sec:dataset}
To obtain a written sample of online populations worldwide, we collect public posts written in any of the 12 chosen languages created by Facebook users who are at least 18 years old and active during a 30-day period between April 20 and May 20, 2020. To ensure that the collected posts represent the writings of individual Facebook users, we exclude posts made by pages, organizational accounts, and public profiles. We also exclude posts that did not contain any text or text that was shorter than 2 characters or longer than 1000 characters, as well as posts that contained URLs as these are more likely to be copied and pasted from other sources rather than composed by users. \swvv{After pre-processing the text, the median length of posts in our dataset is 18 characters, the average is 51 characters, and the 95th percentile is 187 characters.}


All the analysis and statistics were computed over a populational aggregate, based on self-reported attributes such as country and gender. \olle were only calculated for groups with at least 1,000 active users, to minimize the risk of user de-anonymization. The original text, together with the intermediate results, were dropped after populational level \olle and other statics were calculated. 

More details about Facebook public post data collection and use can be found in \smappsec{sec:procedure}.
\color{black}

\subsection{Estimating online language literacy}\label{sec:estimate}
We use the vast amount of text produced by Facebook users to quantify populations' online language literacy skills. This method relies on two assumptions: (i) literacy across populations may be measured collectively by aggregating the observed data within country or regional borders, and (ii) vocabulary usage patterns observed in corpora of written texts can be used as a proxy for language practice in a digital environment. The first assumption comes from prior observations that literacy skill is locally clustered, largely determined by the socioeconomic status of the local community and by the abundance of public resources such as public school systems and libraries \cite{kutner2007literacy}. This also alleviates the need to analyze individual-level data, which are harder to de-identify, and often too sparse for reliable estimates. The second assumption is motivated by the relationship between literacy and vocabulary knowledge found in the education domain \cite{lee2011size,curtis2006role,ouellette2006s,national2012improving,schmitt2013introduction}. In practice, vocabulary size was often employed as a proxy for literacy skill, as reading comprehension cannot be achieved unless the reader knows 95\% of the words in the text \cite{lauren1989special}, and a certain vocabulary size is required for unassisted text comprehension \cite{nation2006large}. Vocabulary knowledge of words at various frequency levels has been used to measure total vocabulary size. For example, the Vocabulary Size Test (VST) tests a learner's knowledge of 140 or 200 words, with 10 words sampled from each 1000 word in frequency levels based on British National Corpus \cite{beglar2007vocabulary}.

Analogous to the VST method, we \swepj{measure the aggregated use of lower-frequency words (``\bigword'') - secondary vocabulary words outside the high-frequency everyday lexicons - in public Facebook posts as a proxy for online populations' literacy skills in a given language.} \yrepj{Intentionally designed as a population-level, aggregated measure, our approach does not collect any personally identifiable information or any personal/private content (see \smappsec{sec:procedure}).} For maximal geographical coverage, we pick the twelve (12) most widely used languages (in terms of countries) and algorithmically define a set of lower-frequency words for each language. The 12 selected languages are: Arabic (ar), German (de), English (en), Spanish (es), French (fr), Italian (it),  Malay (ms), Dutch (nl), Portuguese (pt), Russian (ru), Turkish (tr), and Chinese (zh). The next section will detail our method to determine the sets of \bigword based on a multi-lingual reference corpus.

\subsection{Detecting \bigword from a reference language corpus}\label{sec:bigword}
\swepj{While the VST method relies on the British National Corpus (BNC) for baseline word-frequencies,} we use the fastText unigram data \cite{grave2018learning} as our reference corpus. \swepj{Comparing to other popular language corpora such as BNC and Google Books Ngram~\cite{google-ngrams}, fastText has two major advantages:} (i) coverage: The distributed fastText data currently supports 157 languages and covers all 12 languages considered in our study. (ii) \swepj{up-to-date representation of vocabularies used by online population: fastText is based on texts collected from Wikipedia and Common Crawl~\footnote{https://commoncrawl.org/}, containing petabytes of web page data collected since 2008.}  
 
Using fastText data, we are able to retrieve up 200,000 most frequent words in each language as the candidates for \bigword in that language. \swepj{Understanding that the exact range of \bigword may vary depending on the language, we first try to understand the use of these top 200,000 words by Facebook users through language-specific ``word popularity curves'':} a scatter plot where the x-axis represents the rank of a word by its frequency (0 for the most frequent), and the y-axis represents ``word popularity'', as measured by the percentage of unique users in the language population who have used that word in our data. Fig.~\ref{fig:word_curve} (\smapp) shows the word popularity curves for the 12 chosen languages in our study. 

 \swepj{As illustrated in Fig.~\ref{fig:word_curve}, the word popularity curves first decline sharply for the most frequent words before settling into a relatively flat region. The transitional area - the ``elbow'' (or ``knee'') region of the smoothed word popularity curve - corresponds to the words that are neither too popular nor too unpopular, thus ideally covering the set of \bigword for that language. Mathematically, the curvature is a mathematical measure of how much a function differs from a straight line \cite{satopaa2011finding,antunes2018knee}, and the elbow areas start at the point of maximum curvature in a popularity curve. Estimating the knee/elbow point for a continuous function is straightforward since the curvature is well-defined for continuous functions; however, it is a challenging task for discrete data. We leverage the ``Kneedle'' detection \cite{satopaa2011finding}, an efficient algorithm that can efficiently detect knee points in discrete data, and the standard maximum curvature approach on a smoothed function learned from the discrete points. We found this hybrid approach is more robust to rescaling and small fluctuations in our data. Details of this elbow detection is given in \smappsec{sec:elbow}.}

Fig.~\ref{fig:eval} A-C highlight the elbow range detected from the word popularity curve for each of the three most used languages (English, Spanish, and Arabic). Words falling into the elbow range are defined as the ``\bigword'' in the language. Fig.~\ref{fig:word_curve} shows the detected elbow ranges for all 12 languages.

\subsection{Calibration for language use and bias}\label{sec:adjust}
After determining the \bigword per language, we can calculate the relative frequency of \bigword, among all the public text posted by Facebook users from a given country in a given language. We denote the relative frequency of \bigword as $\bar{w}_{c,l}$, with $l$ representing the 12 languages considered in this study and $c$ representing the 167 countries whose official or dominant languages are among these languages. 

While $\bar{w}_{c,l}$ makes it possible to track online language literacy for multiple languages in parallel, we decide to use one representative language per country in all our analyses for simplicity. For most countries, the official language is chosen as the representative language. For countries with no or multiple official languages, we use the language that is used by most users in that country. For example, for India, a country that uses both Hindi and English as official languages, we use English as the representative language for India since English was used by the largest number of Facebook users in India based on our data. Fig.~\ref{fig:countries_covered} (\smapp) shows the number of countries broken down by dominant language.

\swepj{As validation, we group countries by the representative language and compare the value of $\bar{w}_{c}$ with the official literacy rate data for each country. After excluding countries where the official literacy rate is not available or where the Internet penetration is lower than 25\% percent, the three most widely used languages (in terms of the number of countries covered) in our data are English (41 countries), Spanish (19 countries), and Arabic (15 countries). Fig.~\ref{fig:eval} D-F shows the relationship between the officially reported literacy rates and $\bar{w}_{c}$, for English, Spanish, and Arabic, respectively: each dot represents a country, with $x$ value corresponding to the rank-based quantile normalization\cite{beasley2009rank} of $\bar{w}_{c}$, and $y$ value corresponding to the rank-based quantile normalization of its official literacy rate. The shadow areas in Fig.~\ref{fig:eval} D-F visualize Spearman's rank correlation coefficient between the two variables, with 95\% CIs\footnote{\yrrr{Without specification otherwise, the confidence interval for all the correlation coefficients are produced using a nonparametric bootstrap procedure based on the percentile method (with 1000 bootstrap replicates).
}}. As seen in Fig.~\ref{fig:eval} D-F, all three sets of countries exhibit strong positive correlations, though the Spanish-speaking countries show a larger variance than countries mostly using English or Arabic. The positive correlations between $\bar{w}_c$ and the reported literacy rate in the three top languages suggest the efficacy of using the relative \bigw count as a measure of literacy across the different most-used languages.}
The correlations for other languages are not reported because none of the remaining languages cover more than 7 countries.

To generate a global online literacy estimate that is directly comparable across languages, we also calibrate $\bar{w}_{c}$ for all countries using the official literacy data collected by UNESCO~\cite{unesco-literacy-data}. 
Given that the two measures have different distributions, both $\bar{w}_{c}$ and the official literacy data were transformed to better fit a normal distribution. A rank-based ordered quantile normalization transformation \cite{beasley2009rank} was used for both $\bar{w}_c$ and the literacy rates, where the transformation $g(\cdot)$ is given by $g(x_i)=\Phi^{-1}\left(\frac{r_i-0.5}{n}\right)$ where $\Phi$ refers to the standard normal CDF, $x_i$ is a continuous measurement observed for each object $i$, $r_i$ refers to the sample rank of $i$ when the measurements are placed in ascending order, and $n$ refers to the number of observations. In the case of new values that fall outside the observed domain of $x$, we adopt the standard procedure and use generalized linear models to estimate the ranks beyond the bounds of the original domain of $x$. 

A linear fixed effect model is employed to quantify the systematic differences across languages {\it in relation to the offline literacy rates}. Let $l^{(i)}$ be the representative language for a (country) population $i$, the calibrated estimate, denoted as $\hat{w}_i$, is given as 
$\hat{w}_i \propto {y}_{il}$, where
\begin{align*}
{y}_{il} &= \beta x_{il} + \alpha_l + \epsilon_{il},
\end{align*}
where $x_{il} = g(\bar{w}_i)$ is the direct literacy estimate of population $i$, $\alpha_l$ is the language-specific effect, $\epsilon_{il}$ is the idiosyncratic error term with $\epsilon_{il} \sim \mathcal{N}(0,\,\hat{\sigma}_{\epsilon}^{2})$, and $\beta$ is the parameter. \wam{After $\beta$ is learned, \textbf{the global online language literacy estimate (\olle) is the calibrated estimate $\hat{w}_i$}, obtained by rescaling ${y}_{il}$ between 0 and 1.} This rescaling step is to make the \olle value more interpretable and to facilitate the comparison across populations and subpopulations by a single index. 

Since the information about the official literacy rates has been used in the calibration, to get an unbiased evaluation, we use the leave-one-out procedure to obtain an out-of-sample evaluation -- for each $i$, the calibrated estimate was generated by using all countries other than $i$ in the model (the estimated parameters can be found in Model (a) in Table \ref{tab:model-lit-main}). In other words, the language calibration is calculated from the residualized mean aggregated over the rest of the countries using the same language as $i$. 

\swepj{While the official literacy rates data has served an important role in our study to validate and calibrate \olle, they are not suitable as target variables for building a predictive model for online language literacy, for a few reasons: 1) conceptually, we want to distinguish online population literacy and general population literacy in this study; 2) technically, the official literacy rates data were collected for different countries at different time, with methodological variances over the years, introducing extra noise and latent variables for a robust supervised model.}

\subsection{Country-level socioeconomic covariates}\label{sec:socioecon}
To understand the relationship between literacy and other social factors, we collect information about countries' socioeconomic status and technical development from multiple data sources. Tables~\ref{tab:gender_corr} and \ref{tab:region_corr} list all variables used in our study, with definitions and the sources where the variables are gathered. The first-order correlations among these variables are provided in Table~\ref{tab:gender_corr}. Due to the heterogeneous distributions across variables, we report correlations using Spearman's rank correlation coefficient unless otherwise stated.

\swepj{Inspired by previous research showing the effect of a country's income, Internet penetration, and other gender inequality measure on the digital gender gap~\cite{fatehkia2018using}, we study the relationship between social factors and online literacy gaps through regression analysis with the gender or regional differences in \olle as the dependent variable and country-level variables as the independent variables.}
All the regression models presented in this work are based on standard OLS estimates, where all variables are first separately transformed to better fit a normal distribution. For example, the income variable was transformed logarithmically. Considering the geographical clustering of many of the socioeconomic variables, in each of the regression analyses, we provide models with and without controls for geographical groups. The control for geographical groups signifies whether the pattern observed in our study is a global phenomenon or particular to certain areas. For example, when predicting the gender gap in \olle, the coefficient estimates are consistent between the models with and without geographical information. We show in \smapp Tables~\ref{tab:model-gender-main}-\ref{tab:model-gender-3v-geo} the detailed estimates of regression models and their comparisons. The consistent coefficient estimates are also found in predicting regional disparity (\smapp Tables~\ref{tab:model-region-main}-\ref{tab:model-region-4v}).

\section{Results}\label{sec:results}
The results are two-fold. We first validate of our literacy estimates with existing official data and report the global state of online language literacy across 167 countries. Next, we consider within-country demographic segments such as gender and regions to benchmark the differences in online language literacy across subpopulations, and examine the socioeconomic factors that explain those differences. 

\subsection{Online Language Literacy Worldwide}\label{sec:world}

\subsubsection{Significant agreement between \olles and official literacy data}
Following the methodology detailed in \method, we generate \olle, the calibrated online language literacy estimates, in 167 countries or regions whose representative languages are among the 12 selected languages and have at least a thousand adult Facebook users who posted publicly in the representative language during our data collection period. To ensure we obtain a sufficient sample size of the population in each country, we leave out five countries---China, Iran, Russia, Kazakhstan, and Turkmenistan---where Facebook use is curbed by the countries' government regulations or other policy challenges. We show the country and world population coverage breaking down by languages in \smapp Fig~\ref{fig:countries_covered}. The raw and calibrated values of online language literacy estimates for all 167 countries can be found in \smapp Table~\ref{tab:all_country}. 

Fig~\ref{fig:eval} illustrates the key steps of our methodology, as well as the correlation between \olles and official literacy rate data in~\ref{fig:eval}G. As visualized in Fig.~\ref{fig:eval}G, we find a strong and positive correlation between \olle with the reported literacy rates (Spearman's rank correlation $\rho=0.78$, \yrepj{95\% CI [0.69, 0.84]}, $p<0.001$ based on out-of-sample evaluation). Similar results are found when validating our estimates with global educational attainment statistics: \olle is highly correlated with a country's average schooling years ($\rho=0.78$, \yrepj{95\% CI [0.65, 0.87]}, $p<0.001$; see details in \smapp Fig.~\ref{fig:corr_lit_edu}). These findings indicate that our estimates do reflect populations' literacy skills and can be used as a reliable proxy for official literacy and educational attainment statistics when such data are unavailable or outdated.

\subsubsection{Understand global online literacy inequalities through \olle}
\swepj{One direct application of \olle is to track the state of literacy for online populations across the globe. Mapping \olles by country in Fig.~\ref{fig:lit_map}B, we see significant inequalities in online language literacy skills across geographical regions, with the ``global south'' countries collectively lagging behind in terms of online population literacy skills.}  Fig.~\ref{fig:lit_map}A summarizes the aggregated statistics for seven geographical groups and benchmarks the online literacy gaps across regions. The bottom 10\% of countries with the lowest \olles are primarily located in Sub-Saharan Africa (13 countries), plus one in Latin America \& the Caribbean (Haiti), and one in Northern Africa \& Western Asia (Algeria). \swepj{While our result is consistent with the geographical patterns observed in official literacy rate data~\cite{unesco2017}, it also highlights the persistence of literacy gaps across offline and online populations, calling out for additional literacy support for a substantial percentage of the online population in today's digital environment.}

\swepj{On the other end of the spectrum, the top-ranked countries in terms of \olle are all located in the Europe \& North American region, as well as Oceania, with the top 3 countries being Belarus, Ukraine, and San Marino. While this result is again largely consistent with the official literacy rate data by UNESCO, it also suggests potential biases introduced by language-based calibration. For example, countries with Russian as the representative language (e.g. Belarus, Ukraine) could get an extra boost during calibration due to the overall high literacy rates in Russian-speaking countries reported in the official data.}

\subsection{Gender Difference in Online Language Literacy}\label{sec:gender}

Although the gender gap in literacy has been shrinking globally in recent decades, women are still facing obstacles when accessing school and the Internet~\cite{unesco2017,owidinternet}. As a result, serious male-favoring gender gaps in literacy skills still persist in the Middle East and North Africa, South Asia, and Sub-Saharan Africa regions \cite{world2020global}. An earlier study showed that, in low-income countries, female Internet penetration is 24\% lower than that for males \cite{fatehkia2018using}. To track the gender literacy gap in the online population, we calculate the standardized difference in \olle between female and male Facebook users in each country. This measure captures both the direction and the size of the gender gap in online language literacy, where a positive value indicates a female-favoring gap, and vice versa. To ensure a sufficient sample size of the male and female subpopulations, we drop countries (17 out of 167) with fewer than a thousand adult users in either gender group in our dataset for this analysis. While the gender digital divide has generally referred to the gaps in access and use of digital technology, our measure calls for attention to the disparity in language skills between women and men who are already online. Understanding that gender is non-binary, we only present the female-male gender gap here for two reasons: (a) it enables us to correlate our estimates with existing gender gap data; (b) in our dataset, we do not have sufficient data from users with self-reported non-binary gender information to deliver a reliable estimation for this sub-population.

\subsubsection{Women collectively scored higher than men on Facebook in most countries, with substantial \yrepj{male-favoring} gaps in two regions}
Fig.~\ref{fig:gender_gap}A shows the gender differences in \olle captured in our data.  Among the 160 countries where the gender gap in \olle is calculated, 69 countries (43.1\%) have significant female-favoring gaps and 54 countries (33.8\%) have significant male-favoring gaps. The remaining 37 countries do not have a significant gender gap in \olle. \yrepj{The significance is determined based on whether a country's male-favoring gap (or female-favoring gap) falls above the 95\% confidence limits of the expected male-favoring gaps (or female-favoring gaps).}
Overall, we observe more countries having female-favoring gaps in our measure, suggesting on average higher language literacy skills for women than men among today's online population. 

Fig.~\ref{fig:gender_gap}A highlights countries with the most and least substantial female-favoring gaps. As observed in Fig.~\ref{fig:gender_gap}A, almost all of the countries with the world's largest advanced economies (the G7) have a significant female-favoring gap, with Italy the only exception. \swepj{This finding is generally consistent with the data collected in recent PISA tests, which showed that girls outperformed boys in reading in all the OCED countries and regions~\cite{pisa-2018-gender}. Despite progress, gender-based inequalities are still pervasive in today's society.} To summarize the global state of online literacy difference between male and female subpopulations, a world map of \olle gender gap is provided in \smapp Fig.~\ref{fig:gender_map}. Notably, while most regions on average appear to have female-favoring gaps, two regions (Sub-Saharan Africa and Northern Africa \& Western Asia) still show substantial gaps favoring men. In Sub-Saharan Africa, there are 22 countries with male-favoring gaps, compared to 8 with female-favoring gaps; in Northern Africa \& Western Asia, there are 15 countries with male-favoring gaps,  compared to \yrepj{only} two countries with female-favoring gaps. \swepj{The regional patterns here are generally consistent with what was reported by UNESCO in the official literacy rate data~\cite{unesco2017}. Our measure, however, characterizes more countries with female-favoring literacy gaps than the official data, indicating a potential populational difference between women on Facebook and the general female population in a country. We will further explore the relationship between populational factors and the gender gap in \olle in the next subsection.}

\subsubsection{Understand the societal context for gender online literacy gap}
We first compare the observed gender gap in \olle with other country-level measures such as overall \olle, income per capita, Gini index, average education attainment, and Internet penetration rate. As shown in Fig.~\ref{fig:gender_gap} 
\yrepj{(B,C,D) the \olle gender gaps are positively correlated with the countries' \olle (Spearman's rank correlation $\rho=0.59$, 95\% CI [0.47,0.68], $p<0.001$), overall education ($\rho=0.59$, 95\% CI [0.44,0.71], $p<0.001$), and Internet penetration ($\rho=0.30$, 95\% CI [0.14, 0.46], $p<0.001$), suggesting that women are disproportionally disadvantaged in low-resourced, low-literacy countries. Fig.~\ref{fig:gender_gap} (E,F,G) show that countries' \olle gender gaps significantly correlate with other gender parity measures, including a positive association with the female-male difference in offline literacy rates ($\rho=0.43$, 95\% CI [0.26, 0.58], $p<0.001$), women's civic participation ($\rho=0.48$, 95\% CI [0.3, 0.62], $p<0.001$), and a negative association with the countries' Gender Inequality Index or GII ($\rho=-0.4$, 95\% CI [-0.57, -0.23], $p<0.001$).}
The GII reflects how women are disadvantaged in multiple dimensions of human development and thus a negative association is expected~\cite{gender-inequality-index}. Interestingly, among all gender parity or empowerment measurements, women's civic participation -- the extent to which women have the ability to express themselves and to participate in civil society \cite{coppedge2019v} -- appears to have the strongest association with the \olle gender gap. This could suggest that the offline structural barriers to women's civic participation are strongly associated with their literacy relative to men in the online space.

\swepj{To better understand the societal context for online literacy gender gap globally, we further examine the relationship between country-level variables and the observed gender gap in \olle through multiple regression Ordinary Least Squares (OLS) model.} Given that many of the country-level variables are highly correlated (see \smapp Table~\ref{tab:gender_corr}), we only include the most relevant variables in this analysis. Fig.~\ref{fig:gender_model}A summarizes the estimated effect of these variables, where the effect of geographical grouping is further detailed in \smapp (see Tables \ref{tab:model-gender-main}--\ref{tab:model-gender-3v-geo}). 
Based on the OLS estimation, the \olle gender gap remains significantly and positively associated with overall education status and women's civic participation while controlling for other variables.
\yrepj{The overall Internet penetration rate is negatively associated with the \olle gender gap. This may look counterintuitive. One possible interpretation is that a lower level of Internet penetration rate excludes groups from lower socioeconomic status to participate in the digital world, and women in those groups also tend to lack opportunities in education and many other developmental aspects. Hence, a lower level of Internet penetration rate ironically serves as an equalizer for the \olle gender gap.}
Interestingly, the OLS model also reveals an interaction effect between the overall Internet penetration rate and women's civic participation on the \olle gender gap. When a country's Internet penetration rate is high, the country's \olle gap may be either high or low -- depending on whether the country has a high level of women's civic participation (Fig.~\ref{fig:gender_model}B). This could suggest that technological advancements -- i.e., the adoption of the Internet -- are not necessarily associated with more opportunities or higher skills for women relative to men unless such a relationship appears in a society where women have the chance to actively participate in civic processes.

\subsection{Within-Country Regional Disparity in Online Language Literacy}\label{sec:regional}

While it is widely acknowledged that the disparity in education resources and technology infrastructure has contributed to the digital divide between developed and developing countries~\cite{warf2020geographies,wdr:2016}, there have been only a few studies that examined the digital disparities within a country -- often limited to studying a single country with the digital divide among gender and ethnicity groups~\cite{pew:2012,Hilbert:2011}. Here we intend to provide insights into the within-country regional digital disparities for a large number of countries across the world. Extending our methodology, we measure the within-country regional disparity in online language literacy by quantifying the variability of regional \olles for a given country. More specifically, the variability of regional \olles is calculated as the standard deviation of \olles aggregated across available regions within a country, thus a larger value indicates a higher variability observed in \olles at a sub-national level. We define a region as a sub-national administrative division (self-government or jurisdiction under a country's national laws), such as a state or a province. Countries with less than two regions having a minimum of a thousand active adult Facebook users in our dataset are excluded, which resulted in 119 countries in our analysis. Fig.~\ref{fig:region_map} in \smapp provides summary statistics and a map of the within-country regional disparity in \olle, as well as the representative countries with a relatively high or low level of regional disparity from each geographical group.  

\subsubsection{Within-country regional disparity in \olle is associated with multiple inequality measures}
We examine the societal backdrop for the observed regional disparities in online language literacy. Our particular interest is in the link between the regional disparity in \olle and countries' resource distribution, such as the inequalities in education and income within a country, as well as its overall education and socio-technical development. Using multiple regression analysis, we find that, after controlling for all other variables, inequality in education and the Internet penetration rate have a strong and positive association with regional disparities in \olle (Fig.~\ref{fig:gender_model}C). Not surprisingly, a higher level of overall educational attainment predicts a smaller regional variation in online language literacy skills, which is converse to the effect of inequality in education. The inequality in income, as captured by a country's Gini index, however, appears to have a negative relationship with the \olle regional disparity, indicating a greater income inequality is associated with smaller regional \olle disparity within the country.  

\subsubsection{Inequality paradox}
\yrrr{The interaction between inequalities in education and income is also observed (Fig.~\ref{fig:gender_model}D), where a greater level of within-country regional disparity in \olle is predicted for countries with one of the two conditions: either the country has a relatively high level of income and education inequalities, or has a relatively low level of both inequalities.} This finding suggests an ``{\it inequality paradox}'' -- a paradoxical pattern we notice that links the offline socioeconomic inequalities to online language skill disparity in surprising ways. \yrrr{For example, in countries with a higher level of either economic or educational inequalities, access to social media is more likely to be reserved for the more socio-economically advantaged groups, and therefore show a less regional disparity in \olle (corresponding to the top-left or bottom-right corner of Fig.~\ref{fig:gender_model}D). In contrast, in countries where education inequality is low but economic inequality is high, a higher level of regional disparity in \olle is observed (corresponding to the bottom-right corner of Fig.~\ref{fig:gender_model}D).} Similar patterns are observed when taking the geographical grouping into account, suggesting that the observed patterns are common across societies (see \smapp Tables \ref{tab:model-region-main}--\ref{tab:model-region-4v} for more details).



\section{Discussion}\label{sec:discussion}

\paragraph{A data-driven, cross-language, cross-country online literacy estimation}
Taking advantage of the abundance of user-generated text online, our proposed methodology of measuring online language literacy \yrv{can be scaled across languages and subpopulations, as long as population-level text corpora are available. \olle complements the traditional sources and makes it possible to} monitor future progress and answer questions such as: whether the online language skills improve faster (slower), or whether the literacy gap is closing (widening), particularly in low-literacy countries. \swepj{While our current dataset only contains text generated within a 30-day period, further collection of similar data over a longer period of time will offer new insights on the temporal evolution of \olles across the world.}

\paragraph{Tracking global trend in online language literacy}
This study reveals the current state of global online language literacy. Based on our study, the estimated global online language literacy has remarkably high correlation with documented country-level literacy rates and educational attainments data (with $\rho=0.78$ in both correlations; see Fig.~\ref{fig:eval} and \smapp Fig.~\ref{fig:corr_lit_edu}). This finding has two implications. First, given that 86\% of the world population are now reportedly literate (i.e., able to read and write)~\cite{unesco2017}, our study suggests the variation in language skills remains {\it within the literate, online population}. Even though many countries now have more than 95\% literate populations, our online literacy map (Fig.~\ref{fig:lit_map}) has revealed the nuanced differences among the online population's language skills in these countries. Second, beyond a few options in assessing a country's digital advancement, such as the Internet penetration measure, \olle's robust correlation with offline literacy and educational data make it a more relevant alternative for tracking the {\it outcome} of a country's access to education and resources for global literacy development. 

\paragraph{Women's empowerment, social inequalities, and online language literacy disparities}
In our study, we show that the gender difference in \olle is significantly correlated with various offline gender parity metrics, including gender gaps in literacy rate, education, GII (which also considers the economic standing across gender groups), and women's civic participation index. This suggests that a country's offline gender equity progress is crucially relevant to how literate populations across genders participate online. In contrast to existing studies that exposed the well-known correlation between the gender gaps and a country's economic and technical development \cite{world2020global,fatehkia2018using}, we find that the link is not trivial. The relationship between countries' Internet penetration and the gender gap in \olle is {\it not monotonic}, and only when there is a sufficiently high level of women participation in civic society does the \olle gender gap align with countries' Internet access. In countries with a low level of women's civic participation, the \olle gap favoring men persists even with the rise of the overall Internet penetration (Fig.~\ref{fig:gender_model}B). This finding highlights the crucial social condition that allows more literate women to participate online. We also observe non-trivial relationships among multiple inequalities in our analysis of within-country regional disparity. The regional disparity in \olle is positively associated with unequal education, but the relationship is not simple when comparing countries with different levels of income inequality -- for example, a lower disparity \olle may reflect the homogenized language skills from only the economically advantaged subpopulations, or those from only more educated subpopulations. Our study explicates the complex relationship between the multidimensional inequality measurements and their manifestation on digital populations' online literacy skills.

\subsection{Limitations and future research opportunities}\label{sec:discussion}
We discuss the limitations of this study and highlight where study results must be interpreted with caution, as well as future research opportunities.

\paragraph{Self-selection bias in Facebook data}
Traditional literacy surveys are expensive to implement and many areas of the world have limited resources for survey research. The challenge for gathering nationally representative samples is not unique to traditional survey research; more recent assessments -- for example, PIAAC, which was predominantly administered on computers, were subject to selection effects and therefore required additional adjustment \cite{yamamoto2013scaling}. Our analyses are not immune from self-selection bias where the use of Facebook varies in popularity across different demographics and the differences also vary with countries and regions \cite{Facebook2020Q2earning:online}, as well as user subcultures. \yrv{For example, the observed gender gaps may be due to the over-representation of more privileged women online  \cite{magno2014international,kashyap2021analysing}.}
On Facebook, a user's comfort level of posting likely depends on their language skills; those with very limited vocabulary may not be in the data, or may choose to communicate via other modalities, e.g., images or videos. 
This study only considers users' text-based interactions on Facebook and thus the estimates likely miss out on people at the low end of vocabulary skills. To some extent, sampling bias may be mitigated by post-sampling weightings with demographic information, as has been demonstrated in recent data-driven studies \cite{park2019global}. Such an approach nevertheless depends on sufficiently rich demographic information in the data. In our study, only data disaggregated by gender and coarse-grained geographical grouping are available. Future work may consider tackling the selection bias by separately collecting users' information on demographics and their social media interaction practice. 

\paragraph{Representative languages and language-based calibration}
\yrepj{In the current study, a country's online language literacy was measured based on a single representative language (either the official language or the most used language). One potential risk of relying on a single representative language is that the regional disparity measure in a multilingual country may simply capture the distribution of languages, rather than the diversity of language skills. To address this concern, we perform robust checks in the \smappsec{sec:domi} and Fig.~\ref{fig:domi} and do not see systematic biases associated with different penetration rates of the representative language. }

\yrepj{Another potential risk is to underestimate the language literacy of countries that have sizable language minorities (including people who use/speak a dialect), multilingual communities, or multiple monolingual subpopulations, since their data are largely excluded from our methodology.}  For example, an English-majority country with a larger Spanish-speaking population may score lower in a measure of English-language literacy skills. For such countries, focusing on improving a dominant-language literacy measure can be potentially harmful, since more resources may be allocated in favor of the dominant language.  
\swepj{In \smappsec{sec:india}, we present a case study using India as an example of a multilingual country and show that literacy estimation based on multiple languages has neglectable improvement over English-based estimation in its correlation with the official literacy data (see Fig.~\ref{fig:domi}). However, we acknowledge the official literacy data often have a bias against language minorities and recommend future work consider measuring the online language skills separately for all languages used by sizable populations within a country, to better understand the literacy skills and needs across diverse communities.}

\swepj{The use of official literacy data for cross-language \olle calibration also introduces potential biases and noises.} As mentioned in Section \ref{sec:world}, the post-calibration \olles for the Russian-speaking countries are likely to be overestimated due to their historically high literacy rates in the official data. On the other hand, \olles for Arabic-speaking countries be underestimated due to the fact that the alternative learning (e.g., religious education) provided in those countries was not included in the official literacy data. Languages concentrated in only one or a few countries, such as Japanese and Korean, are not considered in our study due to the lack of benchmark data that can be used for validation or calibration. Therefore, to establish an adequate common scale for more languages, future research will benefit from more comprehensive and up-to-date data for literacy skills across languages.

\paragraph{Thresholding vs. continuum measurement}
Our measure relies on thresholding the observed word frequency bands -- i.e., the set of \bigword was identified by the automatically determined word frequency cut-offs -- but one may also consider the continuum of the word frequency range. Our choice of focusing on particular word frequency bands is aligned with the existing literature in language comprehension research. For example, studies from English language comprehension distinguish the utility of high-, mid-, and low-frequency vocabulary: the high-frequency vocabulary (e.g., the most frequent 2000- or 3000-word families from a particular English corpus) provides the largest lexical coverage of any text but is not sufficient for adequate reading comprehension, while the low-frequency vocabulary (including the words over the 9000-word families) is too infrequent and thus has very limited utility; only the mid-frequency vocabulary gives the important range of words required for reading authentic materials \cite{nation2006large,masrai2019vocabulary}. \yrepj{However, different vocabulary sets may serve significant functions for different populations; for example, high-frequency vocabulary has been shown as an important source of knowledge for second-language learners \cite{masrai2019vocabulary}.} Future work may take into account the continuum of the word frequency range and investigate the level of contribution provided by the various word frequency bands to online language skills.

\paragraph{Heterogeneity in social media texts}
A potential concern about using social media text to measure the language skills of a population is how to deal with social media users' heterogeneous behaviors, e.g., some users may post more than others, and some tend to copy content from elsewhere, which could disproportionally impact the population-level measurement.
\swepj{We adopt a few methods to address this concern, including counting each unique unigram once per user, and leaving out posts that are likely to be copy-pasted (see \smappsec{sec:procedure} for more details). However, we did not perform efficacy evaluation for these methods, and would encourage} future work further examine the impact of text recycling and text production disparities for online literacy assessment. 

\paragraph{Aggregate vs. individual measures, and correlations}~\olles are generated based on aggregate data, which inherently poses risks of ecological fallacy compared to other literacy data collected through individual-level tests and surveys. 
In our study, the between-country correlations only involve the between-country differences in aggregate statistics of the within-country distributions, and the unmeasured within-country measures could be uncorrelated, or could even be correlated in the opposite direction. Taking into account individual assessment in a multi-level analysis \yrrr{with a proper privacy protection mechanism} may be a fruitful direction to reduce aggregation bias and the ecological fallacy in future research.
In the case of the observed association between the gender gap in \olle and women's civic engagement, a less ambiguous interpretation -- whether higher literacy empowers women for civic engagement, or civic engagement leads to legal and institutional changes that enhance literacy, or other cultural, religious, political, and socio-economic conditions influence both women's civic engagement and progress in online language literacy -- requires further research to carefully examine the causal pathways.

\subsection{Conclusions}
This work develops a scalable language literacy measurement to monitor the collective language literacy of the online population using social media data from more than 160 countries. The measure then allows for tracking the trends and inequalities in online language literacy and their relationships with various socioeconomic conditions. Our findings identify key regions and populations disproportionally impacted by literacy challenges, and suggest that education or technical infrastructure alone is not sufficient to explain the variance in online population language literacy skills. Our study calls out the need for more attention and resources to be allocated to populations with limited online literacy skills -- especially those who also suffer from poverty, low resource, and other structural discrimination, to empower them through global challenges such as misinformation and social inequality, and to sustain the overall progress in democratic and socioeconomic development.


\begin{backmatter}
\section*{Abbreviations}\label{sec:abbr}
OLLE: online language literacy estimate \\
LoFF words: lower-frequency words \\
UNESCO: United Nations Educational, Scientific and Cultural Organization \\
NALS: National Adult Literacy Survey \\
PIAAC: Program for International Assessment of Adult Competencies \\
ELLs: English Language Learners \\
VST: Vocabulary Size Test \\
CDF: cumulative distribution function \\
OLS: ordinary least squares \\
GII: Gender Inequality Index \\

\section*{Availability of data and materials}\label{sec:data_availability}
Data aggregated at the country level (country-level literacy estimates and summary statistics) will be made available in the Open Science Framework (OSF, at \url{https://osf.io/zcpej/}) upon publication of this manuscript. Facebook requires that this work was to be done in compliance with Facebook's Data Policy and research ethics review process (\url{www.facebook.com/policy.php}). Restrictions apply to the availability of the disaggregated data (user- or post-level data), so they are not publicly available. Data aggregated at the country level and other datasets that support the findings of this study will be available from the OSF repository with the permission of the authors, upon reasonable request. The analysis code used to derive the main results are available in the Open Science Framework (OSF) upon publication of this manuscript.

\section*{Competing interests}
  The authors declare that they have no competing interests.

\section*{Author's contributions}
YRL and SW conceived and designed the research; YRL and SW developed the analysis tools; YRL performed the experiments and analyzed the data; YRL, SW, and WM wrote the paper.

\section*{Acknowledgements}
We thank Lada Adamic, Michael Macy, Mike Bailey, Pablo Barbera, Devra Moehler, Logan Schmid, Alex Pompe, Niki Ramchandani, Alex Leavitt, James Lo, and Edouard Grave, and anonymous reviewers for their valuable feedback.




\bibliographystyle{bmc-mathphys} 
\bibliography{reference.bib,reference_supp.bib}


\begin{thebibliography}{76}
\ifx \bisbn   \undefined \def \bisbn  #1{ISBN #1}\fi
\ifx \binits  \undefined \def \binits#1{#1}\fi
\ifx \bauthor  \undefined \def \bauthor#1{#1}\fi
\ifx \batitle  \undefined \def \batitle#1{#1}\fi
\ifx \bjtitle  \undefined \def \bjtitle#1{#1}\fi
\ifx \bvolume  \undefined \def \bvolume#1{\textbf{#1}}\fi
\ifx \byear  \undefined \def \byear#1{#1}\fi
\ifx \bissue  \undefined \def \bissue#1{#1}\fi
\ifx \bfpage  \undefined \def \bfpage#1{#1}\fi
\ifx \blpage  \undefined \def \blpage #1{#1}\fi
\ifx \burl  \undefined \def \burl#1{\textsf{#1}}\fi
\ifx \doiurl  \undefined \def \doiurl#1{\textsf{#1}}\fi
\ifx \betal  \undefined \def \betal{\textit{et al.}}\fi
\ifx \binstitute  \undefined \def \binstitute#1{#1}\fi
\ifx \binstitutionaled  \undefined \def \binstitutionaled#1{#1}\fi
\ifx \bctitle  \undefined \def \bctitle#1{#1}\fi
\ifx \beditor  \undefined \def \beditor#1{#1}\fi
\ifx \bpublisher  \undefined \def \bpublisher#1{#1}\fi
\ifx \bbtitle  \undefined \def \bbtitle#1{#1}\fi
\ifx \bedition  \undefined \def \bedition#1{#1}\fi
\ifx \bseriesno  \undefined \def \bseriesno#1{#1}\fi
\ifx \blocation  \undefined \def \blocation#1{#1}\fi
\ifx \bsertitle  \undefined \def \bsertitle#1{#1}\fi
\ifx \bsnm \undefined \def \bsnm#1{#1}\fi
\ifx \bsuffix \undefined \def \bsuffix#1{#1}\fi
\ifx \bparticle \undefined \def \bparticle#1{#1}\fi
\ifx \barticle \undefined \def \barticle#1{#1}\fi
\ifx \bconfdate \undefined \def \bconfdate #1{#1}\fi
\ifx \botherref \undefined \def \botherref #1{#1}\fi
\ifx \url \undefined \def \url#1{\textsf{#1}}\fi
\ifx \bchapter \undefined \def \bchapter#1{#1}\fi
\ifx \bbook \undefined \def \bbook#1{#1}\fi
\ifx \bcomment \undefined \def \bcomment#1{#1}\fi
\ifx \oauthor \undefined \def \oauthor#1{#1}\fi
\ifx \citeauthoryear \undefined \def \citeauthoryear#1{#1}\fi
\ifx \endbibitem  \undefined \def \endbibitem {}\fi
\ifx \bconflocation  \undefined \def \bconflocation#1{#1}\fi
\ifx \arxivurl  \undefined \def \arxivurl#1{\textsf{#1}}\fi
\csname PreBibitemsHook\endcsname

\bibitem{NCES:2002}
\begin{botherref}
\oauthor{\bsnm{Kirsch}, \binits{I.S.}},
\oauthor{\bsnm{Jungeblut}, \binits{A.}},
\oauthor{\bsnm{Jenkins}, \binits{L.}},
\oauthor{\bsnm{Kolstad}, \binits{A.}}:
Adult literacy in america: A first look at the findings of the national adult
  literacy survey
(2002).
NATIONAL CENTER FOR EDUCATION STATISTICS. Access on 09/23/2022 at
  \url{https://nces.ed.gov/pubs93/93275.pdf}
\end{botherref}
\endbibitem

\bibitem{kutner2007literacy}
\begin{botherref}
\oauthor{\bsnm{Kutner}, \binits{M.}},
\oauthor{\bsnm{Greenberg}, \binits{E.}},
\oauthor{\bsnm{Jin}, \binits{Y.}},
\oauthor{\bsnm{Boyle}, \binits{B.}},
\oauthor{\bsnm{Hsu}, \binits{Y.-c.}},
\oauthor{\bsnm{Dunleavy}, \binits{E.}}:
{Literacy in Everyday Life: Results from the 2003 National Assessment of Adult
  Literacy. NCES 2007-490.}
Report,
U.S. Department of Education. Washington, DC: National Center for Education
  Statistics
(2007)
\end{botherref}
\endbibitem

\bibitem{schutz2008education}
\begin{barticle}
\bauthor{\bsnm{Sch{\"u}tz}, \binits{G.}},
\bauthor{\bsnm{Ursprung}, \binits{H.W.}},
\bauthor{\bsnm{W{\"o}{\ss}mann}, \binits{L.}}:
\batitle{Education policy and equality of opportunity}.
\bjtitle{Kyklos}
\bvolume{61}(\bissue{2}),
\bfpage{279}--\blpage{308}
(\byear{2008})
\end{barticle}
\endbibitem

\bibitem{NCES:2007}
\begin{botherref}
\oauthor{\bsnm{Kutner}, \binits{M.}},
\oauthor{\bsnm{Greenberg}, \binits{E.}},
\oauthor{\bsnm{Jin}, \binits{Y.}},
\oauthor{\bsnm{Boyle}, \binits{B.}},
\oauthor{\bsnm{Hsu}, \binits{Y.-c.}},
\oauthor{\bsnm{Dunleavy}, \binits{E.}},
\oauthor{\bsnm{White}, \binits{S.}}:
Literacy in everyday life: Results from the 2003 national assessment of adult
  literacy
(2017).
(Accessed on 09/23/2022)
\end{botherref}
\endbibitem

\bibitem{ferrer2006effect}
\begin{barticle}
\bauthor{\bsnm{Ferrer}, \binits{A.}},
\bauthor{\bsnm{Green}, \binits{D.A.}},
\bauthor{\bsnm{Riddell}, \binits{W.C.}}:
\batitle{The effect of literacy on immigrant earnings}.
\bjtitle{Journal of Human Resources}
\bvolume{41}(\bissue{2}),
\bfpage{380}--\blpage{410}
(\byear{2006})
\end{barticle}
\endbibitem

\bibitem{bonikowska2008literacy}
\begin{bbook}
\bauthor{\bsnm{Bonikowska}, \binits{A.}},
\bauthor{\bsnm{Green}, \binits{D.A.}},
\bauthor{\bsnm{Riddell}, \binits{W.C.}}:
\bbtitle{Literacy and the Labour Market: Cognitive Skills and Immigrant
  Earnings}.
\bpublisher{Statistics Canada},
\blocation{Ottawa}
(\byear{2008})
\end{bbook}
\endbibitem

\bibitem{schwerdt2018literacy}
\begin{botherref}
\oauthor{\bsnm{Schwerdt}, \binits{G.}},
\oauthor{\bsnm{Wiederhold}, \binits{S.}},
\oauthor{\bsnm{Murray}, \binits{T.S.}}:
{Literacy and growth: New evidence from PIAAC}.
Retrieved from PIAAC Gateway website: \url{http://piaacgateway.com/}.
(Accessed on 06/30/2020)
(2020)
\end{botherref}
\endbibitem

\bibitem{dewalt2005health}
\begin{barticle}
\bauthor{\bsnm{Dewalt}, \binits{D.}},
\bauthor{\bsnm{Berkman}, \binits{N.}},
\bauthor{\bsnm{Sheridan}, \binits{S.}},
\bauthor{\bsnm{Lohr}, \binits{K.}},
\bauthor{\bsnm{Pignone}, \binits{M.}}:
\batitle{Literacy and health outcomes: A systematic review of the literature}.
\bjtitle{Journal of general internal medicine}
\bvolume{19},
\bfpage{1228}--\blpage{39}
(\byear{2005}).
doi:\doiurl{10.1111/j.1525–1497.2004.40153.x}
\end{barticle}
\endbibitem

\bibitem{OECD:2013}
\begin{botherref}
\oauthor{\bsnm{Desjardins}, \binits{R.}},
\oauthor{\bsnm{Thorn}, \binits{W.}},
\oauthor{\bsnm{Schleicher}, \binits{A.}},
\oauthor{\bsnm{Quintini}, \binits{G.}},
\oauthor{\bsnm{Pellizzari}, \binits{M.}},
\oauthor{\bsnm{Kis}, \binits{V.}},
\oauthor{\bsnm{Chung}, \binits{J.E.}}:
{OECD Skills Outlook 2013: First Results from the Survey of Adult Skills}.
\url{http://dx.doi.org/10.1787/9789264204256-en}.
{Accessed on 11/23/2020}
(2013)
\end{botherref}
\endbibitem

\bibitem{gerger2008}
\begin{botherref}
\oauthor{\bsnm{Gerger}, \binits{C.}}:
Social linguistics and literacies: Ideology in discourses. social linguistics
  and literacies: Ideology in discourses.
Ilha do Desterro
(2008)
\end{botherref}
\endbibitem

\bibitem{unesco2017}
\begin{botherref}
\oauthor{\bsnm{{UNESCO Institute for Statistics}}}:
Literacy Rates Continue to Rise from One Generation to the Next.
Retrieved September 22, 2022 from
  \url{http://uis.unesco.org/sites/default/files/documents/fs45-literacy-rates-continue-rise-generation-to-next-en-2017_0.pdf}
(2017)
\end{botherref}
\endbibitem

\bibitem{mundial2016education}
\begin{botherref}
\oauthor{\bsnm{Mundial}, \binits{G.B.}},
\oauthor{\bsnm{UNICEF}}, et al.:
Education 2030: Incheon declaration and framework for action: towards inclusive
  and equitable quality education and lifelong learning for all
(2016)
\end{botherref}
\endbibitem

\bibitem{bach2018poverty}
\begin{barticle}
\bauthor{\bsnm{Bach}, \binits{A.J.}},
\bauthor{\bsnm{Wolfson}, \binits{T.}},
\bauthor{\bsnm{Crowell}, \binits{J.K.}}:
\batitle{Poverty, literacy, and social transformation: An interdisciplinary
  exploration of the digital divide.}
\bjtitle{Journal of Media Literacy Education}
\bvolume{10}(\bissue{1}),
\bfpage{22}--\blpage{41}
(\byear{2018})
\end{barticle}
\endbibitem

\bibitem{Meta_2022Q3_report}
\begin{botherref}
\oauthor{\bsnm{Relations}, \binits{M.I.}}:
{Meta Reports Third Quarter 2022 Results}.
\url{https://investor.fb.com/investor-news/press-release-details/2022/Meta-Reports-Third-Quarter-2022-Results/default.aspx}
(2022)
\end{botherref}
\endbibitem

\bibitem{Meta_2022Q3_earnings}
\begin{botherref}
\oauthor{\bsnm{Relations}, \binits{M.I.}}:
{Meta Earnings Presentation Q3 2022}.
\url{https://s21.q4cdn.com/399680738/files/doc_financials/2022/q3/Q3-2022_Earnings-Presentation.pdf}
(2022)
\end{botherref}
\endbibitem

\bibitem{rammstedt2016introduction}
\begin{barticle}
\bauthor{\bsnm{Rammstedt}, \binits{B.}},
\bauthor{\bsnm{Maehler}, \binits{D.B.}}:
\batitle{{Introduction: PIAAC and its Methodological Challenges}}.
\bjtitle{methods, data, analyses}
\bvolume{8}(\bissue{2}),
\bfpage{12}
(\byear{2016})
\end{barticle}
\endbibitem

\bibitem{SDG17}
\begin{botherref}
\oauthor{\bsnm{{SDG17}}}:
{United Nations Sustainable Development Goals}.
{\url{https://sdgs.un.org/goals}}.
{Accessed on 10/15/2020}
(2015)
\end{botherref}
\endbibitem

\bibitem{50Yearso58:online}
\begin{botherref}
\oauthor{\bsnm{Montoya}, \binits{S.}}:
{50 Years of International Literacy Day: Time to Develop New Literacy Data |
  UNESCO UIS}.
\url{http://uis.unesco.org/en/blog/50-years-international-literacy-day-time-develop-new-literacy-data}.
(Accessed on 06/30/2020)
(2016)
\end{botherref}
\endbibitem

\bibitem{unesco-literacy-data}
\begin{botherref}
\oauthor{\bsnm{{UNESCO Institute for Statistics}}}:
UIS Statistics.
Retrieved September 22, 2022 from
  \url{http://data.uis.unesco.org/index.aspx?queryid=3445##}
(2022)
\end{botherref}
\endbibitem

\bibitem{PIAAC}
\begin{botherref}
\oauthor{\bsnm{NCES}}:
{Program for the International Assessment of Adult Competencies (PIAAC)}.
\url{https://nces.ed.gov/surveys/piaac/}.
(Accessed on 09/23/2022)
(2012)
\end{botherref}
\endbibitem

\bibitem{hargittai2009update}
\begin{barticle}
\bauthor{\bsnm{Hargittai}, \binits{E.}}:
\batitle{An update on survey measures of web-oriented digital literacy}.
\bjtitle{Social science computer review}
\bvolume{27}(\bissue{1}),
\bfpage{130}--\blpage{137}
(\byear{2009})
\end{barticle}
\endbibitem

\bibitem{dimaggio2001digital}
\begin{barticle}
\bauthor{\bsnm{DiMaggio}, \binits{P.}},
\bauthor{\bsnm{Hargittai}, \binits{E.}}, \betal:
\batitle{From the ``digital divide'' to ``digital inequality'': Studying
  internet use as penetration increases}.
\bjtitle{Princeton: Center for Arts and Cultural Policy Studies, Woodrow Wilson
  School, Princeton University}
\bvolume{4}(\bissue{1}),
\bfpage{4}--\blpage{2}
(\byear{2001})
\end{barticle}
\endbibitem

\bibitem{mckinsey2014}
\begin{botherref}
\oauthor{\bsnm{Sprague}, \binits{K.}},
\oauthor{\bsnm{Grijpink}, \binits{F.}},
\oauthor{\bsnm{Manyika}, \binits{J.}},
\oauthor{\bsnm{Moodley}, \binits{L.}},
\oauthor{\bsnm{Chappuis}, \binits{B.}},
\oauthor{\bsnm{Pattabiraman}, \binits{K.}},
\oauthor{\bsnm{Bughin}, \binits{J.}}:
Offline and falling behind: Barriers to Internet adoption.
Retrieved September 22, 2022 from
  \url{https://www.mckinsey.com/~/media/mckinsey/dotcom/client_service/high\%20tech/pdfs/offline_and_falling_behind_full_report.ashx}
(2014)
\end{botherref}
\endbibitem

\bibitem{Theriseo30:online}
\begin{botherref}
\oauthor{\bsnm{Ortiz-Ospina}, \binits{E.}}:
{The rise of social media - Our World in Data}.
\url{https://ourworldindata.org/rise-of-social-media}.
(Accessed on 11/20/2021)
(2019)
\end{botherref}
\endbibitem

\bibitem{Erstad-2007}
\begin{barticle}
\bauthor{\bsnm{Erstad}, \binits{O.}},
\bauthor{\bsnm{Gilje}, \binits{N.}},
\bauthor{\bparticle{de} \bsnm{Lange}, \binits{T.}}:
\batitle{Re-mixing multimodal resources: Multiliteracies and digital production
  in norwegian media education}.
\bjtitle{Learning, Media and Technology}
\bvolume{32},
\bfpage{183}--\blpage{198}
(\byear{2007}).
doi:\doiurl{10.1080/17439880701343394}
\end{barticle}
\endbibitem

\bibitem{Alvermann-2008}
\begin{barticle}
\bauthor{\bsnm{Alvermann}, \binits{D.E.}}:
\batitle{Why bother theorizing adolescents' online literacies for classroom
  practice and research?}
\bjtitle{Journal of Adolescent \& Adult Literacy}
\bvolume{52}(\bissue{1}),
\bfpage{8}--\blpage{19}
(\byear{2008}).
Accessed 2022-10-01
\end{barticle}
\endbibitem

\bibitem{Greenhow-2009}
\begin{barticle}
\bauthor{\bsnm{Greenhow}, \binits{C.}},
\bauthor{\bsnm{Robelia}, \binits{B.}}:
\batitle{Old communication, new literacies: Social network sites as social
  learning resources}.
\bjtitle{J. Computer-Mediated Communication}
\bvolume{14},
\bfpage{1130}--\blpage{1161}
(\byear{2009}).
doi:\doiurl{10.1111/j.1083-6101.2009.01484.x}
\end{barticle}
\endbibitem

\bibitem{Perkel-2010}
\begin{botherref}
\oauthor{\bsnm{Perkel}, \binits{D.}}:
Copy and paste literacy? literacy practices in the production of a myspace
  profile.
Informal Learning and Digital Media
\textbf{49}
(2010).
doi:\doiurl{10.1590/S0103-18132010000200011}
\end{botherref}
\endbibitem

\bibitem{Greenhow-2012}
\begin{barticle}
\bauthor{\bsnm{Greenhow}, \binits{C.}},
\bauthor{\bsnm{Gleason}, \binits{B.}}:
\batitle{Twitteracy: Tweeting as a new literacy practice}.
\bjtitle{The Educational Forum}
\bvolume{76}(\bissue{4}),
\bfpage{464}--\blpage{478}
(\byear{2012}).
doi:\doiurl{10.1080/00131725.2012.709032}
\end{barticle}
\endbibitem

\bibitem{Davies-2012}
\begin{barticle}
\bauthor{\bsnm{Davies}, \binits{J.}}:
\batitle{Facework on facebook as a new literacy practice}.
\bjtitle{Computers \& Education}
\bvolume{59}(\bissue{1}),
\bfpage{19}--\blpage{29}
(\byear{2012}).
doi:\doiurl{10.1016/j.compedu.2011.11.007}.
\bcomment{CAL 2011}
\end{barticle}
\endbibitem

\bibitem{Kress-2003}
\begin{botherref}
\oauthor{\bsnm{Kress}, \binits{G.}}:
Literacy in the new media age,
1--190
(2003).
doi:\doiurl{10.4324/9780203299234}
\end{botherref}
\endbibitem

\bibitem{Clark-2009}
\begin{botherref}
\oauthor{\bsnm{Clark}, \binits{C.}},
\oauthor{\bsnm{Dugdale}, \binits{G.}}:
{People's Writing: Attitudes, behaviour and the role of technology}.
\url{https://files.eric.ed.gov/fulltext/ED510271.pdf}.
(Accessed on 09/30/2022)
(2009)
\end{botherref}
\endbibitem

\bibitem{Sabaruddin-2019}
\begin{barticle}
\bauthor{\bsnm{Sabaruddin}}:
\batitle{Facebook utilisation to enhance english writing skill}.
\bjtitle{English Language Teaching}
\bvolume{12}(\bissue{8}),
\bfpage{37}--\blpage{43}
(\byear{2019})
\end{barticle}
\endbibitem

\bibitem{Black-2008}
\begin{bchapter}
\bauthor{\bsnm{Black}, \binits{R.W.}}:
\bctitle{Just don't call them cartoons: The new literacy spaces of anime,
  manga, and fanfiction}.
In: \beditor{\bsnm{Coiro}, \binits{J.}},
\beditor{\bsnm{Knobel}, \binits{M.}},
\beditor{\bsnm{Lankshear}, \binits{C.}},
\beditor{\bsnm{Leu}, \binits{D.J.}} (eds.)
\bbtitle{Handbook of Research on New Literacies},
pp. \bfpage{583}--\blpage{610}.
\bpublisher{Taylor \& Francis},
\blocation{New York}
(\byear{2008})
\end{bchapter}
\endbibitem

\bibitem{Kojo-2018}
\begin{barticle}
\bauthor{\bsnm{Kojo}, \binits{D.B.}},
\bauthor{\bsnm{Agyekum}, \binits{B.O.}},
\bauthor{\bsnm{Arthur}, \binits{B.}}:
\batitle{Exploring the effects of social media on the reading culture of
  students in tamale technical university}.
\bjtitle{Journal of Education and Practice}
\bvolume{9}(\bissue{7}),
\bfpage{47}--\blpage{56}
(\byear{2018})
\end{barticle}
\endbibitem

\bibitem{Vosoughi-2018-science}
\begin{barticle}
\bauthor{\bsnm{Vosoughi}, \binits{S.}},
\bauthor{\bsnm{Roy}, \binits{D.}},
\bauthor{\bsnm{Aral}, \binits{S.}}:
\batitle{The spread of true and false news online}.
\bjtitle{Science}
\bvolume{359}(\bissue{6380}),
\bfpage{1146}--\blpage{1151}
(\byear{2018}).
doi:\doiurl{10.1126/science.aap9559}.
\arxivurl{https://www.science.org/doi/pdf/10.1126/science.aap9559}
\end{barticle}
\endbibitem

\bibitem{Edelson-2021}
\begin{bchapter}
\bauthor{\bsnm{Edelson}, \binits{L.}},
\bauthor{\bsnm{Nguyen}, \binits{M.-K.}},
\bauthor{\bsnm{Goldstein}, \binits{I.}},
\bauthor{\bsnm{Goga}, \binits{O.}},
\bauthor{\bsnm{McCoy}, \binits{D.}},
\bauthor{\bsnm{Lauinger}, \binits{T.}}:
\bctitle{Understanding engagement with u.s. (mis)information news sources on
  facebook}.
In: \bbtitle{Proceedings of the 21st ACM Internet Measurement Conference}.
\bsertitle{IMC '21},
pp. \bfpage{444}--\blpage{463}.
\bpublisher{Association for Computing Machinery},
\blocation{New York, NY, USA}
(\byear{2021}).
doi:\doiurl{10.1145/3487552.3487859}.
\burl{https://doi.org/10.1145/3487552.3487859}
\end{bchapter}
\endbibitem

\bibitem{OECD-2021-pisa}
\begin{bbook}
\bauthor{\bsnm{OECD}}:
\bbtitle{21st-Century Readers: Developing Literacy Skills in a Digital World}.
\bpublisher{{OECD Publishing}},
\blocation{\url{https://www.oecd-ilibrary.org/content/publication/a83d84cb-en}}
(\byear{2021})
\end{bbook}
\endbibitem

\bibitem{lee2011size}
\begin{barticle}
\bauthor{\bsnm{Lee}, \binits{J.}}:
\batitle{Size matters: Early vocabulary as a predictor of language and literacy
  competence}.
\bjtitle{Applied Psycholinguistics}
\bvolume{32}(\bissue{1}),
\bfpage{69}
(\byear{2011})
\end{barticle}
\endbibitem

\bibitem{curtis2006role}
\begin{barticle}
\bauthor{\bsnm{Curtis}, \binits{M.E.}}:
\batitle{The role of vocabulary instruction in adult basic education}.
\bjtitle{Comings, J., Garner, B., Smith, C., Review of Adult Learning and
  Literacy}
\bvolume{6},
\bfpage{43}--\blpage{69}
(\byear{2006})
\end{barticle}
\endbibitem

\bibitem{ouellette2006s}
\begin{barticle}
\bauthor{\bsnm{Ouellette}, \binits{G.P.}}:
\batitle{What's meaning got to do with it: The role of vocabulary in word
  reading and reading comprehension.}
\bjtitle{Journal of educational psychology}
\bvolume{98}(\bissue{3}),
\bfpage{554}
(\byear{2006})
\end{barticle}
\endbibitem

\bibitem{national2012improving}
\begin{bbook}
\bauthor{\bsnm{{National Research Council}}}:
\bbtitle{Improving Adult Literacy Instruction: Options for Practice and
  Research}.
\bpublisher{National Academies Press},
\blocation{Washington, DC}
(\byear{2012}).
doi:\doiurl{10.17226/13242}
\end{bbook}
\endbibitem

\bibitem{schmitt2013introduction}
\begin{bbook}
\bauthor{\bsnm{Schmitt}, \binits{N.}}:
\bbtitle{An Introduction to Applied Linguistics}.
\bpublisher{Routledge},
\blocation{New York}
(\byear{2013})
\end{bbook}
\endbibitem

\bibitem{lauren1989special}
\begin{bbook}
\bauthor{\bsnm{Laur{\'e}n}, \binits{C.}},
\bauthor{\bsnm{Nordman}, \binits{M.}}:
\bbtitle{{Special Language: From Humans Thinking to Thinking Machines}}.
\bpublisher{Multilingual Matters},
\blocation{Clevedon, Philadelphia}
(\byear{1989})
\end{bbook}
\endbibitem

\bibitem{nation2006large}
\begin{barticle}
\bauthor{\bsnm{Nation}, \binits{I.}}:
\batitle{How large a vocabulary is needed for reading and listening?}
\bjtitle{Canadian modern language review}
\bvolume{63}(\bissue{1}),
\bfpage{59}--\blpage{82}
(\byear{2006})
\end{barticle}
\endbibitem

\bibitem{beglar2007vocabulary}
\begin{barticle}
\bauthor{\bsnm{Beglar}, \binits{D.}},
\bauthor{\bsnm{Nation}, \binits{P.}}:
\batitle{A vocabulary size test}.
\bjtitle{The language teacher}
\bvolume{31}(\bissue{7}),
\bfpage{9}--\blpage{13}
(\byear{2007})
\end{barticle}
\endbibitem

\bibitem{grave2018learning}
\begin{bchapter}
\bauthor{\bsnm{Grave}, \binits{{\'E}.}},
\bauthor{\bsnm{Bojanowski}, \binits{P.}},
\bauthor{\bsnm{Gupta}, \binits{P.}},
\bauthor{\bsnm{Joulin}, \binits{A.}},
\bauthor{\bsnm{Mikolov}, \binits{T.}}:
\bctitle{Learning word vectors for 157 languages}.
In: \bbtitle{Proceedings of the Eleventh International Conference on Language
  Resources and Evaluation (LREC 2018)}
(\byear{2018})
\end{bchapter}
\endbibitem

\bibitem{google-ngrams}
\begin{bchapter}
\bauthor{\bsnm{Goldberg}, \binits{Y.}},
\bauthor{\bsnm{Orwant}, \binits{J.}}:
\bctitle{A dataset of syntactic-ngrams over time from a very large corpus of
  english books}.
In: \bbtitle{Second Joint Conference on Lexical and Computational Semantics},
\bconflocation{Atlanta, Georgia, USA},
pp. \bfpage{241}--\blpage{247}
(\byear{2013})
\end{bchapter}
\endbibitem

\bibitem{satopaa2011finding}
\begin{bchapter}
\bauthor{\bsnm{Satopaa}, \binits{V.}},
\bauthor{\bsnm{Albrecht}, \binits{J.}},
\bauthor{\bsnm{Irwin}, \binits{D.}},
\bauthor{\bsnm{Raghavan}, \binits{B.}}:
\bctitle{{Finding a ``kneedle'' in a haystack: Detecting knee points in system
  behavior}}.
In: \bbtitle{2011 31st {ICDCSW}},
pp. \bfpage{166}--\blpage{171}
(\byear{2011}).
\bcomment{IEEE}
\end{bchapter}
\endbibitem

\bibitem{antunes2018knee}
\begin{bchapter}
\bauthor{\bsnm{Antunes}, \binits{M.}},
\bauthor{\bsnm{Gomes}, \binits{D.}},
\bauthor{\bsnm{Aguiar}, \binits{R.L.}}:
\bctitle{Knee/elbow estimation based on first derivative threshold}.
In: \bbtitle{2018 IEEE Fourth International Conference on Big Data Computing
  Service and Applications (BigDataService)},
pp. \bfpage{237}--\blpage{240}
(\byear{2018}).
\bcomment{IEEE}
\end{bchapter}
\endbibitem

\bibitem{beasley2009rank}
\begin{barticle}
\bauthor{\bsnm{Beasley}, \binits{T.M.}},
\bauthor{\bsnm{Erickson}, \binits{S.}},
\bauthor{\bsnm{Allison}, \binits{D.B.}}:
\batitle{Rank-based inverse normal transformations are increasingly used, but
  are they merited?}
\bjtitle{Behavior genetics}
\bvolume{39}(\bissue{5}),
\bfpage{580}
(\byear{2009})
\end{barticle}
\endbibitem

\bibitem{fatehkia2018using}
\begin{barticle}
\bauthor{\bsnm{Fatehkia}, \binits{M.}},
\bauthor{\bsnm{Kashyap}, \binits{R.}},
\bauthor{\bsnm{Weber}, \binits{I.}}:
\batitle{{Using Facebook ad data to track the global digital gender gap}}.
\bjtitle{World Development}
\bvolume{107},
\bfpage{189}--\blpage{209}
(\byear{2018})
\end{barticle}
\endbibitem

\bibitem{owidinternet}
\begin{botherref}
\oauthor{\bsnm{Roser}, \binits{M.}},
\oauthor{\bsnm{Ritchie}, \binits{H.}},
\oauthor{\bsnm{Ortiz-Ospina}, \binits{E.}}:
Internet.
Our World in Data
(2022).
https://ourworldindata.org/internet
\end{botherref}
\endbibitem

\bibitem{world2020global}
\begin{bchapter}
\bauthor{\bsnm{{World Economic Forum}}}:
\bctitle{The global gender gap report}.
(\byear{2020}).
\bcomment{World Economic Forum Genebra}
\end{bchapter}
\endbibitem

\bibitem{pisa-2018-gender}
\begin{bbook}
\bauthor{\bsnm{OECD}}:
\bbtitle{PISA 2018 Results (Volume II)},
p. \bfpage{376}
(\byear{2019}).
doi:\doiurl{10.1787/b5fd1b8f-en}.
\burl{https://www.oecd-ilibrary.org/content/publication/b5fd1b8f-en}
\end{bbook}
\endbibitem

\bibitem{gender-inequality-index}
\begin{botherref}
\oauthor{\bsnm{{United Nations Development Programme}}}:
{Technical Notes: Calculating the human development indices—graphical
  presentation}.
Retrieved from UNDP website:
  \url{https://hdr.undp.org/sites/default/files/2021-22_HDR/hdr2021-22_technical_notes.pdf}.
(Accessed on 10/26/2022)
(2021)
\end{botherref}
\endbibitem

\bibitem{coppedge2019v}
\begin{barticle}
\bauthor{\bsnm{Coppedge}, \binits{M.}},
\bauthor{\bsnm{Gerring}, \binits{J.}},
\bauthor{\bsnm{Knutsen}, \binits{C.H.}},
\bauthor{\bsnm{Krusell}, \binits{J.}},
\bauthor{\bsnm{Medzihorsky}, \binits{J.}},
\bauthor{\bsnm{Pernes}, \binits{J.}},
\bauthor{\bsnm{Skaaning}, \binits{S.-E.}},
\bauthor{\bsnm{Stepanova}, \binits{N.}},
\bauthor{\bsnm{Teorell}, \binits{J.}},
\bauthor{\bsnm{Tzelgov}, \binits{E.}}, \betal:
\batitle{The methodology of “varieties of democracy”(v-dem)}.
\bjtitle{Bulletin of Sociological Methodology/Bulletin de M{\'e}thodologie
  Sociologique}
\bvolume{143}(\bissue{1}),
\bfpage{107}--\blpage{133}
(\byear{2019})
\end{barticle}
\endbibitem

\bibitem{warf2020geographies}
\begin{bbook}
\bauthor{\bsnm{Warf}, \binits{B.}}:
\bbtitle{{Geographies of the Internet}}.
\bpublisher{Routledge},
\blocation{New York}
(\byear{2020})
\end{bbook}
\endbibitem

\bibitem{wdr:2016}
\begin{bchapter}
\bauthor{\bsnm{{World Bank}}}:
\bctitle{World development report 2016: Digital dividends}.
(\byear{2016}).
\bcomment{Washington, DC: World Bank}
\end{bchapter}
\endbibitem

\bibitem{pew:2012}
\begin{botherref}
\oauthor{\bsnm{Zickuhr}, \binits{K.}},
\oauthor{\bsnm{Smith}, \binits{A.}}:
Digital differences.
Retrieved December 7, 2020 from
  \url{https://www.pewresearch.org/internet/2012/04/13/digital-differences/}
(2012)
\end{botherref}
\endbibitem

\bibitem{Hilbert:2011}
\begin{barticle}
\bauthor{\bsnm{Hilbert}, \binits{M.}}:
\batitle{Digital gender divide or technologically empowered women in developing
  countries? a typical case of lies, damned lies, and statistics}.
\bjtitle{Women's Studies International Forum}
\bvolume{34}(\bissue{6}),
\bfpage{479}--\blpage{489}
(\byear{2011}).
doi:\doiurl{10.1016/j.wsif.2011.07.001}
\end{barticle}
\endbibitem

\bibitem{yamamoto2013scaling}
\begin{botherref}
\oauthor{\bsnm{Yamamoto}, \binits{K.}},
\oauthor{\bsnm{Khorramdel}, \binits{L.}},
\oauthor{\bsnm{Von~Davier}, \binits{M.}}, et al.:
{Scaling PIAAC Cognitive Data}
\end{botherref}
\endbibitem

\bibitem{Facebook2020Q2earning:online}
\begin{botherref}
\oauthor{\bsnm{{Facebook}}}:
{Facebook Q2 2020 Earnings}.
\url{https://s21.q4cdn.com/399680738/files/doc_financials/2020/q2/Q2-2020-FB-Earnings-Presentation.pdf}.
(Accessed on 10/15/2020)
(2020)
\end{botherref}
\endbibitem

\bibitem{magno2014international}
\begin{bchapter}
\bauthor{\bsnm{Magno}, \binits{G.}},
\bauthor{\bsnm{Weber}, \binits{I.}}:
\bctitle{International gender differences and gaps in online social networks}.
In: \bbtitle{International Conference on Social Informatics},
pp. \bfpage{121}--\blpage{138}
(\byear{2014}).
\bcomment{Springer}
\end{bchapter}
\endbibitem

\bibitem{kashyap2021analysing}
\begin{barticle}
\bauthor{\bsnm{Kashyap}, \binits{R.}},
\bauthor{\bsnm{Verkroost}, \binits{F.C.}}:
\batitle{Analysing global professional gender gaps using linkedin advertising
  data}.
\bjtitle{EPJ Data Science}
\bvolume{10}(\bissue{1}),
\bfpage{39}
(\byear{2021})
\end{barticle}
\endbibitem

\bibitem{park2019global}
\begin{barticle}
\bauthor{\bsnm{Park}, \binits{M.}},
\bauthor{\bsnm{Thom}, \binits{J.}},
\bauthor{\bsnm{Mennicken}, \binits{S.}},
\bauthor{\bsnm{Cramer}, \binits{H.}},
\bauthor{\bsnm{Macy}, \binits{M.}}:
\batitle{Global music streaming data reveal diurnal and seasonal patterns of
  affective preference}.
\bjtitle{Nature human behaviour}
\bvolume{3}(\bissue{3}),
\bfpage{230}--\blpage{236}
(\byear{2019})
\end{barticle}
\endbibitem

\bibitem{masrai2019vocabulary}
\begin{barticle}
\bauthor{\bsnm{Masrai}, \binits{A.}}:
\batitle{Vocabulary and reading comprehension revisited: Evidence for high-,
  mid-, and low-frequency vocabulary knowledge}.
\bjtitle{Sage Open}
\bvolume{9}(\bissue{2}),
\bfpage{2158244019845182}
(\byear{2019})
\end{barticle}
\endbibitem

\bibitem{hastie1990generalized}
\begin{bbook}
\bauthor{\bsnm{Hastie}, \binits{T.J.}},
\bauthor{\bsnm{Tibshirani}, \binits{R.J.}}:
\bbtitle{Generalized Additive Models}
vol. \bseriesno{43}.
\bpublisher{CRC press}, \blocation{???}
(\byear{1990})
\end{bbook}
\endbibitem

\bibitem{Censusof39:online}
\begin{botherref}
\oauthor{\bsnm{{Office of the Registrar General \& Census Commissioner,
  India}}}:
Census of India: Literacy And Level of Education.
\url{https://censusindia.gov.in/census_and_you/literacy_and_level_of_education.aspx}.
(Accessed on 10/28/2020)
(2011)
\end{botherref}
\endbibitem

\bibitem{Facebook15:online}
\begin{botherref}
\oauthor{\bsnm{{World Population}}}:
{Facebook Users by Country 2022}.
\url{https://worldpopulationreview.com/country-rankings/facebook-users-by-country}.
(Accessed on 10/25/2022)
(2022)
\end{botherref}
\endbibitem

\bibitem{unesco2019produce}
\begin{botherref}
How to produce and use the global and thematic education indicators.
UNESCO Institute for Statistics, Montreal, Quebec.
(Accessed on 07/01/2020)
(2019)
\end{botherref}
\endbibitem

\bibitem{BarroLee73:online}
\begin{botherref}
{Barro-Lee Educational Attainment Dataset}.
\url{http://www.barrolee.com/}.
(Accessed on 07/01/2020)
(2010)
\end{botherref}
\endbibitem

\bibitem{Aboutthe43:online}
\begin{botherref}
{Using Digital Traces to Measure Digital Gender Inequality in Real-Time}.
\url{https://www.digitalgendergaps.org/project/}.
(Accessed on 07/01/2020)
(2020)
\end{botherref}
\endbibitem

\bibitem{HDROAPII26:online}
\begin{botherref}
{HDRO API Information}.
\url{http://ec2-54-174-131-205.compute-1.amazonaws.com/API/Information.php}.
(Accessed on 07/01/2020)
(2018)
\end{botherref}
\endbibitem

\bibitem{GlobalDa32:online}
\begin{botherref}
{Global Data Lab - Innovative Instruments for Turning Data into Knowledge}.
\url{https://globaldatalab.org/}.
(Accessed on 07/01/2020)
(2018)
\end{botherref}
\endbibitem

\bibitem{HumanDev90:online}
\begin{botherref}
{Human Development Reports}.
\url{http://hdr.undp.org/en/composite/IHDI}.
(Accessed on 07/01/2020)
(2017)
\end{botherref}
\endbibitem

\end{thebibliography}

\newcommand{\BMCxmlcomment}[1]{}

\BMCxmlcomment{

<refgrp>

<bibl id="B1">
  <title><p>Adult Literacy in America: A First Look at the Findings of the
  National Adult Literacy Survey</p></title>
  <aug>
    <au><snm>Kirsch</snm><fnm>IS</fnm></au>
    <au><snm>Jungeblut</snm><fnm>A</fnm></au>
    <au><snm>Jenkins</snm><fnm>L</fnm></au>
    <au><snm>Kolstad</snm><fnm>A</fnm></au>
  </aug>
  <pubdate>2002</pubdate>
  <note>Access on 09/23/2022 at
  \url{https://nces.ed.gov/pubs93/93275.pdf}</note>
</bibl>

<bibl id="B2">
  <title><p>{Literacy in Everyday Life: Results from the 2003 National
  Assessment of Adult Literacy. NCES 2007-490.}</p></title>
  <aug>
    <au><snm>Kutner</snm><fnm>M</fnm></au>
    <au><snm>Greenberg</snm><fnm>E</fnm></au>
    <au><snm>Jin</snm><fnm>Y</fnm></au>
    <au><snm>Boyle</snm><fnm>B</fnm></au>
    <au><snm>Hsu</snm><fnm>Yc</fnm></au>
    <au><snm>Dunleavy</snm><fnm>E</fnm></au>
  </aug>
  <source>Report</source>
  <pubdate>2007</pubdate>
</bibl>

<bibl id="B3">
  <title><p>Education policy and equality of opportunity</p></title>
  <aug>
    <au><snm>Sch{\"u}tz</snm><fnm>G</fnm></au>
    <au><snm>Ursprung</snm><fnm>HW</fnm></au>
    <au><snm>W{\"o}{\ss}mann</snm><fnm>L</fnm></au>
  </aug>
  <source>Kyklos</source>
  <publisher>Wiley Online Library</publisher>
  <pubdate>2008</pubdate>
  <volume>61</volume>
  <issue>2</issue>
  <fpage>279</fpage>
  <lpage>-308</lpage>
</bibl>

<bibl id="B4">
  <title><p>Literacy in Everyday Life: Results From the 2003 National
  Assessment of Adult Literacy</p></title>
  <aug>
    <au><snm>Kutner</snm><fnm>M</fnm></au>
    <au><snm>Greenberg</snm><fnm>E</fnm></au>
    <au><snm>Jin</snm><fnm>Y</fnm></au>
    <au><snm>Boyle</snm><fnm>B</fnm></au>
    <au><snm>Hsu</snm><fnm>Y</fnm></au>
    <au><snm>Dunleavy</snm><fnm>E</fnm></au>
    <au><snm>White</snm><fnm>S</fnm></au>
  </aug>
  <source>\url{https://nces.ed.gov/Pubs2007/2007480.pdf}</source>
  <pubdate>2017</pubdate>
  <note>(Accessed on 09/23/2022)</note>
</bibl>

<bibl id="B5">
  <title><p>The effect of literacy on immigrant earnings</p></title>
  <aug>
    <au><snm>Ferrer</snm><fnm>A</fnm></au>
    <au><snm>Green</snm><fnm>DA</fnm></au>
    <au><snm>Riddell</snm><fnm>WC</fnm></au>
  </aug>
  <source>Journal of Human Resources</source>
  <publisher>University of Wisconsin Press</publisher>
  <pubdate>2006</pubdate>
  <volume>41</volume>
  <issue>2</issue>
  <fpage>380</fpage>
  <lpage>-410</lpage>
</bibl>

<bibl id="B6">
  <title><p>Literacy and the labour market: Cognitive skills and immigrant
  earnings</p></title>
  <aug>
    <au><snm>Bonikowska</snm><fnm>A</fnm></au>
    <au><snm>Green</snm><fnm>DA</fnm></au>
    <au><snm>Riddell</snm><fnm>WC</fnm></au>
  </aug>
  <publisher>Ottawa: Statistics Canada</publisher>
  <pubdate>2008</pubdate>
</bibl>

<bibl id="B7">
  <title><p>{Literacy and growth: New evidence from PIAAC}</p></title>
  <aug>
    <au><snm>Schwerdt</snm><fnm>G</fnm></au>
    <au><snm>Wiederhold</snm><fnm>S</fnm></au>
    <au><snm>Murray</snm><fnm>TS</fnm></au>
  </aug>
  <source>Retrieved from PIAAC Gateway website:
  \url{http://piaacgateway.com/}</source>
  <pubdate>2020</pubdate>
  <note>(Accessed on 06/30/2020)</note>
</bibl>

<bibl id="B8">
  <title><p>Literacy and health outcomes: A systematic review of the
  literature</p></title>
  <aug>
    <au><snm>Dewalt</snm><fnm>D</fnm></au>
    <au><snm>Berkman</snm><fnm>N</fnm></au>
    <au><snm>Sheridan</snm><fnm>S</fnm></au>
    <au><snm>Lohr</snm><fnm>K</fnm></au>
    <au><snm>Pignone</snm><fnm>M</fnm></au>
  </aug>
  <source>Journal of general internal medicine</source>
  <pubdate>2005</pubdate>
  <volume>19</volume>
  <fpage>1228</fpage>
  <lpage>39</lpage>
</bibl>

<bibl id="B9">
  <title><p>{OECD Skills Outlook 2013: First Results from the Survey of Adult
  Skills}</p></title>
  <aug>
    <au><snm>Desjardins</snm><fnm>R</fnm></au>
    <au><snm>Thorn</snm><fnm>W</fnm></au>
    <au><snm>Schleicher</snm><fnm>A</fnm></au>
    <au><snm>Quintini</snm><fnm>G</fnm></au>
    <au><snm>Pellizzari</snm><fnm>M</fnm></au>
    <au><snm>Kis</snm><fnm>V</fnm></au>
    <au><snm>Chung</snm><fnm>JE</fnm></au>
  </aug>
  <source>\url{http://dx.doi.org/10.1787/9789264204256-en}</source>
  <pubdate>2013</pubdate>
  <note>{Accessed on 11/23/2020}</note>
</bibl>

<bibl id="B10">
  <title><p>Social Linguistics and Literacies: Ideology in Discourses. Social
  Linguistics and Literacies: Ideology in Discourses.</p></title>
  <aug>
    <au><snm>Gerger</snm><fnm>C</fnm></au>
  </aug>
  <source>Ilha do Desterro</source>
  <pubdate>2008</pubdate>
</bibl>

<bibl id="B11">
  <title><p>Literacy Rates Continue to Rise from One Generation to the
  Next</p></title>
  <aug>
    <au><cnm>{UNESCO Institute for Statistics}</cnm></au>
  </aug>
  <source>UIS Fact Sheet No. 45</source>
  <pubdate>2017</pubdate>
  <note>Retrieved September 22, 2022 from
  \url{http://uis.unesco.org/sites/default/files/documents/fs45-literacy-rates-continue-rise-generation-to-next-en-2017_0.pdf}</note>
</bibl>

<bibl id="B12">
  <title><p>Education 2030: Incheon declaration and framework for action:
  towards inclusive and equitable quality education and lifelong learning for
  all</p></title>
  <aug>
    <au><snm>Mundial</snm><fnm>GB</fnm></au>
    <au><cnm>UNICEF</cnm></au>
    <au><cnm>others</cnm></au>
  </aug>
  <publisher>{UNESCO}, 2016</publisher>
  <pubdate>2016</pubdate>
</bibl>

<bibl id="B13">
  <title><p>Poverty, Literacy, and Social Transformation: An Interdisciplinary
  Exploration of the Digital Divide.</p></title>
  <aug>
    <au><snm>Bach</snm><fnm>AJ</fnm></au>
    <au><snm>Wolfson</snm><fnm>T</fnm></au>
    <au><snm>Crowell</snm><fnm>JK</fnm></au>
  </aug>
  <source>Journal of Media Literacy Education</source>
  <publisher>ERIC</publisher>
  <pubdate>2018</pubdate>
  <volume>10</volume>
  <issue>1</issue>
  <fpage>22</fpage>
  <lpage>-41</lpage>
</bibl>

<bibl id="B14">
  <title><p>{Meta Reports Third Quarter 2022 Results}</p></title>
  <aug>
    <au><snm>Relations</snm><fnm>MI</fnm></au>
  </aug>
  <source>\url{https://investor.fb.com/investor-news/press-release-details/2022/Meta-Reports-Third-Quarter-2022-Results/default.aspx}</source>
  <pubdate>2022</pubdate>
</bibl>

<bibl id="B15">
  <title><p>{Meta Earnings Presentation Q3 2022}</p></title>
  <aug>
    <au><snm>Relations</snm><fnm>MI</fnm></au>
  </aug>
  <source>\url{https://s21.q4cdn.com/399680738/files/doc_financials/2022/q3/Q3-2022_Earnings-Presentation.pdf}</source>
  <pubdate>2022</pubdate>
</bibl>

<bibl id="B16">
  <title><p>{Introduction: PIAAC and its Methodological Challenges}</p></title>
  <aug>
    <au><snm>Rammstedt</snm><fnm>B</fnm></au>
    <au><snm>Maehler</snm><fnm>DB</fnm></au>
  </aug>
  <source>methods, data, analyses</source>
  <pubdate>2016</pubdate>
  <volume>8</volume>
  <issue>2</issue>
  <fpage>12</fpage>
</bibl>

<bibl id="B17">
  <title><p>{United Nations Sustainable Development Goals}</p></title>
  <aug>
    <au><cnm>{SDG17}</cnm></au>
  </aug>
  <source>{\url{https://sdgs.un.org/goals}}</source>
  <pubdate>2015</pubdate>
  <note>{Accessed on 10/15/2020}</note>
</bibl>

<bibl id="B18">
  <title><p>{50 Years of International Literacy Day: Time to Develop New
  Literacy Data | UNESCO UIS}</p></title>
  <aug>
    <au><snm>Montoya</snm><fnm>S</fnm></au>
  </aug>
  <source>\url{http://uis.unesco.org/en/blog/50-years-international-literacy-day-time-develop-new-literacy-data}</source>
  <pubdate>2016</pubdate>
  <note>(Accessed on 06/30/2020)</note>
</bibl>

<bibl id="B19">
  <title><p>UIS Statistics</p></title>
  <aug>
    <au><cnm>{UNESCO Institute for Statistics}</cnm></au>
  </aug>
  <source>Retrieved September 22, 2022 from
  \url{http://data.uis.unesco.org/index.aspx?queryid=3445##}</source>
  <pubdate>2022</pubdate>
</bibl>

<bibl id="B20">
  <title><p>{Program for the International Assessment of Adult Competencies
  (PIAAC)}</p></title>
  <aug>
    <au><cnm>NCES</cnm></au>
  </aug>
  <source>\url{https://nces.ed.gov/surveys/piaac/}</source>
  <pubdate>2012</pubdate>
  <note>(Accessed on 09/23/2022)</note>
</bibl>

<bibl id="B21">
  <title><p>An update on survey measures of web-oriented digital
  literacy</p></title>
  <aug>
    <au><snm>Hargittai</snm><fnm>E</fnm></au>
  </aug>
  <source>Social science computer review</source>
  <publisher>Sage Publications Sage CA: Los Angeles, CA</publisher>
  <pubdate>2009</pubdate>
  <volume>27</volume>
  <issue>1</issue>
  <fpage>130</fpage>
  <lpage>-137</lpage>
</bibl>

<bibl id="B22">
  <title><p>From the ``digital divide'' to ``digital inequality'': Studying
  Internet use as penetration increases</p></title>
  <aug>
    <au><snm>DiMaggio</snm><fnm>P</fnm></au>
    <au><snm>Hargittai</snm><fnm>E</fnm></au>
    <au><cnm>others</cnm></au>
  </aug>
  <source>Princeton: Center for Arts and Cultural Policy Studies, Woodrow
  Wilson School, Princeton University</source>
  <pubdate>2001</pubdate>
  <volume>4</volume>
  <issue>1</issue>
  <fpage>4</fpage>
  <lpage>-2</lpage>
</bibl>

<bibl id="B23">
  <title><p>Offline and falling behind: Barriers to Internet
  adoption</p></title>
  <aug>
    <au><snm>Sprague</snm><fnm>K</fnm></au>
    <au><snm>Grijpink</snm><fnm>F</fnm></au>
    <au><snm>Manyika</snm><fnm>J</fnm></au>
    <au><snm>Moodley</snm><fnm>L</fnm></au>
    <au><snm>Chappuis</snm><fnm>B</fnm></au>
    <au><snm>Pattabiraman</snm><fnm>K</fnm></au>
    <au><snm>Bughin</snm><fnm>J</fnm></au>
  </aug>
  <source>Technology, Media, and Telecom Practice</source>
  <pubdate>2014</pubdate>
  <note>Retrieved September 22, 2022 from
  \url{https://www.mckinsey.com/~/media/mckinsey/dotcom/client_service/high\%20tech/pdfs/offline_and_falling_behind_full_report.ashx}</note>
</bibl>

<bibl id="B24">
  <title><p>{The rise of social media - Our World in Data}</p></title>
  <aug>
    <au><snm>Ortiz Ospina</snm><fnm>E</fnm></au>
  </aug>
  <source>\url{https://ourworldindata.org/rise-of-social-media}</source>
  <pubdate>2019</pubdate>
  <note>(Accessed on 11/20/2021)</note>
</bibl>

<bibl id="B25">
  <title><p>Re-mixing multimodal resources: Multiliteracies and digital
  production in Norwegian media education</p></title>
  <aug>
    <au><snm>Erstad</snm><fnm>O</fnm></au>
    <au><snm>Gilje</snm><fnm>N</fnm></au>
    <au><snm>Lange</snm><fnm>T</fnm></au>
  </aug>
  <source>Learning, Media and Technology</source>
  <pubdate>2007</pubdate>
  <volume>32</volume>
  <fpage>183</fpage>
  <lpage>198</lpage>
</bibl>

<bibl id="B26">
  <title><p>Why Bother Theorizing Adolescents' Online Literacies for Classroom
  Practice and Research?</p></title>
  <aug>
    <au><snm>Alvermann</snm><fnm>DE</fnm></au>
  </aug>
  <source>Journal of Adolescent \& Adult Literacy</source>
  <publisher>[Wiley, International Reading Association]</publisher>
  <pubdate>2008</pubdate>
  <volume>52</volume>
  <issue>1</issue>
  <fpage>8</fpage>
  <lpage>-19</lpage>
  <url>http://www.jstor.org/stable/30139646</url>
</bibl>

<bibl id="B27">
  <title><p>Old Communication, New Literacies: Social Network Sites as Social
  Learning Resources</p></title>
  <aug>
    <au><snm>Greenhow</snm><fnm>C</fnm></au>
    <au><snm>Robelia</snm><fnm>B</fnm></au>
  </aug>
  <source>J. Computer-Mediated Communication</source>
  <pubdate>2009</pubdate>
  <volume>14</volume>
  <fpage>1130</fpage>
  <lpage>1161</lpage>
</bibl>

<bibl id="B28">
  <title><p>Copy and paste literacy? Literacy practices in the production of a
  MySpace profile</p></title>
  <aug>
    <au><snm>Perkel</snm><fnm>D</fnm></au>
  </aug>
  <source>Informal Learning and Digital Media</source>
  <pubdate>2010</pubdate>
  <volume>49</volume>
</bibl>

<bibl id="B29">
  <title><p>Twitteracy: Tweeting as a New Literacy Practice</p></title>
  <aug>
    <au><snm>Greenhow</snm><fnm>C</fnm></au>
    <au><snm>Gleason</snm><fnm>B</fnm></au>
  </aug>
  <source>The Educational Forum</source>
  <publisher>Routledge</publisher>
  <pubdate>2012</pubdate>
  <volume>76</volume>
  <issue>4</issue>
  <fpage>464</fpage>
  <lpage>478</lpage>
</bibl>

<bibl id="B30">
  <title><p>Facework on Facebook as a new literacy practice</p></title>
  <aug>
    <au><snm>Davies</snm><fnm>J</fnm></au>
  </aug>
  <source>Computers \& Education</source>
  <pubdate>2012</pubdate>
  <volume>59</volume>
  <issue>1</issue>
  <fpage>19</fpage>
  <lpage>29</lpage>
  <url>https://www.sciencedirect.com/science/article/pii/S0360131511002776</url>
  <note>CAL 2011</note>
</bibl>

<bibl id="B31">
  <title><p>Literacy in the New Media Age</p></title>
  <aug>
    <au><snm>Kress</snm><fnm>G</fnm></au>
  </aug>
  <pubdate>2003</pubdate>
  <fpage>1</fpage>
  <lpage>190</lpage>
</bibl>

<bibl id="B32">
  <title><p>{People's Writing: Attitudes, behaviour and the role of
  technology}</p></title>
  <aug>
    <au><snm>Clark</snm><fnm>C</fnm></au>
    <au><snm>Dugdale</snm><fnm>G</fnm></au>
  </aug>
  <source>\url{https://files.eric.ed.gov/fulltext/ED510271.pdf}</source>
  <pubdate>2009</pubdate>
  <note>(Accessed on 09/30/2022)</note>
</bibl>

<bibl id="B33">
  <title><p>Facebook Utilisation to Enhance English Writing Skill</p></title>
  <aug>
    <au><cnm>Sabaruddin</cnm></au>
  </aug>
  <source>English Language Teaching</source>
  <publisher>Canadian Center of Science and Education</publisher>
  <pubdate>2019</pubdate>
  <volume>12</volume>
  <issue>8</issue>
  <fpage>37</fpage>
  <lpage>43</lpage>
</bibl>

<bibl id="B34">
  <title><p>Just Don't Call Them Cartoons: The New Literacy Spaces of Anime,
  Manga, and Fanfiction</p></title>
  <aug>
    <au><snm>Black</snm><fnm>RW</fnm></au>
  </aug>
  <source>Handbook of Research on New Literacies</source>
  <publisher>New York: Taylor \& Francis</publisher>
  <editor>J. Coiro and M. Knobel and C. Lankshear and D. J. Leu</editor>
  <pubdate>2008</pubdate>
  <fpage>583–610</fpage>
</bibl>

<bibl id="B35">
  <title><p>Exploring the Effects of Social Media on the Reading Culture of
  Students in Tamale Technical University</p></title>
  <aug>
    <au><snm>Kojo</snm><fnm>DB</fnm></au>
    <au><snm>Agyekum</snm><fnm>BO</fnm></au>
    <au><snm>Arthur</snm><fnm>B</fnm></au>
  </aug>
  <source>Journal of Education and Practice</source>
  <pubdate>2018</pubdate>
  <volume>9</volume>
  <issue>7</issue>
  <fpage>47</fpage>
  <lpage>56</lpage>
</bibl>

<bibl id="B36">
  <title><p>The spread of true and false news online</p></title>
  <aug>
    <au><snm>Vosoughi</snm><fnm>S</fnm></au>
    <au><snm>Roy</snm><fnm>D</fnm></au>
    <au><snm>Aral</snm><fnm>S</fnm></au>
  </aug>
  <source>Science</source>
  <pubdate>2018</pubdate>
  <volume>359</volume>
  <issue>6380</issue>
  <fpage>1146</fpage>
  <lpage>1151</lpage>
  <url>https://www.science.org/doi/abs/10.1126/science.aap9559</url>
</bibl>

<bibl id="B37">
  <title><p>Understanding Engagement with U.S. (Mis)Information News Sources on
  Facebook</p></title>
  <aug>
    <au><snm>Edelson</snm><fnm>L</fnm></au>
    <au><snm>Nguyen</snm><fnm>MK</fnm></au>
    <au><snm>Goldstein</snm><fnm>I</fnm></au>
    <au><snm>Goga</snm><fnm>O</fnm></au>
    <au><snm>McCoy</snm><fnm>D</fnm></au>
    <au><snm>Lauinger</snm><fnm>T</fnm></au>
  </aug>
  <source>Proceedings of the 21st ACM Internet Measurement Conference</source>
  <publisher>New York, NY, USA: Association for Computing Machinery</publisher>
  <series><title><p>IMC '21</p></title></series>
  <pubdate>2021</pubdate>
  <fpage>444–463</fpage>
  <url>https://doi.org/10.1145/3487552.3487859</url>
</bibl>

<bibl id="B38">
  <title><p>21st-Century Readers: Developing Literacy Skills in a Digital
  World</p></title>
  <aug>
    <au><cnm>OECD</cnm></au>
  </aug>
  <publisher>\url{https://www.oecd-ilibrary.org/content/publication/a83d84cb-en}:
  {OECD Publishing}</publisher>
  <pubdate>2021</pubdate>
</bibl>

<bibl id="B39">
  <title><p>Size matters: Early vocabulary as a predictor of language and
  literacy competence</p></title>
  <aug>
    <au><snm>Lee</snm><fnm>J</fnm></au>
  </aug>
  <source>Applied Psycholinguistics</source>
  <publisher>Cambridge University Press</publisher>
  <pubdate>2011</pubdate>
  <volume>32</volume>
  <issue>1</issue>
  <fpage>69</fpage>
</bibl>

<bibl id="B40">
  <title><p>The role of vocabulary instruction in adult basic
  education</p></title>
  <aug>
    <au><snm>Curtis</snm><fnm>ME</fnm></au>
  </aug>
  <source>Comings, J., Garner, B., Smith, C., Review of Adult Learning and
  Literacy</source>
  <pubdate>2006</pubdate>
  <volume>6</volume>
  <fpage>43</fpage>
  <lpage>-69</lpage>
</bibl>

<bibl id="B41">
  <title><p>What's meaning got to do with it: The role of vocabulary in word
  reading and reading comprehension.</p></title>
  <aug>
    <au><snm>Ouellette</snm><fnm>GP</fnm></au>
  </aug>
  <source>Journal of educational psychology</source>
  <publisher>American Psychological Association</publisher>
  <pubdate>2006</pubdate>
  <volume>98</volume>
  <issue>3</issue>
  <fpage>554</fpage>
</bibl>

<bibl id="B42">
  <title><p>Improving adult literacy instruction: Options for practice and
  research</p></title>
  <aug>
    <au><cnm>{National Research Council}</cnm></au>
  </aug>
  <publisher>Washington, DC: National Academies Press</publisher>
  <pubdate>2012</pubdate>
</bibl>

<bibl id="B43">
  <title><p>An introduction to applied linguistics</p></title>
  <aug>
    <au><snm>Schmitt</snm><fnm>N</fnm></au>
  </aug>
  <publisher>New York: Routledge</publisher>
  <pubdate>2013</pubdate>
</bibl>

<bibl id="B44">
  <title><p>{Special language: From humans thinking to thinking
  machines}</p></title>
  <aug>
    <au><snm>Laur{\'e}n</snm><fnm>C</fnm></au>
    <au><snm>Nordman</snm><fnm>M</fnm></au>
  </aug>
  <publisher>Clevedon, Philadelphia: Multilingual Matters</publisher>
  <pubdate>1989</pubdate>
</bibl>

<bibl id="B45">
  <title><p>How large a vocabulary is needed for reading and
  listening?</p></title>
  <aug>
    <au><snm>Nation</snm><fnm>I</fnm></au>
  </aug>
  <source>Canadian modern language review</source>
  <publisher>University of Toronto Press</publisher>
  <pubdate>2006</pubdate>
  <volume>63</volume>
  <issue>1</issue>
  <fpage>59</fpage>
  <lpage>-82</lpage>
</bibl>

<bibl id="B46">
  <title><p>A vocabulary size test</p></title>
  <aug>
    <au><snm>Beglar</snm><fnm>D</fnm></au>
    <au><snm>Nation</snm><fnm>P</fnm></au>
  </aug>
  <source>The language teacher</source>
  <pubdate>2007</pubdate>
  <volume>31</volume>
  <issue>7</issue>
  <fpage>9</fpage>
  <lpage>-13</lpage>
</bibl>

<bibl id="B47">
  <title><p>Learning Word Vectors for 157 Languages</p></title>
  <aug>
    <au><snm>Grave</snm><fnm>{\'E}</fnm></au>
    <au><snm>Bojanowski</snm><fnm>P</fnm></au>
    <au><snm>Gupta</snm><fnm>P</fnm></au>
    <au><snm>Joulin</snm><fnm>A</fnm></au>
    <au><snm>Mikolov</snm><fnm>T</fnm></au>
  </aug>
  <source>Proceedings of the Eleventh International Conference on Language
  Resources and Evaluation (LREC 2018)</source>
  <pubdate>2018</pubdate>
</bibl>

<bibl id="B48">
  <title><p>A Dataset of Syntactic-Ngrams over Time from a Very Large Corpus of
  English Books</p></title>
  <aug>
    <au><snm>Goldberg</snm><fnm>Y</fnm></au>
    <au><snm>Orwant</snm><fnm>J</fnm></au>
  </aug>
  <source>Second Joint Conference on Lexical and Computational
  Semantics</source>
  <publisher>Atlanta, Georgia, USA</publisher>
  <pubdate>2013</pubdate>
  <fpage>241</fpage>
  <lpage>247</lpage>
</bibl>

<bibl id="B49">
  <title><p>{Finding a ``kneedle'' in a haystack: Detecting knee points in
  system behavior}</p></title>
  <aug>
    <au><snm>Satopaa</snm><fnm>V</fnm></au>
    <au><snm>Albrecht</snm><fnm>J</fnm></au>
    <au><snm>Irwin</snm><fnm>D</fnm></au>
    <au><snm>Raghavan</snm><fnm>B</fnm></au>
  </aug>
  <source>2011 31st {ICDCSW}</source>
  <pubdate>2011</pubdate>
  <fpage>166</fpage>
  <lpage>-171</lpage>
</bibl>

<bibl id="B50">
  <title><p>Knee/elbow estimation based on first derivative
  threshold</p></title>
  <aug>
    <au><snm>Antunes</snm><fnm>M</fnm></au>
    <au><snm>Gomes</snm><fnm>D</fnm></au>
    <au><snm>Aguiar</snm><fnm>RL</fnm></au>
  </aug>
  <source>2018 IEEE Fourth International Conference on Big Data Computing
  Service and Applications (BigDataService)</source>
  <pubdate>2018</pubdate>
  <fpage>237</fpage>
  <lpage>-240</lpage>
</bibl>

<bibl id="B51">
  <title><p>Rank-based inverse normal transformations are increasingly used,
  but are they merited?</p></title>
  <aug>
    <au><snm>Beasley</snm><fnm>TM</fnm></au>
    <au><snm>Erickson</snm><fnm>S</fnm></au>
    <au><snm>Allison</snm><fnm>DB</fnm></au>
  </aug>
  <source>Behavior genetics</source>
  <publisher>Springer</publisher>
  <pubdate>2009</pubdate>
  <volume>39</volume>
  <issue>5</issue>
  <fpage>580</fpage>
</bibl>

<bibl id="B52">
  <title><p>{Using Facebook ad data to track the global digital gender
  gap}</p></title>
  <aug>
    <au><snm>Fatehkia</snm><fnm>M</fnm></au>
    <au><snm>Kashyap</snm><fnm>R</fnm></au>
    <au><snm>Weber</snm><fnm>I</fnm></au>
  </aug>
  <source>World Development</source>
  <publisher>Elsevier</publisher>
  <pubdate>2018</pubdate>
  <volume>107</volume>
  <fpage>189</fpage>
  <lpage>-209</lpage>
</bibl>

<bibl id="B53">
  <title><p>Internet</p></title>
  <aug>
    <au><snm>Roser</snm><fnm>M</fnm></au>
    <au><snm>Ritchie</snm><fnm>H</fnm></au>
    <au><snm>Ortiz Ospina</snm><fnm>E</fnm></au>
  </aug>
  <source>Our World in Data</source>
  <pubdate>2022</pubdate>
  <note>https://ourworldindata.org/internet</note>
</bibl>

<bibl id="B54">
  <title><p>The global gender gap report</p></title>
  <aug>
    <au><cnm>{World Economic Forum}</cnm></au>
  </aug>
  <pubdate>2020</pubdate>
</bibl>

<bibl id="B55">
  <title><p>PISA 2018 Results (Volume II)</p></title>
  <aug>
    <au><cnm>OECD</cnm></au>
  </aug>
  <pubdate>2019</pubdate>
  <fpage>376</fpage>
  <url>https://www.oecd-ilibrary.org/content/publication/b5fd1b8f-en</url>
</bibl>

<bibl id="B56">
  <title><p>{Technical Notes: Calculating the human development
  indices—graphical presentation}</p></title>
  <aug>
    <au><cnm>{United Nations Development Programme}</cnm></au>
  </aug>
  <source>Retrieved from UNDP website:
  \url{https://hdr.undp.org/sites/default/files/2021-22_HDR/hdr2021-22_technical_notes.pdf}</source>
  <pubdate>2021</pubdate>
  <note>(Accessed on 10/26/2022)</note>
</bibl>

<bibl id="B57">
  <title><p>The Methodology of “Varieties of Democracy”(V-Dem)</p></title>
  <aug>
    <au><snm>Coppedge</snm><fnm>M</fnm></au>
    <au><snm>Gerring</snm><fnm>J</fnm></au>
    <au><snm>Knutsen</snm><fnm>CH</fnm></au>
    <au><snm>Krusell</snm><fnm>J</fnm></au>
    <au><snm>Medzihorsky</snm><fnm>J</fnm></au>
    <au><snm>Pernes</snm><fnm>J</fnm></au>
    <au><snm>Skaaning</snm><fnm>SE</fnm></au>
    <au><snm>Stepanova</snm><fnm>N</fnm></au>
    <au><snm>Teorell</snm><fnm>J</fnm></au>
    <au><snm>Tzelgov</snm><fnm>E</fnm></au>
    <au><cnm>others</cnm></au>
  </aug>
  <source>Bulletin of Sociological Methodology/Bulletin de M{\'e}thodologie
  Sociologique</source>
  <publisher>SAGE Publications Sage UK: London, England</publisher>
  <pubdate>2019</pubdate>
  <volume>143</volume>
  <issue>1</issue>
  <fpage>107</fpage>
  <lpage>-133</lpage>
</bibl>

<bibl id="B58">
  <title><p>{Geographies of the Internet}</p></title>
  <aug>
    <au><snm>Warf</snm><fnm>B</fnm></au>
  </aug>
  <publisher>New York: Routledge</publisher>
  <pubdate>2020</pubdate>
</bibl>

<bibl id="B59">
  <title><p>World Development Report 2016: Digital Dividends</p></title>
  <aug>
    <au><cnm>{World Bank}</cnm></au>
  </aug>
  <pubdate>2016</pubdate>
</bibl>

<bibl id="B60">
  <title><p>Digital differences</p></title>
  <aug>
    <au><snm>Zickuhr</snm><fnm>K</fnm></au>
    <au><snm>Smith</snm><fnm>A</fnm></au>
  </aug>
  <source>Pew Research Center</source>
  <pubdate>2012</pubdate>
  <note>Retrieved December 7, 2020 from
  \url{https://www.pewresearch.org/internet/2012/04/13/digital-differences/}</note>
</bibl>

<bibl id="B61">
  <title><p>Digital gender divide or technologically empowered women in
  developing countries? A typical case of lies, damned lies, and
  statistics</p></title>
  <aug>
    <au><snm>Hilbert</snm><fnm>M</fnm></au>
  </aug>
  <source>Women's Studies International Forum</source>
  <pubdate>2011</pubdate>
  <volume>34</volume>
  <issue>6</issue>
  <fpage>479</fpage>
  <lpage>489</lpage>
  <url>http://www.sciencedirect.com/science/article/pii/S0277539511001099</url>
</bibl>

<bibl id="B62">
  <title><p>{Scaling PIAAC cognitive data}</p></title>
  <aug>
    <au><snm>Yamamoto</snm><fnm>K</fnm></au>
    <au><snm>Khorramdel</snm><fnm>L</fnm></au>
    <au><snm>Von Davier</snm><fnm>M</fnm></au>
    <au><cnm>others</cnm></au>
  </aug>
  <source>Technical report of the survey of adult skills (PIAAC)</source>
  <pubdate>2013</pubdate>
  <fpage>408</fpage>
  <lpage>-440</lpage>
</bibl>

<bibl id="B63">
  <title><p>{Facebook Q2 2020 Earnings}</p></title>
  <aug>
    <au><cnm>{Facebook}</cnm></au>
  </aug>
  <source>\url{https://s21.q4cdn.com/399680738/files/doc_financials/2020/q2/Q2-2020-FB-Earnings-Presentation.pdf}</source>
  <pubdate>2020</pubdate>
  <note>(Accessed on 10/15/2020)</note>
</bibl>

<bibl id="B64">
  <title><p>International gender differences and gaps in online social
  networks</p></title>
  <aug>
    <au><snm>Magno</snm><fnm>G</fnm></au>
    <au><snm>Weber</snm><fnm>I</fnm></au>
  </aug>
  <source>International Conference on Social Informatics</source>
  <pubdate>2014</pubdate>
  <fpage>121</fpage>
  <lpage>-138</lpage>
</bibl>

<bibl id="B65">
  <title><p>Analysing global professional gender gaps using LinkedIn
  advertising data</p></title>
  <aug>
    <au><snm>Kashyap</snm><fnm>R</fnm></au>
    <au><snm>Verkroost</snm><fnm>FC</fnm></au>
  </aug>
  <source>EPJ Data Science</source>
  <publisher>Springer Berlin Heidelberg</publisher>
  <pubdate>2021</pubdate>
  <volume>10</volume>
  <issue>1</issue>
  <fpage>39</fpage>
</bibl>

<bibl id="B66">
  <title><p>Global music streaming data reveal diurnal and seasonal patterns of
  affective preference</p></title>
  <aug>
    <au><snm>Park</snm><fnm>M</fnm></au>
    <au><snm>Thom</snm><fnm>J</fnm></au>
    <au><snm>Mennicken</snm><fnm>S</fnm></au>
    <au><snm>Cramer</snm><fnm>H</fnm></au>
    <au><snm>Macy</snm><fnm>M</fnm></au>
  </aug>
  <source>Nature human behaviour</source>
  <publisher>Nature Publishing Group</publisher>
  <pubdate>2019</pubdate>
  <volume>3</volume>
  <issue>3</issue>
  <fpage>230</fpage>
  <lpage>-236</lpage>
</bibl>

<bibl id="B67">
  <title><p>Vocabulary and reading comprehension revisited: Evidence for high-,
  mid-, and low-frequency vocabulary knowledge</p></title>
  <aug>
    <au><snm>Masrai</snm><fnm>A</fnm></au>
  </aug>
  <source>Sage Open</source>
  <publisher>SAGE Publications Sage CA: Los Angeles, CA</publisher>
  <pubdate>2019</pubdate>
  <volume>9</volume>
  <issue>2</issue>
  <fpage>2158244019845182</fpage>
</bibl>

<bibl id="B68">
  <title><p>Generalized additive models</p></title>
  <aug>
    <au><snm>Hastie</snm><fnm>TJ</fnm></au>
    <au><snm>Tibshirani</snm><fnm>RJ</fnm></au>
  </aug>
  <publisher>CRC press</publisher>
  <pubdate>1990</pubdate>
  <volume>43</volume>
</bibl>

<bibl id="B69">
  <title><p>Census of India: Literacy And Level of Education</p></title>
  <aug>
    <au><cnm>{Office of the Registrar General \& Census Commissioner,
  India}</cnm></au>
  </aug>
  <source>\url{https://censusindia.gov.in/census_and_you/literacy_and_level_of_education.aspx}</source>
  <pubdate>2011</pubdate>
  <note>(Accessed on 10/28/2020)</note>
</bibl>

<bibl id="B70">
  <title><p>{Facebook Users by Country 2022}</p></title>
  <aug>
    <au><cnm>{World Population}</cnm></au>
  </aug>
  <source>\url{https://worldpopulationreview.com/country-rankings/facebook-users-by-country}</source>
  <pubdate>2022</pubdate>
  <note>(Accessed on 10/25/2022)</note>
</bibl>

<bibl id="B71">
  <title><p>How to produce and use the global and thematic education
  indicators</p></title>
  <source>\url{http://hdl.voced.edu.au/10707/532231}</source>
  <publisher>UNESCO Institute for Statistics, Montreal, Quebec</publisher>
  <pubdate>2019</pubdate>
  <note>(Accessed on 07/01/2020)</note>
</bibl>

<bibl id="B72">
  <title><p>{Barro-Lee Educational Attainment Dataset}</p></title>
  <source>\url{http://www.barrolee.com/}</source>
  <pubdate>2010</pubdate>
  <note>(Accessed on 07/01/2020)</note>
</bibl>

<bibl id="B73">
  <title><p>{Using Digital Traces to Measure Digital Gender Inequality in
  Real-Time}</p></title>
  <source>\url{https://www.digitalgendergaps.org/project/}</source>
  <pubdate>2020</pubdate>
  <note>(Accessed on 07/01/2020)</note>
</bibl>

<bibl id="B74">
  <title><p>{HDRO API Information}</p></title>
  <source>\url{http://ec2-54-174-131-205.compute-1.amazonaws.com/API/Information.php}</source>
  <pubdate>2018</pubdate>
  <note>(Accessed on 07/01/2020)</note>
</bibl>

<bibl id="B75">
  <title><p>{Global Data Lab - Innovative Instruments for Turning Data into
  Knowledge}</p></title>
  <source>\url{https://globaldatalab.org/}</source>
  <pubdate>2018</pubdate>
  <note>(Accessed on 07/01/2020)</note>
</bibl>

<bibl id="B76">
  <title><p>{Human Development Reports}</p></title>
  <source>\url{http://hdr.undp.org/en/composite/IHDI}</source>
  <pubdate>2017</pubdate>
  <note>(Accessed on 07/01/2020)</note>
</bibl>

</refgrp>
} 




\section*{Figures}
Figure 1-4.\\
\begin{figure*}[t]
\centering
    \includegraphics[width=.9\linewidth]{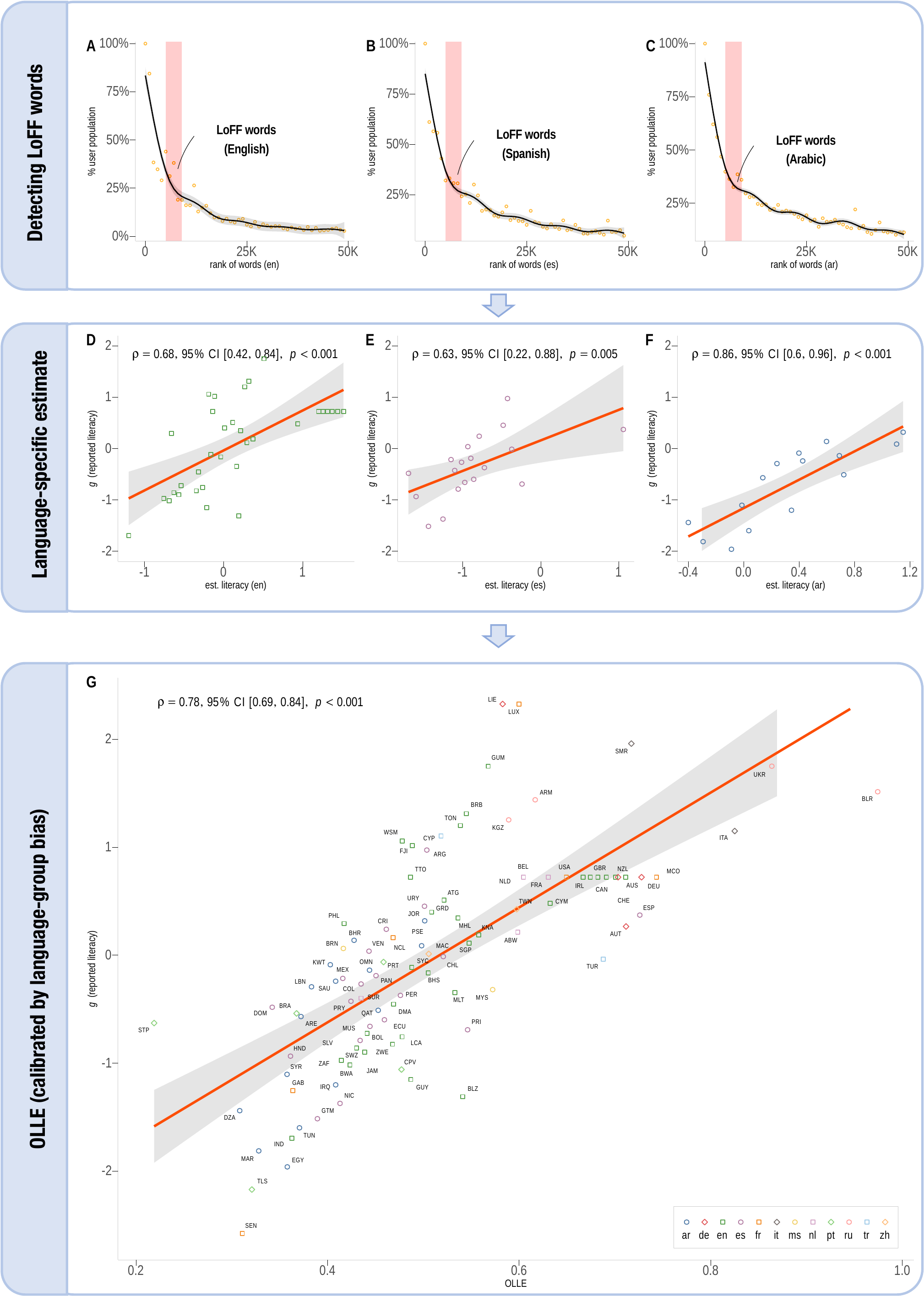}
    \caption{
    {\bf Creating online language literacy estimate.}
Our methodology produces online language literacy estimate (\olle) through three major steps: 1. (A-C) The set of \bigword (lower-frequency frequent words) is algorithmically determined based on the vocabulary popularity in the language corpus; the red bands in (A-C) indicate the selected sets of \bigword in the three most widely used languages in our data (English, Spanish, Arabic). 2. (D-F)
Normalized total occurrence of \bigword in Facebook dataset from each country is used as a language-specific online literacy estimate for that country. (D-F) show the strong correlations found between our estimates and countries' officially reported literacy rates in English, Spanish, and Arabic, respectively. 3. (G) The calibrated global estimates, \olle, are generated after addressing language group bias and shown here with a strong correlation with reported literacy rates (Spearman's rank correlation coefficient $\rho=0.78$, \yrepj{95\% CI [0.69, 0.84]}, $p<0.001$.) Error bounds represent the 95\% confidence intervals. 
\yrepj{In (D-G), each dot represents a country, with $x$ value indicating the country's raw (D-F) or calibrated (G) literacy estimate and $y$ value the country's officially reported literacy rate. }
}
    \label{fig:eval}
\end{figure*}

\begin{figure*}
    \centering
    \includegraphics[width=\linewidth]{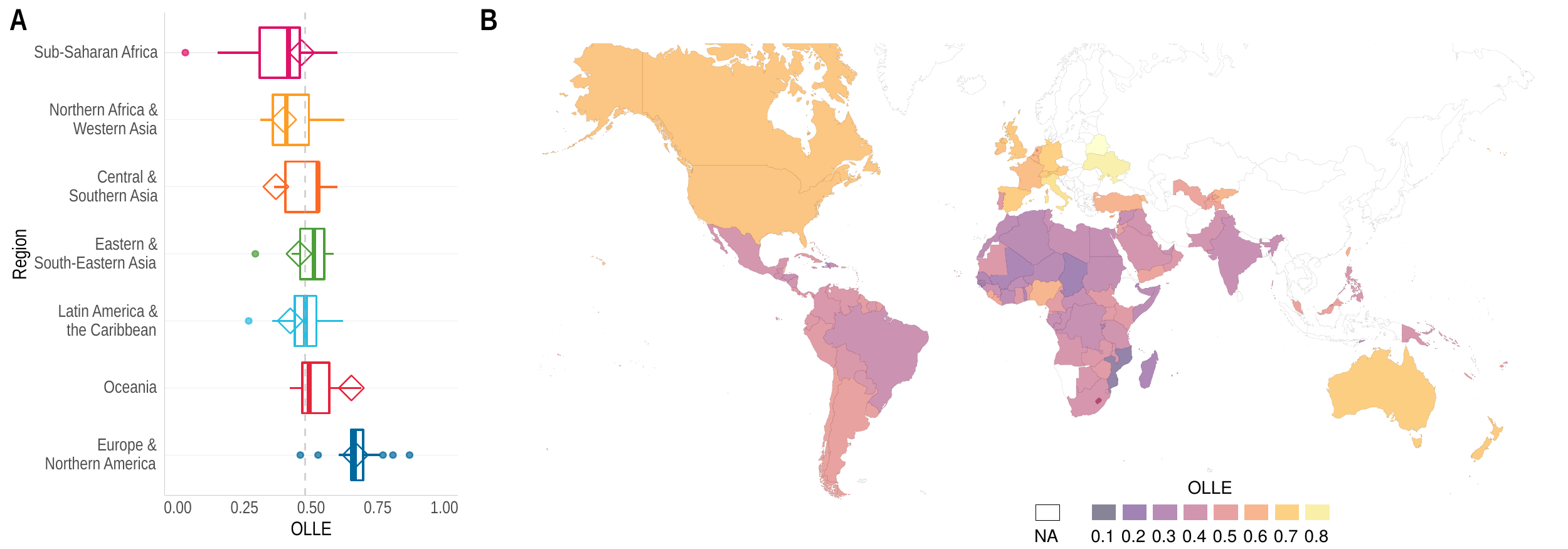}
    \caption{{\bf Online language literacy estimates across the world.} (A) Summary of the \olles across the seven geographical groups. The dashed line marks the global average (0.477, population-weighted). Within each box, a line denotes the median value of the group, and a diamond indicates the population-weighted mean of the group; boxes extend from the first to the third quartile of each group's distribution of values; whiskers (lines extending from the boxes) denote the most extreme values within 1.5 interquartile range of each group; dots denote observations outside the range of extreme values.
 (B) Map of the \olles available in our dataset.}
    \label{fig:lit_map}
\end{figure*}
\begin{figure*}
    \centering
    \includegraphics[width=.95\linewidth]{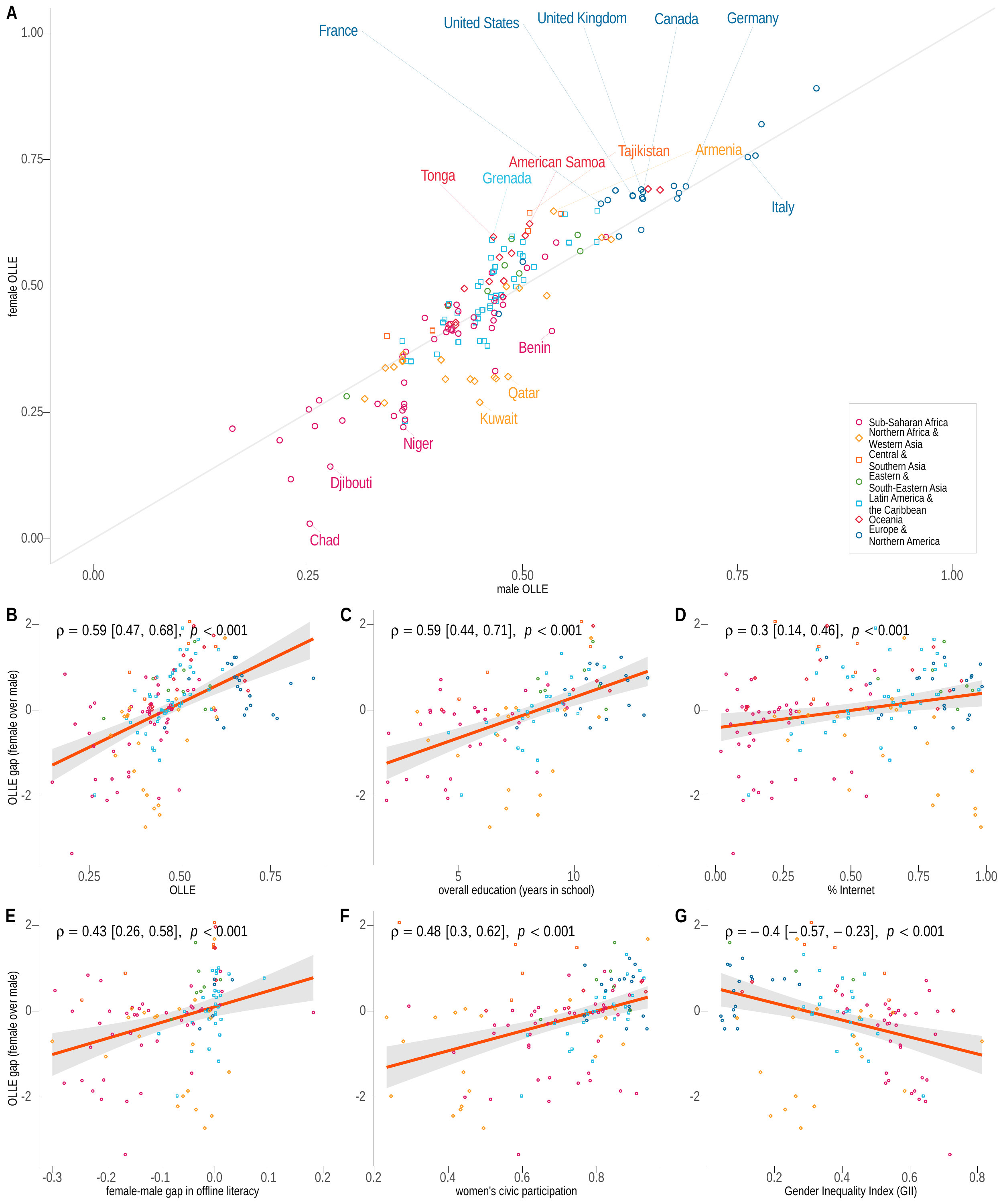}
    \caption{{\bf Gender gap in online language literacy.} (A) Country-level \olles for women and men. (B-G) The standardized online literacy gender difference (women over men) compared with the country's overall \olle, education, Internet penetration, as well as gender parity and empowerment measures. Error bounds represent the 95\% confidence intervals. 
    \yrepj{Spearman's rank correlation coefficients and 95\% CI are indicated in the scatterplots.}
    }
    \label{fig:gender_gap}
\end{figure*}
\begin{figure*}
    \centering
    \includegraphics[width=\linewidth]{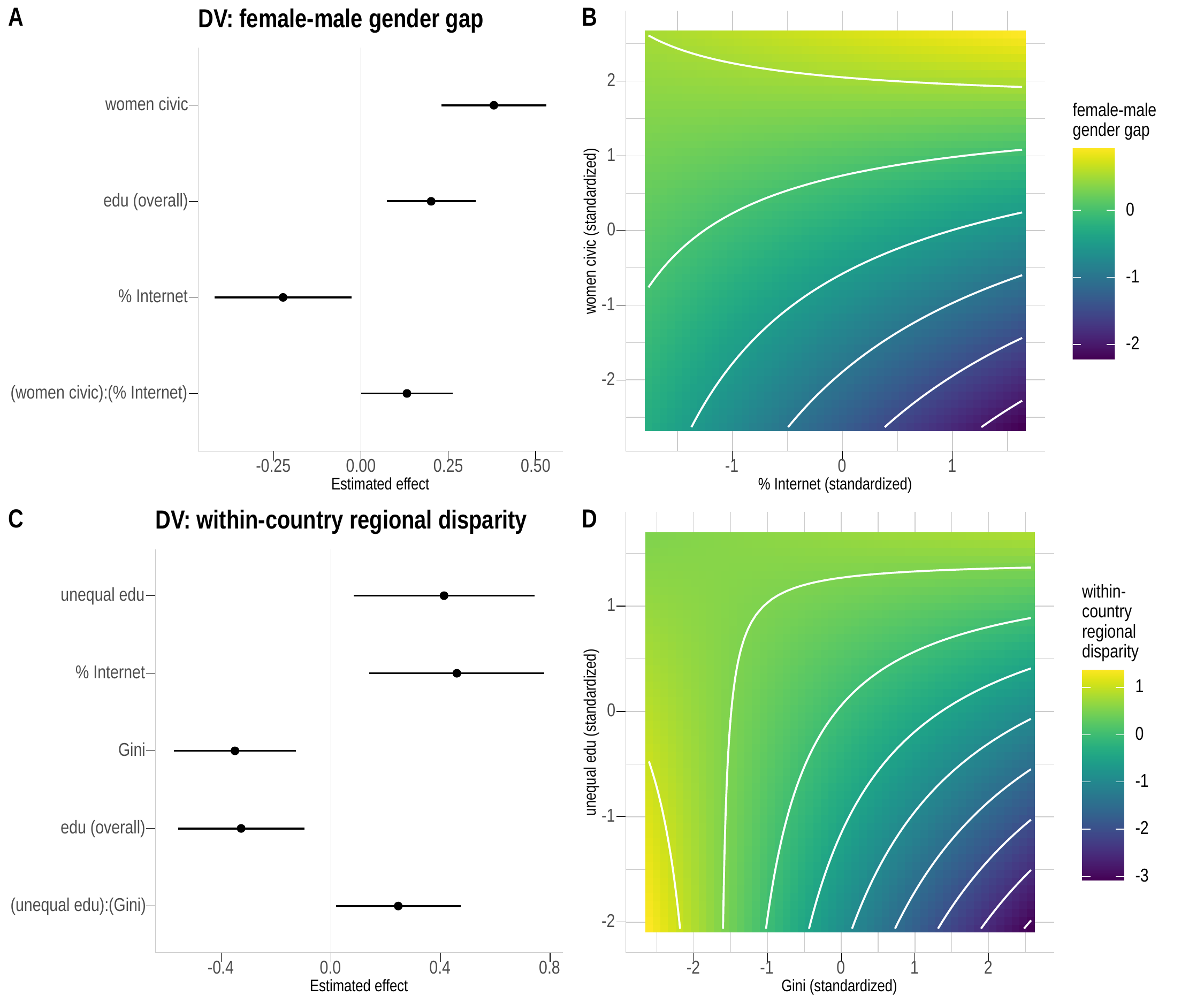}
    \caption{{\bf The societal context and disparities in \olle.}
    (A) Multiple regression model shows robust associations of the gender gap (female over male) in \olle with overall education status and women's civic participation. (B) The significant interaction between women's civic participation and Internet penetration in predicting gender gap in \olle. 
    (C) Multiple regression models show robust associations of regional disparity in \olle with inequality in education, internet penetration, inequality in income (Gini index), and overall educational attainment. (D) Inequalities in education and income appear to have the opposite contribution in predicting a country's regional disparity in \olle. 
    Full regression estimates are provided in \smapp Tables \ref{tab:model-gender-main} and \ref{tab:model-region-main}, respectively.
    (B,D) show the average marginal effects of the two interaction terms. }
    \label{fig:gender_model}
\end{figure*}




\section*{Additional Files}
  \subsection*{Additional file --- Supplemental Information}
  
This document includes the additional information:
(1) Supplementary text;
(2) Figures S1 to S7;
(3) Tables S1 to S12;
(4) SI References. 
In addition, data aggregated at the country level (country-level literacy estimates and summary statistics) and the R analysis code for deriving the main results will be made available in the Open Science Framework (OSF) upon publication of this manuscript. Facebook requires that this work was to be done in compliance with Facebook's Data Policy and research ethics review process (\url{www.facebook.com/policy.php}). Restrictions apply to the availability of the disaggregate data (user- or post-level data), so they are not publicly available. Data aggregated at the country level and other datasets that support the findings of this study will be available from the OSF repository with the permission of the authors, upon reasonable request.

\end{backmatter}

\clearpage
\newcommand{\fulltitle}{Mapping Language Literacy At Scale: A Case Study on Facebook}
\begin{center}
{\Large\bf Supplemental Information -- \fulltitle}
\end{center}
\vspace*{.5in}

\setcounter{table}{0}
\renewcommand{\thetable}{S\arabic{table}}%
\setcounter{figure}{0}
\renewcommand{\thefigure}{S\arabic{figure}}%
\setcounter{section}{0}
\renewcommand{\thesection}{S\arabic{section}}%

\addcontentsline{toc}{section}{Supplementary Text}
\section{Supplementary Text}\label{sec:supp_text}

\addcontentsline{toc}{subsection}{Language literacy and visual information consumption}
\subsection{Language literacy and visual information consumption}\label{sec:vis}

We examine the relationship between populations' language literacy levels and their interest in different types of content. Because our assessment concerns the ability to process textual content, we assume there exists a negative relationship between a population's literacy estimate and their attention toward non-textual (e.g., visual) content. Fig.~\ref{fig:lit_visual} shows the correlations between countries' \olles ($x$-axes) and the relative time spent on visual content by the countries' Facebook population ($y$-axes).  As expected, populations with a lower level of language literacy tend to spend relatively more time on visual content. The global correlation is $-0.38$ (Spearman's rank correlation; $p<0.001$), with correlations over different areas ranging from $-0.29$ (Latin America \& the Caribbean; $p<0.062 $) to $-0.62$ (Europe/Oceania/Northern America; $p<0.001$)\footnote{As two of the seven geographical groups only have few countries, we merge the seven groups into five (based on proximity) to provide an adequate statistical description.}.

\addcontentsline{toc}{subsection}{Elbow range detection in popularity curves}
\subsection{Elbow range detection in popularity curves}\label{sec:elbow}
\Bigword are determined based on ``elbow'' range on the word popularity curve for each language, where the relative word frequencies begin to have a systematic decline. 
\yrepj{Fig.~\ref{fig:word_curve} shows the popularity of words in decreasing order, i.e., from the most to the least popular word, as popularity curves. It can be seen that a systematic decline in the word popularity appears at the point of maximum curvature in a popularity curve. In other words, the interest region associated with \bigword corresponds to the ``elbow'' (or ``knee'') point on the smoothed word popularity curve. Mathematically, the curvature is a mathematical measure of how much a function differs from a straight line \cite{satopaa2011finding,antunes2018knee}. Estimating the knee/elbow point for a continuous function is straightforward since the curvature is well-defined for continuous functions; however, it is a challenging task for discrete data. It is also an inherently heuristic process \cite{antunes2018knee}. To reliably detect the elbow range, we leverage the ``Kneedle'' detection \cite{satopaa2011finding}, an efficient algorithm that can efficiently detect knee points in discrete data, and the standard maximum curvature approach on a smoothed function learned from the discrete points. First, we employ generalized additive models with cross-validation to learn a smooth function for each of the popularity curves \cite{hastie1990generalized}. As shown in Fig.~1 E-G, we define an elbow range as an area between two points $k_0$ and $k_1$ (highlighted in red) that best describe the systematic decline in the curve. The two points were determined by combining two heuristic methods: (i) the standard maximum curvature points that can be calculated from any continuous function, and (ii) the approximate knee points (Kneedle detection method) based on the notion that knee points differ most from the straight line connecting the curve's two endpoints \cite{satopaa2011finding}. Note that while the two notions may be considered to be conceptually similar, the approximate knee points (the second notion) are not necessarily the maximum curvature points especially when the curves are skewed. In a right-skewed curve (as in the case of a word popularity curve), the approximate knee points tend to fall into the right of the maximum curvature points. Thus, we detect an elbow by two points $k_0$ and $k_1$ through maximum curvature measurement and approximate knee point detection method respectively. Unlike other knee/elbow detention methods that are sensitive to noises and rescaling, we found this hybrid approach is more robust to rescaling and small fluctuations in our data.} 
Words with ranks falling into the elbow range are considered to be the ``\bigword'' in the language. Fig.~1 A-C highlights the elbow range detected from the word popularity curve for each of the three most used languages, and Fig.~\ref{fig:word_curve} shows the detected elbow ranges for all 12 languages. 

\addcontentsline{toc}{subsection}{Procedure for estimating online language literacy}
\subsection{Procedure for estimating online language literacy}\label{sec:procedure}

For a given population with a given language, the procedure to measure the collective language literacy involves the following steps:

\begin{enumerate}
\item[(i)] Processing of user-generated texts: We use public posts written in any of the chosen languages created by Facebook users who are at least 18 years old and active during a 30-day period between April 20 and May 20, 2020. We exclude posts that did not contain any text or text that was shorter than 2 characters or longer than 1000 characters, as well as posts that contained URLs as these are more likely to be copied and pasted from other sources rather then composed by users. 

\item[(ii)] Aggregate statistics per user: After tokenizing the public posts, for each user, we quantify the number of unique words (unigrams) that falls in the range of \bigword. We then obtain a relative \bigw count $w_u$ that is normalized by the active level of post creation per user, i.e., $w_u$ is given by (the total number of \bigword observed from $u$'s public posts) / (the total number of $u$'s public posts). We count each unigram once per user, regardless of the frequency used, to avoid overestimating the use of particular words or the inflation from copy-pasted content.

\item[(iii)] Aggregate statistics per population: For each geographically bounded community (e.g. a county or a region) with at least 1000 active users observed in the study period, the population-level estimate is calculated as $\bar{w}$, the average of $w_u$'s over all active users $u$'s in the geographically bounded community. The gender- or region-disaggregate population-level estimates also require a minimum of 1000 active users observed in the study period in any of the disaggregate groups. The threshold of 1000 unique users from any group was chosen to ensure user privacy and the statistical power of our method. We also exclude users who produced a high volume of posts (above 75 percentiles) to avoid a small number of highly productive users dominating the measurement.
\end{enumerate}

Throughout the procedure, none of the personal identifiable information or any personal or private content was used. Only the aggregate statistics $w_u$ and $\bar{w}$ were generated from the process. 

\paragraph{How to retrieve pre-computed \bigword}
\yrrr{\bigword are determined based on the word popularity curves derived from the Facebook users' use of the up 200,000 most frequent words in each language. These pre-computed \bigword can be retrieved through the following steps:}
\begin{itemize}
\item[(i)] Install the fastText\footnote{\url{https://fasttext.cc/docs/en/python-module.html\#installation}}
\item[(ii)] Run \texttt{download\_model.py \$lang} to get the dictionary of a specific language, where \texttt{\$lang} is the language indicator (e.g., {\texttt en} for English, {\texttt es} for Spanish). This script will download the dictionary in a binary file (let \texttt{\$filename} be the filename of the downloaded file).
\item[(ii)] Run \texttt{fasttext dump \$filename dict > \$ofilename} to convert the binary dictionary file to a text file (let \texttt{\$ofilename} be the filename of the output text file). This file contains up to 200,000 lines where each line is a word and its frequency. The frequencies can be used to rank the words from the most to the least frequent. 
\item[(iv)] Extract the \bigword based on the knee points listed in Table~\ref{tab:knees}. For example, the \bigword in English correspond to the words in the fastText dictionary that are ranked between 5,000 to 9,000 in the decreasing order of word frequency.
\end{itemize}

\addcontentsline{toc}{subsection}{D.}
\subsection{Case study: India as a multilingual country}\label{sec:india}

\yrv{We choose India as a case study for countries using multiple languages to study the effect of choosing the most used language as a single representative language for literacy estimate. While India uses Hindi and English as official languages nationwide, it has no single national language. It has over 30 states/union territories, each of which has its own official language(s). There are 22 official languages recognized by country officials, in addition to some other languages recognized as additional official languages at the regional level. In this analysis, we include the additional five most used languages in India according to the India census reported in 2011 \cite{Censusof39:online}: Hindi (43.6\%), Bengali (8.3\%), Marathi (7.1\%), Telugu (6.9\%), and Tamil (5.9\%). Languages with less than 5\% speakers among the Indian national population are not considered. On Facebook, the most used language in Indian users' public posts is English (en), which has about three times the users posting in Hindi (hi), and about 20, 44, 90, and 181 times those posting in Bengali (bn), Marathi (mr), Telugu (te), and Tamil (ta), respectively. 
Across regions, the non-English language using populations on Facebook are sparse. Only 14 (48.2\%) regions have more than 10\% of the number of English posters posting in Hindi, and only 4 (13.8\%) and 1 (3.4\%) regions have more than 1\% of the number of English posters posting in Bengali and Marathi.} 

\yrv{We then create a language literacy estimation for each of the six languages, using the same approach but include the additional languages (Hindi, Bengali, Marathi, Telugu, and Tamil) from fastText unigram data \cite{grave2018learning}. For validation, we gather the regional literacy survey reported in the India census 2011 \cite{Censusof39:online}, which is the most recent data available. 
Fig.~\ref{fig:india} shows the comparison of our language estimation with the officially reported literacy data. 
\yrrr{We first estimate the language literacy for every language. Fig.~\ref{fig:india} A and C-E show the estimation based on posts in a single language. Note that, while there are regional differences, the use of Hindi, Bengali, and Marathi is extremely sparse in most regions. Therefore, the non-English language estimates alone cannot be directly used to create a regional measurement. Due to the sparse use of non-English language on the platform, the correlation between the non-English language estimates and the reported literacy is insignificant. We additionally create a multi-language estimation weighted by the popularity of each language within a region, as shown in Fig~\ref{fig:india} B.}
We observe that both English-only and multi-language estimates (without any additional calibration) have significant correlations with the reported literacy data (Spearman's rank correlations with positive 95\% CIs and $p<0.005$). However, literacy estimation based on multiple languages has neglectable improvement over English-based estimation in terms of the correlations -- from 0.51 to 0.52. This is likely due to the low rate of users posting in non-English languages in many regions. \yrrr{Here, the comparison relies on the officially reported literacy data, which has a limitation: they do not capture the change in regional literacy levels since 2011, and likely do not properly reflect the diverse language skills used by minority populations.} 
This case study illustrates the challenge of obtaining gold-standard literacy measures for multilingual countries. While this does not prevent us to create a multi-language literacy measurement per country, for validation purposes, we simply choose a single representative language for multilingual countries. Thus the correlation should be interpreted with caution -- the officially reported data that guide this choice often has a bias against language minorities.}

\addcontentsline{toc}{subsection}{E.}
\subsection{Robust check: regional disparity and language dominance}\label{sec:domi}

\yrepj{Our \olle is created based on a country's representative language, i.e., the language used by the most Facebook users in the country. One potential risk of relying on a single representative language is that the regional disparity measure in a multilingual country may simply capture the distribution of languages, rather than the diversity of language skills. To test this, we examine the relationship between the user percentage of the representative language in a multilingual country and the country's regional disparity measure. Among the 167 countries studied, there are 20 multilingual countries, but only 13 meet the criteria to have a regional disparity measure. Recall in Section~\ref{sec:procedure} that each geographically bounded community (i.e., in this case, a region within a country) with at least 1000 active users observed in the study period. For these 13 multilingual countries, we plot the countries' percentage of Facebook users using the representative language on the $y$-axis and on the $x$-axis, either (A) \olle, or (B) regional disparity as shown in Fig.~\ref{fig:domi}. If there is a systematic bias, e.g., countries with low representative language user percentages tend to have high a regional disparity measure, we would see a trend in such a plot. However, we do not observe a clear systematic bias. While our sample size is limited, this analysis is helpful for checking whether there is a potential bias in the small sample of multilingual countries.}

\addcontentsline{toc}{section}{Supplementary Figures}

\begin{figure*}
    \centering
    \includegraphics[width=.95\linewidth]{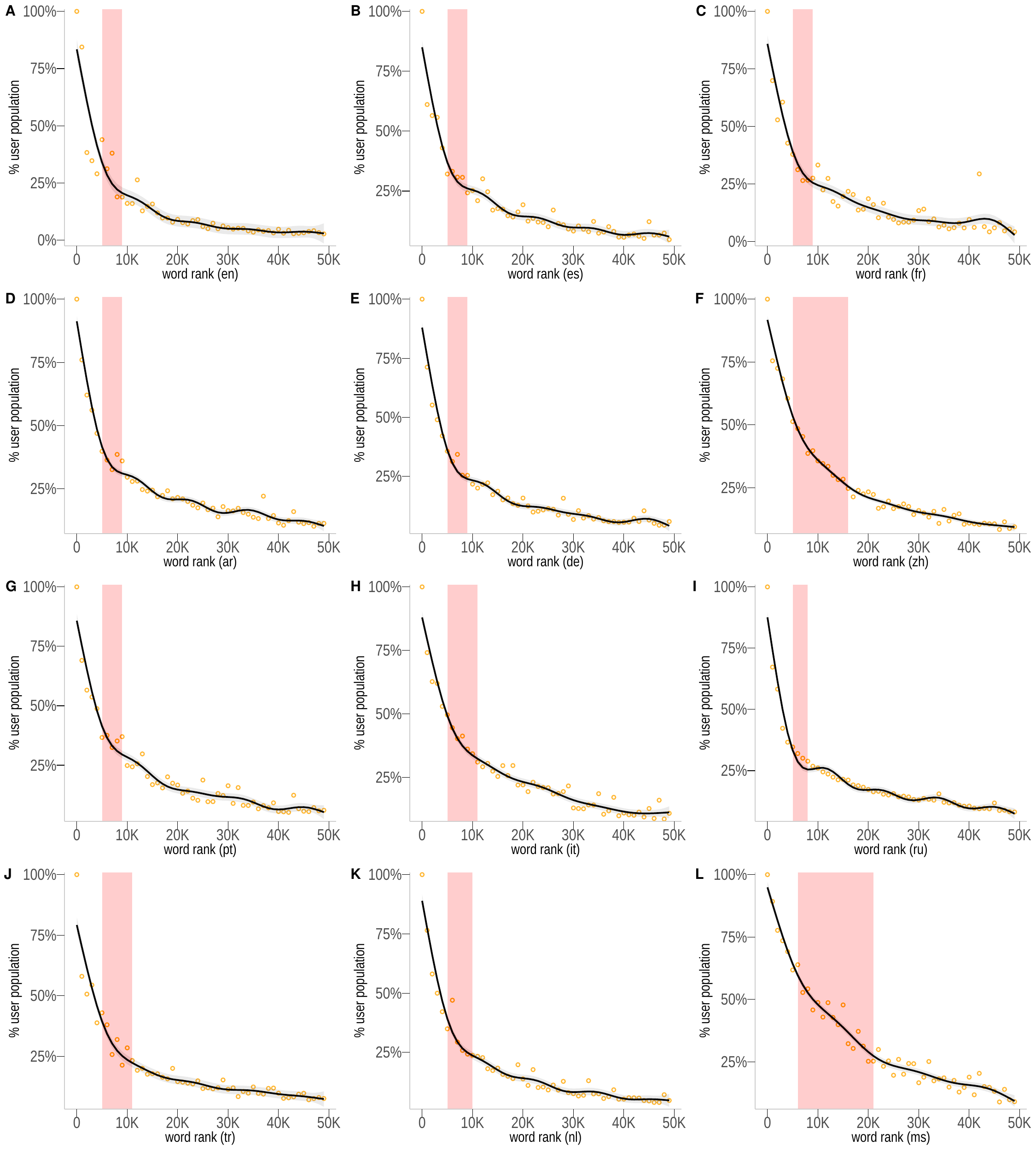}
    \caption{Determining the ``\bigw'' range using curvature and knee points detection. In half of the studied languages, the ranks of the \Bigword range between 5000 and 9000. Others (zh, it, ru, tr, nl, and ms) have wider or narrower ranges.}
    \label{fig:word_curve}
\end{figure*}

\begin{figure*}
    \centering
    \includegraphics[width=.95\linewidth]{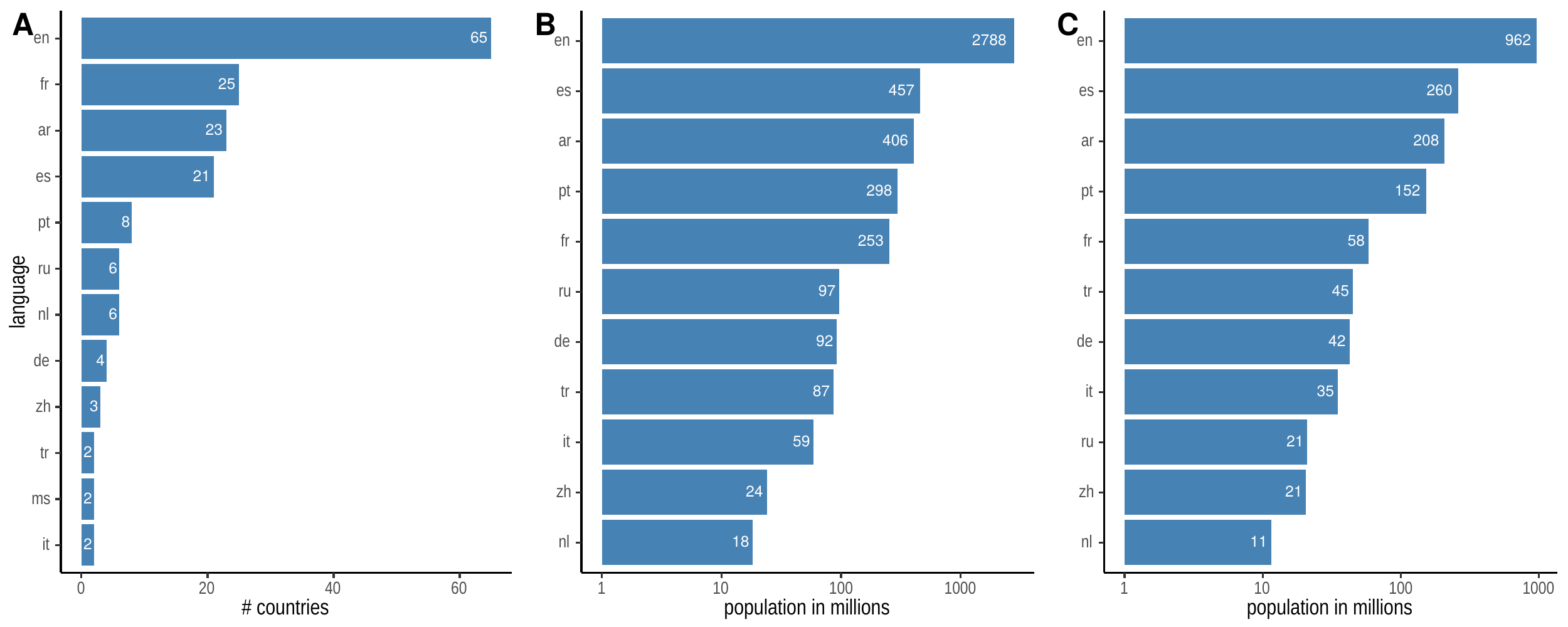}
    \caption{Countries covered in our estimation. There are 167 countries in 12 different languages, including 147 countries with a single or dominant language and 20 multilingual countries. For a multilingual country (having multiple official languages), we use the most used language of the country to estimate its language literacy.
    (A) The number of countries in each language. (B) The total population in each language. (C) The Facebook user count in each language, according to publicly available information about Facebook penetration statistics in 2022 \cite{Facebook15:online}. The populations from the 20 multilingual countries are excluded in (B) and (C) because we do not have the sub-population estimates of different languages within these multilingual countries. 
    }
    \label{fig:countries_covered}
\end{figure*}

\begin{figure*}
    \centering
    \includegraphics[width=.95\linewidth]{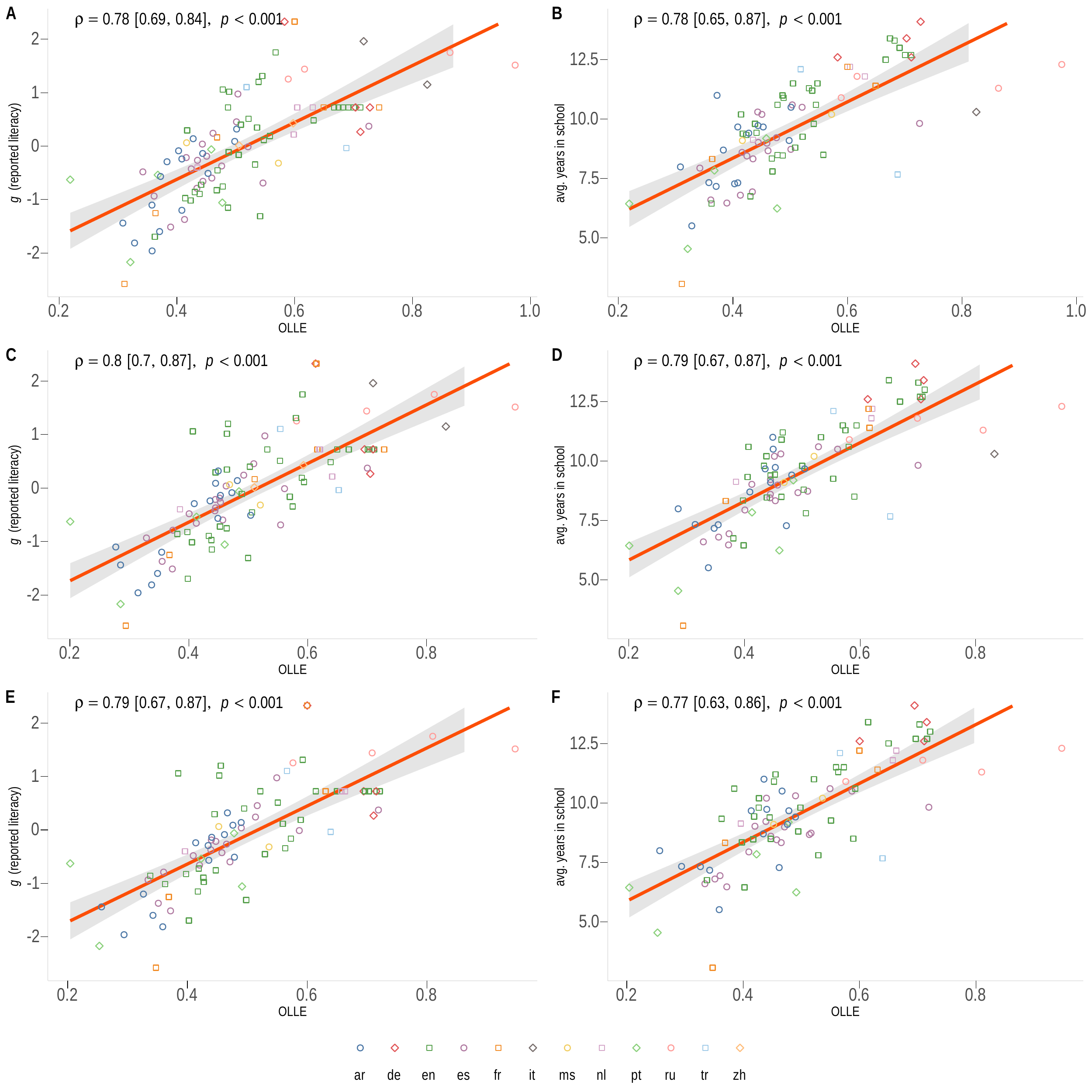}
    \caption{The literacy estimate ($x$-axis) obtained from Models (a,b,c) listed in Table \ref{tab:model-lit-main}, compared with the reported literacy rate and the education in terms of schooling years ($y$-axis). The reported literacy rate was transformed for normality. 
(A,B) Literacy estimate adjusted by Model (a).
(C,D) Literacy estimate adjusted by Model (b).
(E,F) Literacy estimate adjusted by Model (c).}
    \label{fig:corr_lit_edu}
\end{figure*}

\begin{figure*}
    \centering
    \includegraphics[width=.95\linewidth]{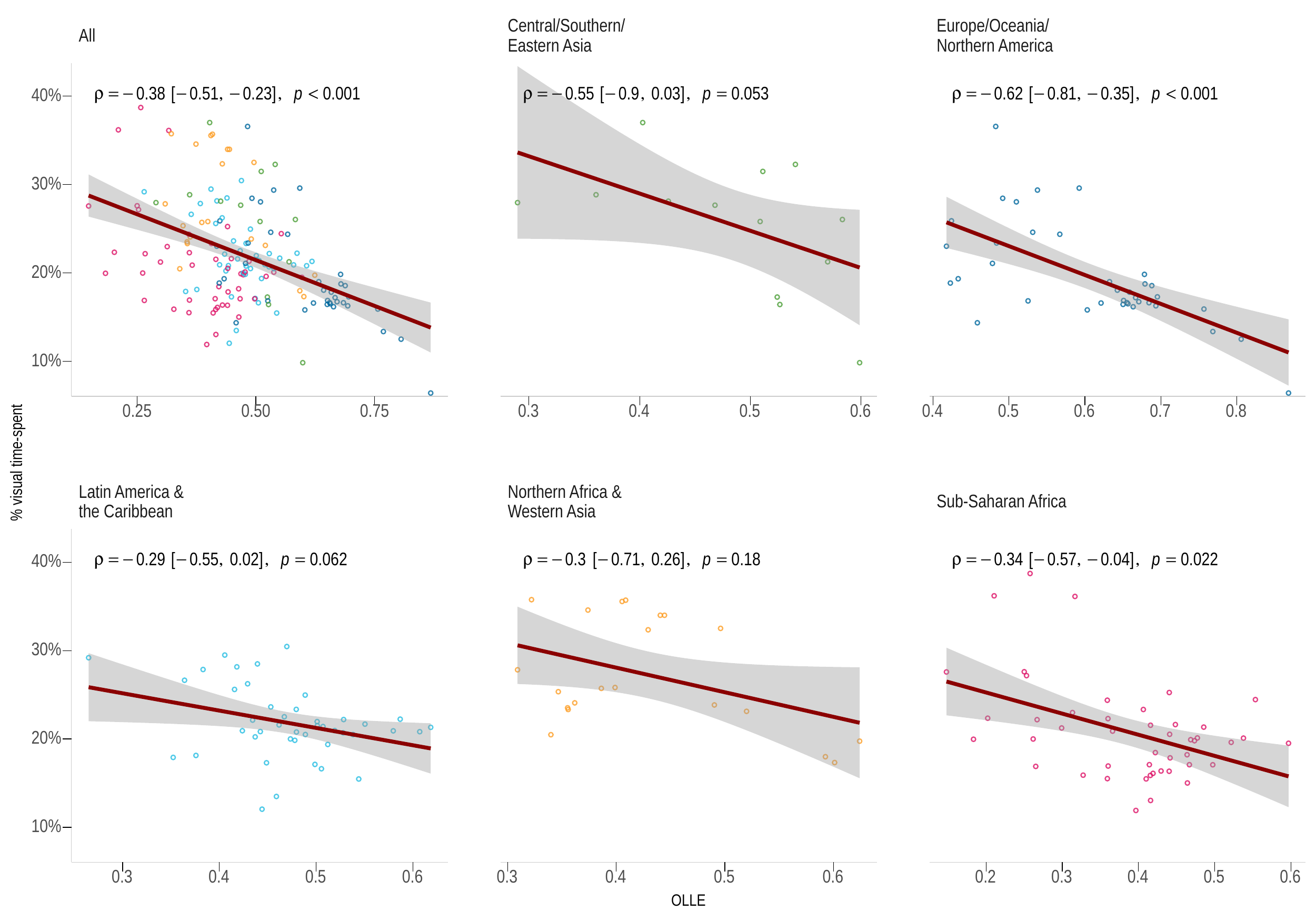}
    \caption{Relationship between countries' \olles ($x$-axes) and the relative visual time-spent ($y$-axes), where the relative visual time-spent is given as the proportion of time spent on photos and videos relative to the time on news feeds, photos and videos combined. Correlations are reported based on Spearman's rank correlation. }
    \label{fig:lit_visual}
\end{figure*}

\begin{figure*}
    \centering
    \includegraphics[width=\linewidth]{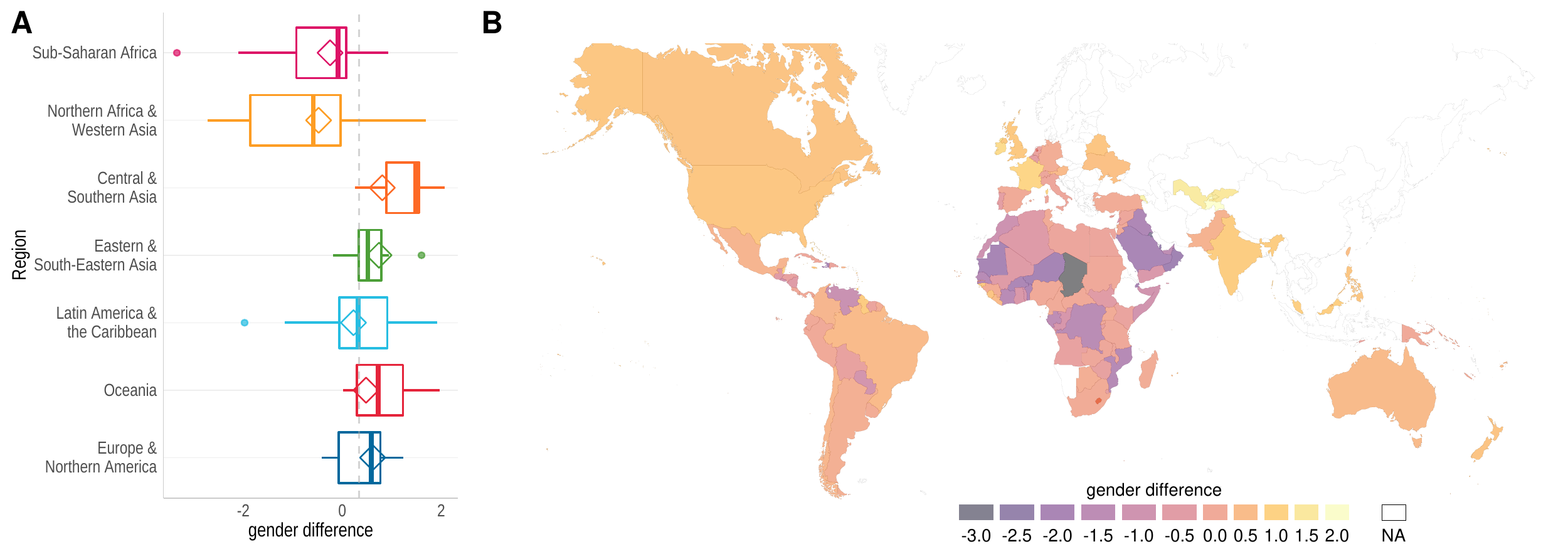}
    \caption{Gender differences in online language literacy across the world. (A) Summary of the standardized female-male differences across the seven geographical groups. Dashed line marks the global average (0.345, population-weighted), and a diamond indicates the population-weighted mean of the group. (B) Map of female-male differences available in our dataset.}
    \label{fig:gender_map}
\end{figure*}

\begin{figure*}
    \centering
    \includegraphics[width=\linewidth]{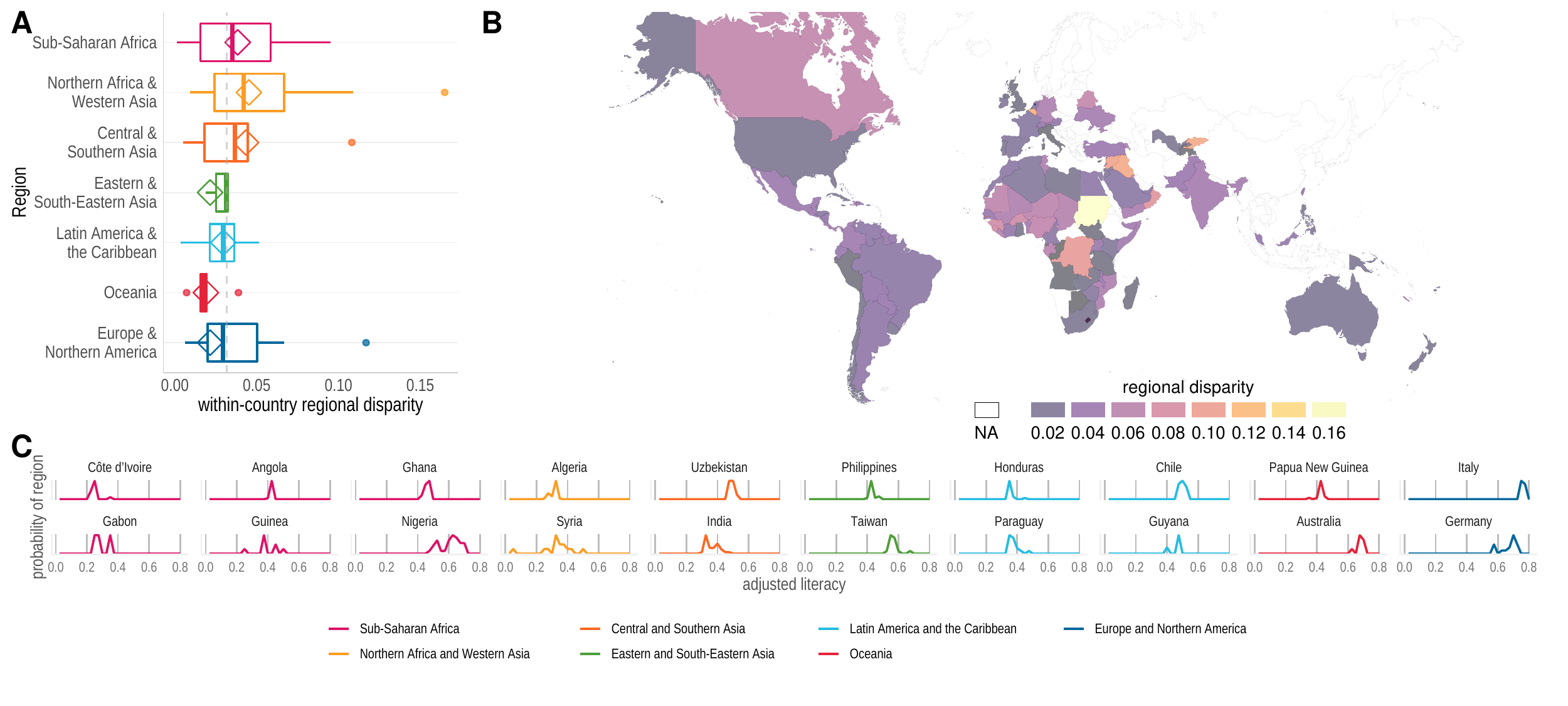}
    \caption{Within-country regional disparity in online language literacy across the world. (A) Summary of the within-country regional disparity across the seven geographical groups. Dashed line marks the global average (0.032, population-weighted), and a diamond indicates the population-weighted mean of the group. (B) Map of country-level regional disparity available in our dataset. (C) Countries with a higher or lower level of regional disparity from each geographical group.}
    \label{fig:region_map}
\end{figure*}

\begin{figure*}
    \centering
    \includegraphics[width=.9\linewidth]{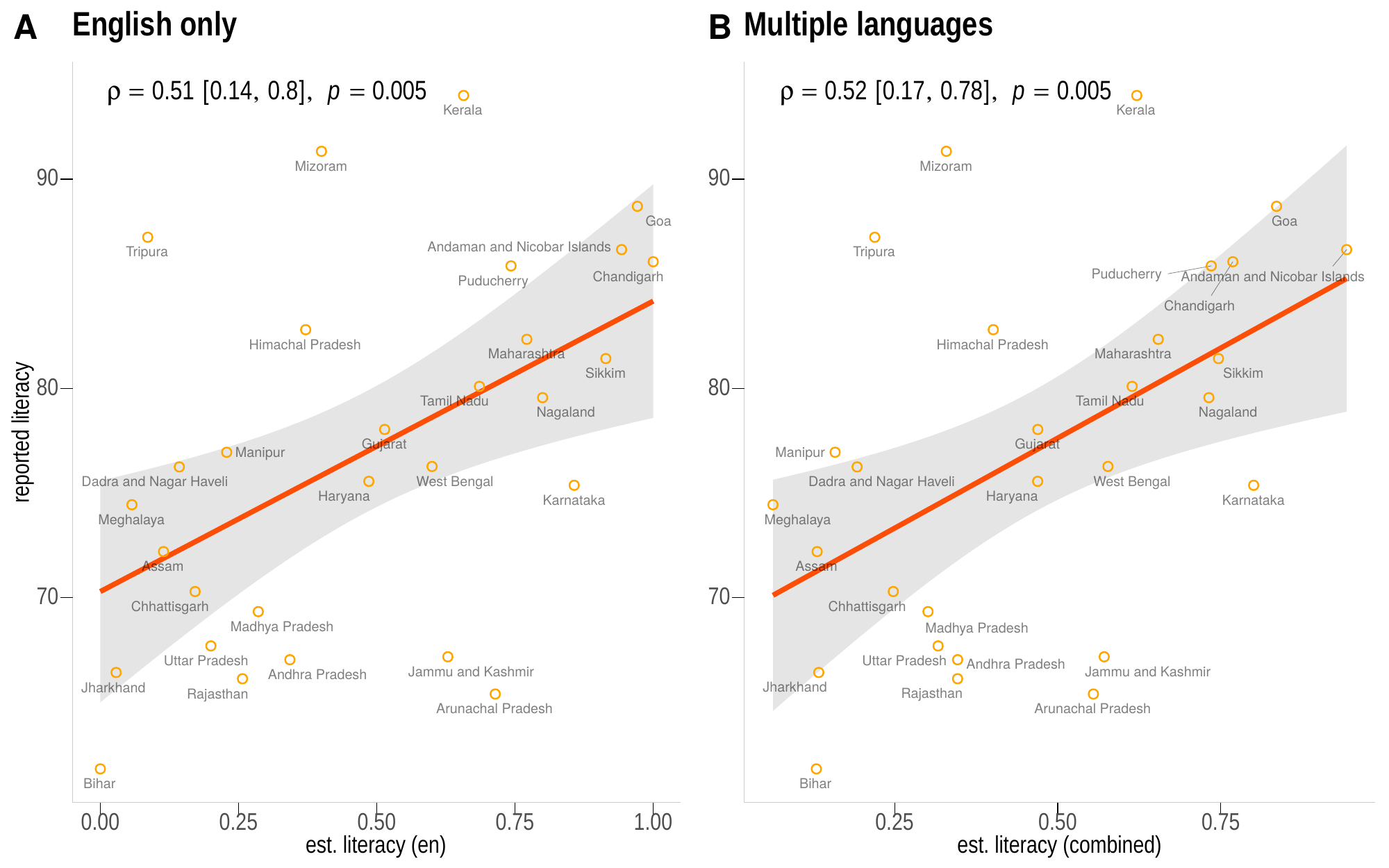}
    \includegraphics[width=.9\linewidth]{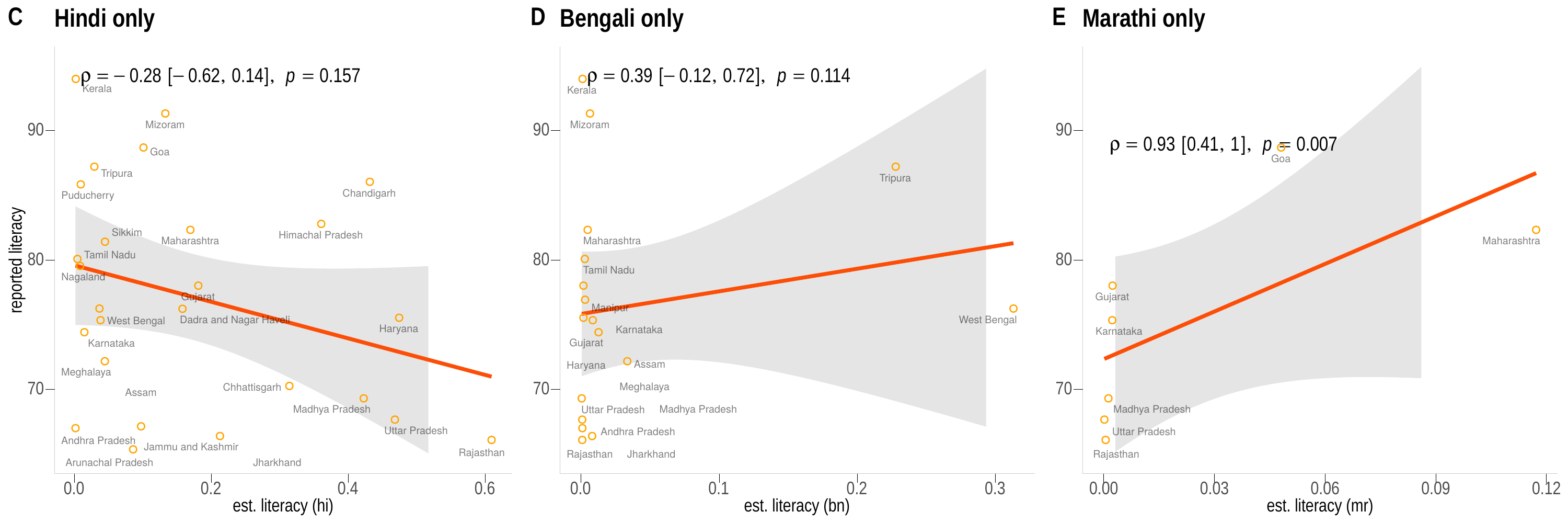}
    \caption{A multilingual country case study with India's user population. We include the six most used languages in India (English, Hindi, Bengali, Marathi, Telugu, and Tamil) to compare the literacy estimation based on the single representative language (English) with the estimation based on multiple languages. The results suggest that literacy estimation based on multiple languages has neglectable improvement over English-based estimation in terms of Spearman's rank correlations. The y-axis indicates the officially reported literacy level, and the x-axes indicate (A) the language literacy estimated using the regions' public posts in English, (B) the estimation using multiple languages combined, and (C-E) the estimation using posts in Hindi, Bengali, and Marathi only, respectively.}
    \label{fig:india}
\end{figure*}

\begin{figure*}
    \centering
    \includegraphics[width=.9\linewidth]{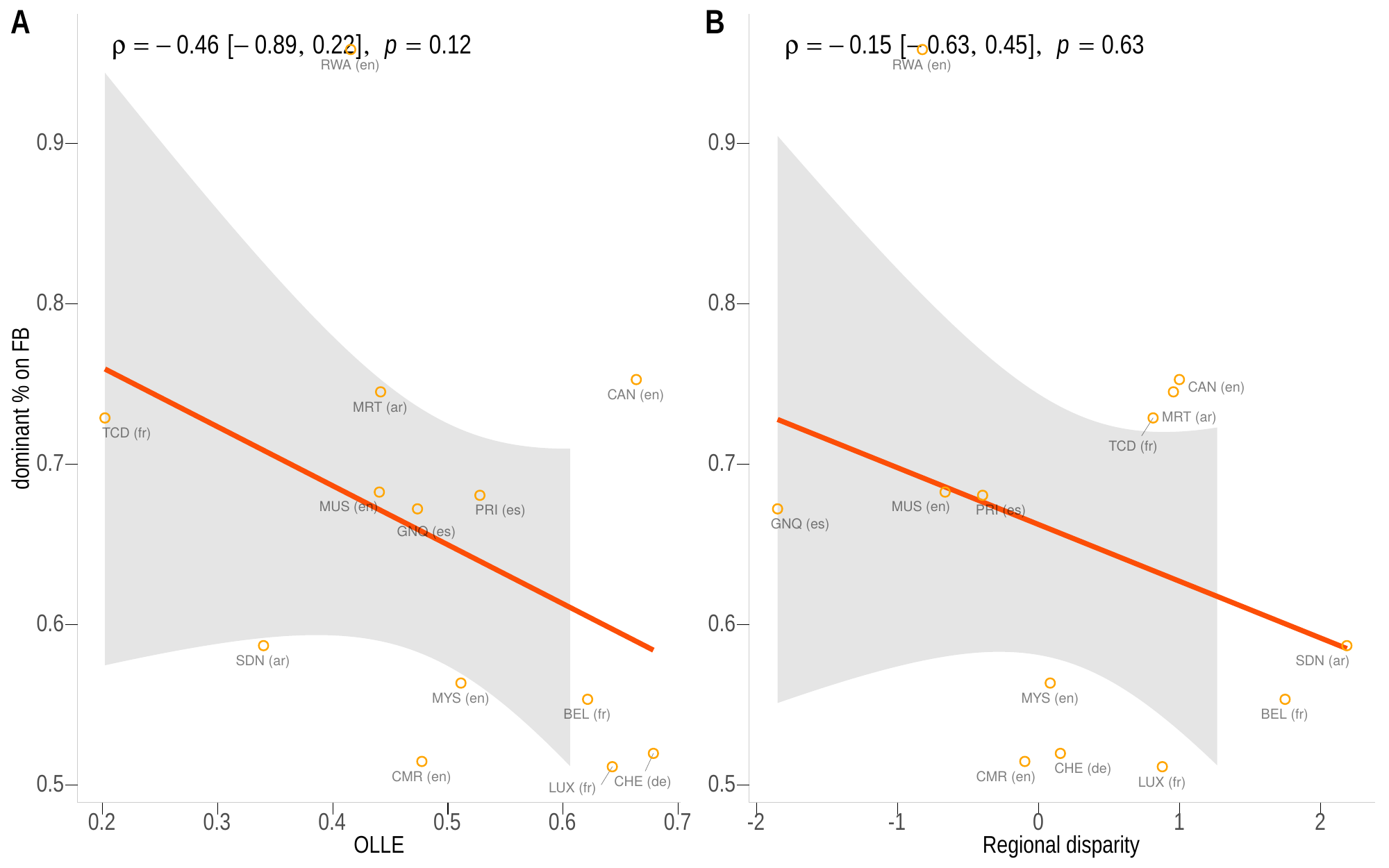}
    \caption{The relationship between the user percentage of the dominant language ($y$-axis) and (A) \olle, or (B) regional disparity, in 13 multilingual countries studied. We do not observe a clear systematic bias. This serves as a robust check to see whether there is a potential bias in the set of multilingual countries. }
    \label{fig:domi}
\end{figure*}
\clearpage
\addcontentsline{toc}{section}{Supplementary Tables}

\begin{table}[]
    \centering
    \caption{Knee points detected based on the Facebook popularity curves shown in Fig.~\ref{fig:word_curve}. The two knee points, measured in 1,000 words, determine the \bigword in each language. For example, the \bigword in English correspond to the words in the fastText `{\texttt en}' dictionary that are ranked between 5,000 to 9,000 in the decreasing order of word frequency.}
    \label{tab:knees}
    \begin{tabular}{c|c|c|c|c |c|c|c|c |c|c|c|c}
    \toprule
        Language & en & es & fr & ar & de & zh & pt & it & ru & tr & ml & ms\\
        \midrule
        $k_0$ & 5 & 5 & 5 & 5 & 5 & 5 & 5 & 5 & 5 & 5 & 5 & 6\\
        $k_1$ & 9 & 9 & 9 & 9 & 9 & 16 & 9 & 11 & 8 & 11 & 10 & 21\\
    \bottomrule
    \end{tabular}
\end{table}
\begin{table*}[!htbp] \centering {\tiny 
  \caption{OLS for predicting the reported literacy rate with online literacy estimates. Model (a) is the fixed-effect model accounting for language-specific bias. The calibrated online literacy estimates (\olles) are produced using model (a). The observations include all countries having online literacy estimates and corresponding predictors. Countries without sufficient Internet penetration ($<25\%$) are excluded to obtain reliable calibrated models. For comparison, models (b,c) include additional predictors, the Internet penetration and income. All variables were transformed to better fit a normal distribution.} 
  \label{tab:model-lit-main} 
\begin{tabular}{@{\extracolsep{-10pt}}lccc} 
\\[-1.8ex]\hline 
\hline \\[-1.8ex] 
\\[-1.8ex] & \multicolumn{3}{c}{DV: reported literacy rate} \\ 
 & (a) & (b) & (c) \\ 
\hline \\[-1.8ex] 
 est. literacy & 0.80$^{***}$ (0.61, 0.99) & 0.52$^{***}$ (0.27, 0.76) & 0.53$^{***}$ (0.27, 0.79) \\ 
  \% Internet &  & 0.32$^{***}$ (0.13, 0.50) & 0.49$^{***}$ (0.19, 0.80) \\ 
  income &  &  & $-$0.19 ($-$0.48, 0.11) \\ 
  language [de] & 1.43$^{***}$ (0.72, 2.14) & 1.22$^{***}$ (0.53, 1.90) & 1.29$^{***}$ (0.59, 1.99) \\ 
  language [en] & 0.94$^{***}$ (0.55, 1.33) & 0.98$^{***}$ (0.61, 1.35) & 0.97$^{***}$ (0.58, 1.37) \\ 
  language [es] & 1.31$^{***}$ (0.82, 1.80) & 1.15$^{***}$ (0.68, 1.62) & 1.25$^{***}$ (0.76, 1.74) \\ 
  language [fr] & 1.43$^{***}$ (0.79, 2.06) & 1.13$^{***}$ (0.51, 1.75) & 1.30$^{***}$ (0.57, 2.03) \\ 
  language [it] & 1.24$^{**}$ (0.26, 2.22) & 1.75$^{***}$ (0.78, 2.73) & 1.57$^{**}$ (0.25, 2.90) \\ 
  language [ms] & $-$0.75 ($-$1.75, 0.25) & $-$0.55 ($-$1.50, 0.40) & $-$0.52 ($-$1.48, 0.44) \\ 
  language [nl] & 0.91$^{**}$ (0.21, 1.62) & 0.78$^{**}$ (0.10, 1.45) & 0.97$^{**}$ (0.21, 1.73) \\ 
  language [pt] & 0.62$^{*}$ ($-$0.06, 1.29) & 0.61$^{*}$ ($-$0.03, 1.26) & 0.66$^{*}$ (0.01, 1.31) \\ 
  language [ru] & 3.41$^{***}$ (2.64, 4.17) & 3.14$^{***}$ (2.41, 3.88) & 3.10$^{***}$ (2.34, 3.86) \\ 
  language [tr] & 0.86$^{*}$ ($-$0.09, 1.81) & 0.96$^{**}$ (0.05, 1.86) & 1.04$^{**}$ (0.12, 1.95) \\ 
  language [zh] & $-$0.62 ($-$1.64, 0.40) & $-$0.35 ($-$1.32, 0.63) &  \\ 
  Constant & $-$0.94$^{***}$ ($-$1.27, $-$0.61) & $-$0.90$^{***}$ ($-$1.21, $-$0.59) & $-$0.94$^{***}$ ($-$1.26, $-$0.61) \\ 
 \hline \\[-1.8ex] 
OOS correlation $\rho$ & 0.78 & 0.8 & 0.79 \\ 
OOS RMSE & 0.7 & 0.66 & 0.68 \\ 
OOS R$^{2}$ & 0.51 & 0.57 & 0.55 \\ 
Observations & 98 & 98 & 86 \\ 
R$^{2}$ & 0.64 & 0.68 & 0.69 \\ 
Adjusted R$^{2}$ & 0.59 & 0.63 & 0.64 \\ 
AIC & 205.52 & 195.32 & 172.75 \\ 
BIC & 241.70 & 234.10 & 209.57 \\ 
Residual Std. Error & 0.64 (df = 85) & 0.61 (df = 84) & 0.61 (df = 72) \\ 
F Statistic & 12.49$^{***}$ (df = 12; 85) & 13.76$^{***}$ (df = 13; 84) & 12.52$^{***}$ (df = 13; 72) \\ 
\hline 
\hline \\[-1.8ex] 
\textit{Note:}  & \multicolumn{3}{r}{$^{*}$p$<$0.1; $^{**}$p$<$0.05; $^{***}$p$<$0.01} \\ 
 & \multicolumn{3}{r}{The out-of-sample (OOS) Spearman correlation $\rho$, RMSE, and R$^{2}$ are obtained using leave-one-out cross-validation.} \\ 
\end{tabular} }
\end{table*} 
\begin{table*}[!htbp] \centering {\tiny 
  \caption{OLS for predicting the reported literacy rate without online literacy estimates. All variables were transformed to better fit a normal distribution.} 
  \label{tab:model-no-lit} 
\begin{tabular}{@{\extracolsep{-10pt}}lccc} 
\\[-1.8ex]\hline 
\hline \\[-1.8ex] 
\\[-1.8ex] & \multicolumn{3}{c}{DV: reported literacy rate} \\ 
 & (a) & (b) & (c) \\ 
\hline \\[-1.8ex] 
 \% Internet & 0.54$^{***}$ (0.37, 0.71) & 0.58$^{***}$ (0.43, 0.73) & 0.74$^{***}$ (0.43, 1.05) \\ 
  income &  &  & $-$0.15 ($-$0.47, 0.18) \\ 
  language [de] &  & 1.14$^{***}$ (0.40, 1.89) & 1.22$^{***}$ (0.45, 1.98) \\ 
  language [en] &  & 0.96$^{***}$ (0.55, 1.36) & 0.95$^{***}$ (0.52, 1.38) \\ 
  language [es] &  & 0.69$^{***}$ (0.23, 1.14) & 0.79$^{***}$ (0.31, 1.27) \\ 
  language [fr] &  & 0.64$^{**}$ (0.02, 1.27) & 0.75$^{*}$ (0.004, 1.50) \\ 
  language [it] &  & 2.55$^{***}$ (1.57, 3.53) & 2.34$^{***}$ (0.95, 3.74) \\ 
  language [ms] &  & 0.08 ($-$0.90, 1.07) & 0.14 ($-$0.86, 1.14) \\ 
  language [nl] &  & 0.70$^{*}$ ($-$0.04, 1.43) & 0.95$^{**}$ (0.11, 1.78) \\ 
  language [pt] &  & 0.33 ($-$0.35, 1.01) & 0.40 ($-$0.30, 1.10) \\ 
  language [ru] &  & 2.51$^{***}$ (1.78, 3.25) & 2.49$^{***}$ (1.72, 3.26) \\ 
  language [tr] &  & 1.17$^{**}$ (0.20, 2.15) & 1.27$^{**}$ (0.27, 2.26) \\ 
  language [zh] &  & 0.42 ($-$0.56, 1.41) &  \\ 
  Constant & 0.00 ($-$0.17, 0.17) & $-$0.78$^{***}$ ($-$1.11, $-$0.44) & $-$0.83$^{***}$ ($-$1.18, $-$0.48) \\ 
 \hline \\[-1.8ex] 
OOS correlation $\rho$ & 0.52 & 0.73 & 0.72 \\ 
OOS RMSE & 0.86 & 0.7 & 0.73 \\ 
OOS R$^{2}$ & 0.26 & 0.51 & 0.48 \\ 
Observations & 98 & 98 & 86 \\ 
R$^{2}$ & 0.29 & 0.62 & 0.62 \\ 
Adjusted R$^{2}$ & 0.28 & 0.56 & 0.56 \\ 
AIC & 249.67 & 211.51 & 188.23 \\ 
BIC & 257.42 & 247.70 & 222.59 \\ 
Residual Std. Error & 0.85 (df = 96) & 0.66 (df = 85) & 0.67 (df = 73) \\ 
F Statistic & 39.04$^{***}$ (df = 1; 96) & 11.32$^{***}$ (df = 12; 85) & 10.10$^{***}$ (df = 12; 73) \\ 
\hline 
\hline \\[-1.8ex] 
\textit{Note:}  & \multicolumn{3}{r}{$^{*}$p$<$0.1; $^{**}$p$<$0.05; $^{***}$p$<$0.01} \\ 
 & \multicolumn{3}{r}{The out-of-sample (OOS) Spearman correlation $\rho$, RMSE, and R$^{2}$ are obtained using leave-one-out cross-validation.} \\ 
\end{tabular} }
\end{table*} 
\clearpage
{
\begingroup
\onecolumn
\tiny \centering
\begin{longtable}[H]{@{\extracolsep{-8pt}}lcrrrrrc}
  \caption{Online language literacy estimates for all countries included in this study. Measures include: average relative \bigword, \olle ($N=167$), female-male gender gap ($N=160$) and regional disparity ($N=119$).} 
  \label{tab:all_country} 
\\
\hline \\[-1.8ex] 
country & code & big word ($\bar{w}$) & \olle & female-male gap & regional disparity & no. regions & rep. language \\ 
\hline \\[-1.8ex] 
Algeria & DZA & 0.584 & 0.309 & -0.584 & 0.022 & 48 & ar \\ 
American Samoa & ASM & 0.811 & 0.593 & 1.744 & -- & -- & en \\ 
Angola & AGO & 0.634 & 0.419 & -0.281 & 0.008 & 17 & pt \\ 
Anguilla & AIA & 0.521 & 0.416 & 0.372 & -- & -- & en \\ 
Antigua \& Barbuda & ATG & 0.715 & 0.519 & 1.422 & -- & -- & en \\ 
Argentina & ARG & 0.584 & 0.505 & 0.166 & 0.03 & 24 & es \\ 
Armenia & ARM & 0.308 & 0.624 & 1.684 & 0.047 & 10 & ru \\ 
Aruba & ABW & 0.785 & 0.58 & 0.476 & -- & -- & nl \\ 
Australia & AUS & 1.004 & 0.688 & 0.456 & 0.019 & 8 & en \\ 
Austria & AUT & 0.786 & 0.667 & 0.481 & 0.03 & 9 & de \\ 
Bahamas & BHS & 0.687 & 0.501 & 1.056 & 0.026 & 2 & en \\ 
Bahrain & BHR & 0.784 & 0.429 & -2.293 & 0.084 & 2 & ar \\ 
Barbados & BRB & 0.743 & 0.544 & 1.325 & -- & -- & en \\ 
Belarus & BLR & 0.709 & 0.868 & 0.746 & 0.067 & 7 & ru \\ 
Belgium & BEL & 0.876 & 0.621 & -0.411 & 0.117 & 3 & nl \\ 
Belize & BLZ & 0.729 & 0.528 & 1.011 & 0.021 & 4 & en \\ 
Benin & BEN & 0.548 & 0.498 & -1.861 & 0.042 & 12 & fr \\ 
Bermuda & BMU & 0.955 & 0.659 & 0.551 & -- & -- & en \\ 
Bolivia & BOL & 0.437 & 0.439 & -0.262 & 0.033 & 9 & es \\ 
Botswana & BWA & 0.535 & 0.422 & 0.144 & 0.004 & 8 & en \\ 
Brazil & BRA & 0.557 & 0.376 & 0.469 & 0.026 & 27 & pt \\ 
British Virgin Islands & VGB & 0.688 & 0.501 & 1.408 & -- & -- & en \\ 
Brunei & BRN & 1.522 & 0.468 & 0.465 & -- & -- & ms \\ 
Burkina Faso & BFA & 0.303 & 0.327 & -1.921 & 0.074 & 6 & fr \\ 
Burundi & BDI & 0.141 & 0.211 & -0.324 & -- & -- & fr \\ 
Cameroon & CMR & 0.638 & 0.477 & 0.016 & 0.027 & 10 & en \\ 
Canada & CAN & 0.965 & 0.664 & 0.695 & 0.065 & 13 & en \\ 
Cape Verde & CPV & 0.682 & 0.44 & 0.379 & 0.029 & 6 & pt \\ 
Caribbean Netherlands & BES & 0.849 & 0.609 & -- & -- & -- & nl \\ 
Cayman Islands & CYM & 0.872 & 0.618 & 0.949 & -- & -- & en \\ 
Central African Republic & CAF & 0.317 & 0.36 & 0 & -- & -- & fr \\ 
Chad & TCD & 0.418 & 0.202 & -3.344 & 0.056 & 2 & ar \\ 
Chile & CHL & 0.585 & 0.512 & 0.37 & 0.015 & 15 & es \\ 
Colombia & COL & 0.436 & 0.434 & 0.328 & 0.039 & 32 & es \\ 
Comoros & COM & 0.238 & 0.027 & -- & -- & -- & ar \\ 
Congo - Brazzaville & COG & 0.316 & 0.36 & -0.791 & 0.001 & 2 & fr \\ 
Congo - Kinshasa & COD & 0.312 & 0.359 & -1.552 & 0.095 & 11 & fr \\ 
Cook Islands & COK & 0.848 & 0.61 & -- & -- & -- & en \\ 
Costa Rica & CRI & 0.486 & 0.462 & -0.039 & 0.043 & 7 & es \\ 
Côte d’Ivoire & CIV & 0.299 & 0.313 & -1.604 & 0.032 & 14 & fr \\ 
Cuba & CUB & 0.475 & 0.459 & -0.078 & 0.018 & 16 & es \\ 
Curaçao & CUW & 0.795 & 0.587 & 0.022 & -- & -- & nl \\ 
Cyprus & CYP & 0.846 & 0.601 & -0.159 & 0.009 & 3 & tr \\ 
Djibouti & DJI & 0.468 & 0.258 & -2.006 & -- & -- & ar \\ 
Dominica & DMA & 0.621 & 0.467 & 0.227 & -- & -- & en \\ 
Dominican Republic & DOM & 0.386 & 0.352 & -0.189 & 0.029 & 31 & es \\ 
Ecuador & ECU & 0.451 & 0.453 & 0.002 & 0.034 & 24 & es \\ 
Egypt & EGY & 0.68 & 0.347 & -0.145 & 0.028 & 27 & ar \\ 
El Salvador & SLV & 0.435 & 0.429 & -0.937 & 0.033 & 14 & es \\ 
Equatorial Guinea & GNQ & 0.536 & 0.474 & 0.119 & 0.007 & 2 & es \\ 
Fiji & FJI & 0.678 & 0.492 & 0.482 & 0.016 & 3 & en \\ 
France & FRA & 0.767 & 0.632 & 1.094 & 0.025 & 22 & fr \\ 
French Guiana & GUF & 0.473 & 0.479 & 0.105 & -- & -- & fr \\ 
French Polynesia & PYF & 0.494 & 0.484 & 0.941 & -- & -- & fr \\ 
Gabon & GAB & 0.311 & 0.359 & -1.445 & 0.059 & 3 & fr \\ 
Gambia & GMB & 0.522 & 0.416 & -0.048 & 0.089 & 2 & en \\ 
Germany & DEU & 0.854 & 0.695 & 0.113 & 0.05 & 16 & de \\ 
Ghana & GHA & 0.576 & 0.448 & -0.697 & 0.014 & 10 & en \\ 
Gibraltar & GIB & 0.945 & 0.651 & 0.774 & -- & -- & en \\ 
Grenada & GRD & 0.704 & 0.507 & 1.918 & 0.051 & 2 & en \\ 
Guadeloupe & GLP & 0.613 & 0.538 & 0.882 & -- & -- & fr \\ 
Guam & GUM & 0.772 & 0.567 & 1.475 & -- & -- & en \\ 
Guatemala & GTM & 0.403 & 0.383 & -0.526 & 0.039 & 22 & es \\ 
Guinea & GIN & 0.348 & 0.366 & 0.085 & 0.076 & 8 & fr \\ 
Guinea-Bissau & GNB & 0.299 & 0.183 & 0.842 & -- & -- & pt \\ 
Guyana & GUY & 0.65 & 0.48 & 0.871 & 0.035 & 4 & en \\ 
Haiti & HTI & 0.28 & 0.265 & -1.976 & 0.047 & 9 & fr \\ 
Honduras & HND & 0.387 & 0.364 & -0.281 & 0.024 & 18 & es \\ 
Hong Kong SAR China & HKG & 2.021 & 0.57 & 0.02 & -- & -- & zh \\ 
India & IND & 0.423 & 0.361 & 0.887 & 0.044 & 34 & en \\ 
Iraq & IRQ & 0.745 & 0.399 & -1.859 & 0.109 & 19 & ar \\ 
Ireland & IRL & 0.944 & 0.65 & 1.235 & 0.02 & 26 & en \\ 
Isle of Man & IMN & 1.086 & 0.693 & 0.335 & -- & -- & en \\ 
Italy & ITA & 1.159 & 0.757 & -0.109 & 0.006 & 20 & it \\ 
Jamaica & JAM & 0.561 & 0.437 & 0.772 & 0.021 & 11 & en \\ 
Jersey & JEY & 0.949 & 0.655 & 1.234 & -- & -- & en \\ 
Jordan & JOR & 0.918 & 0.496 & 0.01 & 0.02 & 12 & ar \\ 
Kenya & KEN & 0.62 & 0.467 & -0.208 & 0.021 & 8 & en \\ 
Kiribati & KIR & 0.557 & 0.433 & 0.751 & -- & -- & en \\ 
Kuwait & KWT & 0.759 & 0.405 & -2.727 & 0.061 & 6 & ar \\ 
Kyrgyzstan & KGZ & 0.299 & 0.599 & 1.485 & 0.108 & 2 & ru \\ 
Lebanon & LBN & 0.734 & 0.386 & -0.771 & 0.052 & 6 & ar \\ 
Lesotho & LSO & 0.485 & 0.41 & -0.03 & 0.088 & 6 & en \\ 
Liberia & LBR & 0.74 & 0.538 & 0.483 & 0.053 & 2 & en \\ 
Libya & LBY & 0.697 & 0.355 & -0.143 & 0.017 & 19 & ar \\ 
Liechtenstein & LIE & 0.779 & 0.656 & -- & -- & -- & de \\ 
Luxembourg & LUX & 0.775 & 0.643 & 1.075 & 0.059 & 2 & fr \\ 
Macau SAR China & MAC & 1.69 & 0.509 & 0.433 & -- & -- & zh \\ 
Madagascar & MDG & 0.252 & 0.253 & 0.078 & 0.012 & 18 & fr \\ 
Malawi & MWI & 0.522 & 0.416 & -0.09 & 0.036 & 5 & en \\ 
Malaysia & MYS & 1.807 & 0.511 & 0.936 & 0.032 & 15 & ms \\ 
Mali & MLI & 0.244 & 0.25 & -0.537 & 0.051 & 6 & fr \\ 
Malta & MLT & 0.718 & 0.525 & 0.733 & 0.029 & 4 & en \\ 
Marshall Islands & MHL & 0.729 & 0.531 & 1.173 & -- & -- & en \\ 
Mauritania & MRT & 0.8 & 0.442 & -2.055 & 0.063 & 2 & ar \\ 
Mauritius & MUS & 0.568 & 0.44 & 0.59 & 0.018 & 9 & en \\ 
Mayotte & MYT & 0.406 & 0.406 & 0.773 & -- & -- & fr \\ 
Mexico & MEX & 0.428 & 0.418 & 0.318 & 0.037 & 32 & es \\ 
Micronesia (Federated States of) & FSM & 0.532 & 0.418 & 0.139 & 0.007 & 2 & en \\ 
Monaco & MCO & 0.883 & 0.7 & -- & -- & -- & fr \\ 
Morocco & MAR & 0.628 & 0.322 & -1.058 & 0.026 & 16 & ar \\ 
Mozambique & MOZ & 0.288 & 0.148 & -1.678 & 0.051 & 10 & pt \\ 
Nauru & NRU & 0.611 & 0.465 & -- & -- & -- & en \\ 
Netherlands & NLD & 0.839 & 0.603 & -0.215 & 0.025 & 12 & nl \\ 
New Caledonia & NCL & 0.469 & 0.478 & 0.032 & 0.039 & 2 & fr \\ 
New Zealand & NZL & 1.001 & 0.679 & 0.686 & 0.016 & 16 & en \\ 
Nicaragua & NIC & 0.42 & 0.406 & -0.553 & 0.031 & 17 & es \\ 
Niger & NER & 0.294 & 0.299 & -2.103 & 0.061 & 4 & fr \\ 
Nigeria & NGA & 0.814 & 0.597 & 0.002 & 0.064 & 37 & en \\ 
Northern Mariana Islands & MNP & 0.708 & 0.51 & 1.28 & -- & -- & en \\ 
Oman & OMN & 0.796 & 0.441 & -2.218 & 0.088 & 3 & ar \\ 
Pakistan & PAK & 0.455 & 0.403 & 0.262 & 0.037 & 8 & en \\ 
Palestinian Territories & PSE & 0.905 & 0.49 & 0.266 & 0.019 & 2 & ar \\ 
Panama & PAN & 0.446 & 0.449 & 0.005 & 0.043 & 8 & es \\ 
Papua New Guinea & PNG & 0.539 & 0.423 & 0.013 & 0.019 & 17 & en \\ 
Paraguay & PRY & 0.429 & 0.424 & -0.883 & 0.035 & 17 & es \\ 
Peru & PER & 0.535 & 0.47 & 0.001 & 0.009 & 25 & es \\ 
Philippines & PHL & 0.551 & 0.426 & 0.735 & 0.019 & 17 & en \\ 
Portugal & PRT & 0.711 & 0.458 & -0.41 & 0.018 & 20 & pt \\ 
Puerto Rico & PRI & 0.632 & 0.528 & 0.375 & 0.022 & 68 & es \\ 
Qatar & QAT & 0.812 & 0.444 & -2.442 & -- & -- & ar \\ 
Réunion & REU & 0.579 & 0.522 & 0.461 & -- & -- & fr \\ 
Rwanda & RWA & 0.521 & 0.416 & -0.037 & 0.015 & 5 & en \\ 
Saint Martin (French part) & MAF & 0.699 & 0.576 & -- & -- & -- & fr \\ 
Samoa & WSM & 0.655 & 0.483 & 0.725 & -- & -- & en \\ 
San Marino & SMR & 1.208 & 0.768 & -0.191 & -- & -- & it \\ 
São Tomé \& Príncipe & STP & 0.399 & 0.265 & 0.168 & -- & -- & pt \\ 
Saudi Arabia & SAU & 0.764 & 0.409 & -1.981 & 0.026 & 13 & ar \\ 
Senegal & SEN & 0.285 & 0.267 & -1.618 & 0.035 & 14 & fr \\ 
Seychelles & SYC & 0.66 & 0.486 & 0.931 & -- & -- & en \\ 
Sierra Leone & SLE & 0.759 & 0.553 & 0.714 & 0.076 & 4 & en \\ 
Singapore & SGP & 0.742 & 0.541 & 1.602 & -- & -- & en \\ 
Solomon Islands & SLB & 0.545 & 0.424 & 0.081 & -- & -- & en \\ 
Somalia & SOM & 0.608 & 0.317 & -0.959 & 0.04 & 5 & ar \\ 
South Africa & ZAF & 0.514 & 0.414 & 0.052 & 0.018 & 9 & en \\ 
South Sudan & SSD & 0.605 & 0.464 & -0.514 & 0.01 & 2 & en \\ 
Spain & ESP & 0.892 & 0.684 & 0.033 & 0.023 & 17 & es \\ 
St. Kitts \& Nevis & KNA & 0.753 & 0.55 & 1.657 & 0.003 & 2 & en \\ 
St. Lucia & LCA & 0.631 & 0.473 & 0.792 & 0.02 & 3 & en \\ 
St. Vincent \& Grenadines & VCT & 0.641 & 0.478 & 0.183 & 0.043 & 2 & en \\ 
Sudan & SDN & 0.673 & 0.34 & -0.032 & 0.165 & 9 & ar \\ 
Suriname & SUR & 0.572 & 0.442 & -0.185 & 0.031 & 2 & nl \\ 
Swaziland & SWZ & 0.555 & 0.43 & -0.326 & 0.016 & 4 & en \\ 
Switzerland & CHE & 0.812 & 0.678 & -0.113 & 0.033 & 22 & de \\ 
Syria & SYR & 0.7 & 0.356 & -0.107 & 0.105 & 14 & ar \\ 
Taiwan & TWN & 2.101 & 0.583 & 0.565 & 0.033 & 16 & zh \\ 
Tajikistan & TJK & 0.229 & 0.527 & 2.07 & 0.005 & 2 & ru \\ 
Tanzania & TZA & 0.444 & 0.397 & -0.037 & 0.012 & 19 & en \\ 
Timor-Leste & TLS & 0.416 & 0.29 & -0.193 & -- & -- & pt \\ 
Togo & TGO & 0.273 & 0.262 & -0.839 & -- & -- & fr \\ 
Tonga & TON & 0.739 & 0.538 & 1.969 & -- & -- & en \\ 
Trinidad \& Tobago & TTO & 0.677 & 0.489 & 0.948 & 0.028 & 13 & en \\ 
Tunisia & TUN & 0.709 & 0.362 & 0.058 & 0.05 & 24 & ar \\ 
Turkey & TUR & 0.835 & 0.593 & 0.055 & 0.037 & 78 & tr \\ 
Turks \& Caicos Islands & TCA & 0.676 & 0.489 & 0.935 & -- & -- & en \\ 
Uganda & UGA & 0.624 & 0.469 & 0.06 & 0.035 & 22 & en \\ 
Ukraine & UKR & 0.584 & 0.806 & 0.625 & 0.044 & 27 & ru \\ 
United Arab Emirates & ARE & 0.718 & 0.374 & -1.421 & 0.025 & 6 & ar \\ 
United Kingdom & GBR & 0.987 & 0.671 & 0.807 & 0.011 & 4 & en \\ 
United States & USA & 0.951 & 0.657 & 0.755 & 0.017 & 51 & en \\ 
Uruguay & URY & 0.572 & 0.499 & 0.104 & 0.02 & 19 & es \\ 
U.S. Virgin Islands & VIR & 0.842 & 0.607 & 1.41 & -- & -- & en \\ 
Uzbekistan & UZB & 0.223 & 0.524 & 1.558 & 0.018 & 9 & ru \\ 
Venezuela & VEN & 0.437 & 0.444 & -1.165 & 0.028 & 24 & es \\ 
Yemen & YEM & 0.979 & 0.52 & -0.703 & 0.048 & 7 & ar \\ 
Zambia & ZMB & 0.568 & 0.441 & -0.083 & 0.016 & 9 & en \\ 
Zimbabwe & ZWE & 0.602 & 0.464 & -0.295 & 0.024 & 8 & en \\ 
\hline \\[-1.8ex] 
\multicolumn{8}{l}{The representative language for each country is chosen as the most used language by the country's population observed on Facebook.} \\ 
\end{longtable} 
\endgroup
}
\begin{table*}[!htbp] \centering {\tiny
  \caption{Variables related to gender gap analysis. Reported $N$ is the number of countries matched with our data.} 
  \label{tab:gender_vars}
\begin{tabular}{@{\extracolsep{-4pt}}lrrrrrrll} 
\\[-1.8ex]\hline 
\hline \\[-1.8ex] 
Statistic & \multicolumn{1}{c}{N} & \multicolumn{1}{c}{Mean} & \multicolumn{1}{c}{St. Dev.} & \multicolumn{1}{c}{Min} & \multicolumn{1}{c}{Max} & \multicolumn{1}{c}{Median} & \multicolumn{1}{l}{Definition} & \multicolumn{1}{l}{Source} \\ 
\hline \\[-1.8ex] 
\olle gap & 160 & 0.053 & 1.002 & $-$3.344 & 2.070 & 0.080 & female-map gap in \olle &\\ 
offline literacy (all) & 143 & 84.181 & 19.263 & 19.100 & 100.000 & 93.464 & literacy rate & UNESCO \cite{unesco2019produce}\\  
offline literacy (gap) & 114 & $-$0.068 & 0.090 & $-$0.301 & 0.182 & $-$0.036 & female-male gap in literacy & UNESCO \cite{unesco2019produce}\\ 
education (all) & 106 & 7.884 & 2.741 & 1.880 & 13.180 & 8.085 & mean schooling years & Barro-Lee Educational Attainment Data \cite{BarroLee73:online}\\ 
education (gap) & 106 & $-$0.517 & 0.968 & $-$3.250 & 1.600 & $-$0.420 & female-map gap in schooling years & Barro-Lee Educational Attainment Data \cite{BarroLee73:online}\\ 
\% Internet (all) & 151 & 0.520 & 0.284 & 0.020 & 0.984 & 0.555 & overall Internet penetration & ITU Internet gender gap \cite{Aboutthe43:online}\\
\% Internet (gap) & 128 & 0.880 & 0.123 & 0.545 & 1.000 & 0.919 & femal-male gap in Internet penetration & Digital gender gap (U. Oxford) \cite{Aboutthe43:online} \\ 
civic (all) & 123 & 0.695 & 0.211 & 0.105 & 0.973 & 0.746 & overall civic society participation & V-Dem Institute \cite{coppedge2019v}\\ 
civic (women) & 123 & 0.720 & 0.173 & 0.234 & 0.937 & 0.775 & women's civic society participation & V-Dem Institute \cite{coppedge2019v}\\ 
GII & 107 & 0.398 & 0.194 & 0.040 & 0.835 & 0.424 & Gender Inequality Index & HDRO \cite{HDROAPII26:online}\\  
\hline \\[-1.8ex] 
\end{tabular} }
\end{table*} 
\begin{table*}[!htbp] \centering {\tiny 
  \caption{Correlations among variables related to gender gap or gender equity. All correlations are reported using Spearman rank correlation coefficients.} 
  \label{tab:gender_corr} 
\begin{tabular}{@{\extracolsep{-4pt}} lllllllllll} 
\\[-1.8ex]\hline 
\hline \\[-1.8ex] 
 & \olle gap & \olle & \shortstack[l]{offline\\[-.5ex]literacy (all)} & \shortstack[l]{offline\\[-.5ex]literacy (gap)} & eduation (all) & education (gap) & \% Internet (all) & \% Internet (gap) & civic (all) & civic (women) \\ 
\hline \\[-1.8ex] 
\olle &  0.585$^*$$^*$$^*$ &  &  &  &  &  &  &  &  &  \\ 
off. literacy (all) &  0.524$^*$$^*$$^*$ &  0.740$^*$$^*$$^*$ &  &  &  &  &  &  &  &  \\ 
off. literacy (gap) &  0.420$^*$$^*$$^*$ &  0.438$^*$$^*$$^*$ &  0.709$^*$$^*$$^*$ &  &  &  &  &  &  &  \\ 
eduation (all) &  0.587$^*$$^*$$^*$ &  0.739$^*$$^*$$^*$ &  0.872$^*$$^*$$^*$ &  0.694$^*$$^*$$^*$ &  &  &  &  &  &  \\ 
education (gap) &  0.225$^*$  &  0.207$^*$  &  0.515$^*$$^*$$^*$ &  0.787$^*$$^*$$^*$ &  0.455$^*$$^*$$^*$ &  &  &  &  &  \\ 
\% Internet (all) &  0.304$^*$$^*$$^*$ &  0.573$^*$$^*$$^*$ &  0.748$^*$$^*$$^*$ &  0.582$^*$$^*$$^*$ &  0.743$^*$$^*$$^*$ &  0.434$^*$$^*$$^*$ &  &  &  &  \\ 
\% Internet (gap) &  0.487$^*$$^*$$^*$ &  0.533$^*$$^*$$^*$ &  0.663$^*$$^*$$^*$ &  0.779$^*$$^*$$^*$ &  0.730$^*$$^*$$^*$ &  0.620$^*$$^*$$^*$ &  0.617$^*$$^*$$^*$ &  &  &  \\ 
civic (all) &  0.287$^*$$^*$  &  0.355$^*$$^*$$^*$ &  0.159  & -0.019  &  0.358$^*$$^*$$^*$ &  0.028  &  0.228$^*$  &  0.328$^*$$^*$$^*$ &  &  \\ 
civic (women) &  0.479$^*$$^*$$^*$ &  0.484$^*$$^*$$^*$ &  0.394$^*$$^*$$^*$ &  0.317$^*$$^*$  &  0.564$^*$$^*$$^*$ &  0.252$^*$  &  0.429$^*$$^*$$^*$ &  0.598$^*$$^*$$^*$ &  0.695$^*$$^*$$^*$ &  \\ 
GII & -0.391$^*$$^*$$^*$ & -0.624$^*$$^*$$^*$ & -0.847$^*$$^*$$^*$ & -0.610$^*$$^*$$^*$ & -0.838$^*$$^*$$^*$ & -0.475$^*$$^*$$^*$ & -0.872$^*$$^*$$^*$ & -0.647$^*$$^*$$^*$ & -0.219$^*$  & -0.414$^*$$^*$$^*$ \\ 
\hline \\[-1.8ex] 
\multicolumn{11}{l}{significance levels: $^{***}p<0.001$; $^{**}p<0.01$; $^*p<0.05$} \\ 
\end{tabular} }
\end{table*} 

\begin{table*}[!htbp] \centering {\tiny
  \caption{Variables related to country resource access and inequality. Reported $N$ is the number of countries matched with our data.} 
  \label{tab:region_vars} 
\begin{tabular}{@{\extracolsep{-4pt}}lrrrrrrll} 
\\[-1.8ex]\hline 
\hline \\[-1.8ex] 
Statistic & \multicolumn{1}{c}{N} & \multicolumn{1}{c}{Mean} & \multicolumn{1}{c}{St. Dev.} & \multicolumn{1}{c}{Min} & \multicolumn{1}{c}{Max} & \multicolumn{1}{c}{Median} & \multicolumn{1}{l}{Definition} & \multicolumn{1}{l}{Source} \\  
\hline \\[-1.8ex] 
regional disparity & 119 & 0.037 & 0.027 & 0.001 & 0.165 & 0.030 & St. Dev. of sub-national \olle \\  
income & 115 & 16.193 & 17.054 & 0.800 & 71.160 & 9.359 & GNI per capita in 1,000 US\$ (2011 PPP) & HDI \cite{GlobalDa32:online}\\ 
Gini index & 94 & 40.004 & 7.973 & 25.000 & 63.000 & 40.200 & Gini coefficient for income & HDR \cite{HumanDev90:online}\\  
education (all) & 95 & 7.864 & 2.753 & 1.880 & 13.180 & 7.970 & mean schooling years & Barro-Lee Educational Attainment Data \cite{BarroLee73:online}\\ 
unequal education & 102 & 21.025 & 14.073 & 0.800 & 49.300 & 17.450 & Inequality in education & HDR \cite{HumanDev90:online}\\  
\% Internet (all) & 119 & 0.507 & 0.275 & 0.020 & 0.980 & 0.508 & overall Internet penetration & ITU Internet gender gap \cite{Aboutthe43:online}\\ 
civic (all) & 110 & 0.704 & 0.213 & 0.105 & 0.973 & 0.764 & overall civic society participation & V-Dem Institute \cite{coppedge2019v}\\ 
\hline \\[-1.8ex] 
\end{tabular} }
\end{table*} 
\begin{table*}[!htbp] \centering {\tiny 
  \caption{Correlations among variables related to country resource access and inequality. All correlations are reported using Spearman rank correlation coefficients.} 
  \label{tab:region_corr} 
\begin{tabular}{@{\extracolsep{-4pt}} llllllll} 
\\[-1.8ex]\hline 
\hline \\[-1.8ex] 
 & regional disparity & \olle & income & Gini index & education (all) & unequal education & \% Internet (all) \\ 
\hline \\[-1.8ex] 
regional disparity &  &  &  &  &  &  &  \\ 
\olle & -0.158  &  &  &  &  &  &  \\ 
income & -0.156  &  0.476$^*$$^*$$^*$ &  &  &  &  &  \\ 
Gini index & -0.271$^*$$^*$  & -0.346$^*$$^*$$^*$ & -0.260$^*$  &  &  &  &  \\ 
eduation (all) & -0.312$^*$$^*$  &  0.706$^*$$^*$$^*$ &  0.745$^*$$^*$$^*$ & -0.303$^*$$^*$  &  &  &  \\ 
unequal education &  0.283$^*$$^*$  & -0.681$^*$$^*$$^*$ & -0.703$^*$$^*$$^*$ &  0.240$^*$  & -0.875$^*$$^*$$^*$ &  &  \\ 
\% Internet (all) & -0.043  &  0.532$^*$$^*$$^*$ &  0.905$^*$$^*$$^*$ & -0.306$^*$$^*$  &  0.751$^*$$^*$$^*$ & -0.735$^*$$^*$$^*$ &  \\ 
civic (all) & -0.012  &  0.344$^*$$^*$$^*$ &  0.192$^*$  & -0.154  &  0.366$^*$$^*$$^*$ & -0.207$^*$  &  0.237$^*$  \\ 
\hline \\[-1.8ex] 
\multicolumn{8}{l}{significance levels: $^{***}p<0.001$; $^{**}p<0.01$; $^*p<0.05$} \\ 
\end{tabular} }
\end{table*}

\begin{table*}[!htbp] \centering {\tiny 
  \caption{OLS for predicting female-male gender gap in online literacy estimates. Models (d) corresponds to the figure in the main text. Models (b,c,e,f) include alternative interaction terms. Values in parentheses are the lower and upper bounds of the 95\% confidence intervals of the estimated effects.} 
  \label{tab:model-gender-main} 
\begin{tabular}{@{\extracolsep{-10pt}}lcccccc} 
\\[-1.8ex]\hline 
\hline \\[-1.8ex] 
\\[-1.8ex] & \multicolumn{6}{c}{DV: female-male gender gap} \\ 
 & (a) & (b) & (c) & (d) & (e) & (f) \\ 
\hline \\[-1.8ex] 
 women civic & 0.32$^{***}$ & 0.33$^{***}$ & 0.35$^{***}$ & 0.38$^{***}$ & 0.37$^{***}$ & 0.40$^{***}$ \\ 
  & (0.15, 0.48) & (0.14, 0.52) & (0.17, 0.52) & (0.20, 0.56) & (0.18, 0.56) & (0.22, 0.57) \\ 
  & & & & & & \\ 
 (women civic):(\% Internet) & 0.25$^{***}$ &  &  & 0.13$^{*}$ &  &  \\ 
  & (0.10, 0.41) &  &  & ($-$0.02, 0.29) &  &  \\ 
  & & & & & & \\ 
 (women civic):(education) &  & 0.28 &  &  & 0.19 &  \\ 
  &  & ($-$0.10, 0.66) &  &  & ($-$0.14, 0.53) &  \\ 
  & & & & & & \\ 
 education (all) & 0.25$^{***}$ & 0.37$^{***}$ & 0.90$^{**}$ & 0.20$^{**}$ & 0.28$^{**}$ & 0.66$^{**}$ \\ 
  & (0.08, 0.42) & (0.13, 0.61) & (0.19, 1.61) & (0.05, 0.35) & (0.06, 0.49) & (0.03, 1.28) \\ 
  & & & & & & \\ 
 (education):(\% Internet) &  &  & 0.51$^{*}$ &  &  & 0.36 \\ 
  &  &  & ($-$0.02, 1.05) &  &  & ($-$0.11, 0.84) \\ 
  & & & & & & \\ 
 \% Internet (all) & $-$0.02 & $-$0.08 & $-$0.23$^{*}$ & $-$0.22$^{*}$ & $-$0.28$^{**}$ & $-$0.41$^{***}$ \\ 
  & ($-$0.21, 0.16) & ($-$0.27, 0.11) & ($-$0.50, 0.03) & ($-$0.45, 0.01) & ($-$0.50, $-$0.06) & ($-$0.69, $-$0.13) \\ 
  & & & & & & \\ 
 Central/Southern/Eastern Asia &  &  &  & 1.66$^{***}$ & 1.77$^{***}$ & 1.80$^{***}$ \\ 
  &  &  &  & (1.05, 2.27) & (1.17, 2.37) & (1.22, 2.39) \\ 
  & & & & & & \\ 
 Europe/Oceania/Northern America &  &  &  & 0.63$^{**}$ & 0.76$^{***}$ & 0.79$^{***}$ \\ 
  &  &  &  & (0.03, 1.22) & (0.20, 1.33) & (0.25, 1.34) \\ 
  & & & & & & \\ 
 Latin America \& the Caribbean &  &  &  & 0.29 & 0.32 & 0.38$^{*}$ \\ 
  &  &  &  & ($-$0.15, 0.74) & ($-$0.12, 0.77) & ($-$0.06, 0.82) \\ 
  & & & & & & \\ 
 Northern Africa \& Western Asia &  &  &  & 0.29 & 0.30 & 0.41 \\ 
  &  &  &  & ($-$0.26, 0.84) & ($-$0.26, 0.85) & ($-$0.14, 0.96) \\ 
  & & & & & & \\ 
 Constant & $-$0.35$^{***}$ & $-$0.30$^{***}$ & $-$0.47$^{***}$ & $-$0.69$^{***}$ & $-$0.72$^{***}$ & $-$0.89$^{***}$ \\ 
  & ($-$0.51, $-$0.18) & ($-$0.48, $-$0.13) & ($-$0.75, $-$0.19) & ($-$1.02, $-$0.37) & ($-$1.05, $-$0.40) & ($-$1.28, $-$0.50) \\ 
  & & & & & & \\ 
\hline \\[-1.8ex] 
Out-of-sample RMSE & 0.8 & 0.83 & 0.83 & 0.72 & 0.73 & 0.73 \\ 
Out-of-sample R2 & 0.24 & 0.21 & 0.2 & 0.39 & 0.38 & 0.38 \\ 
Observations & 101 & 101 & 101 & 101 & 101 & 101 \\ 
R$^{2}$ & 0.34 & 0.29 & 0.30 & 0.51 & 0.51 & 0.51 \\ 
Adjusted R$^{2}$ & 0.31 & 0.26 & 0.27 & 0.47 & 0.46 & 0.47 \\ 
AIC & 240.92 & 248.67 & 247.25 & 217.60 & 219.21 & 218.11 \\ 
BIC & 256.61 & 264.36 & 262.94 & 243.75 & 245.36 & 244.26 \\ 
Residual Std. Error & 0.77 (df = 96) & 0.80 (df = 96) & 0.80 (df = 96) & 0.67 (df = 92) & 0.68 (df = 92) & 0.68 (df = 92) \\ 
F Statistic & 12.26$^{***}$ (df = 4; 96) & 9.58$^{***}$ (df = 4; 96) & 10.06$^{***}$ (df = 4; 96) & 12.19$^{***}$ (df = 8; 92) & 11.82$^{***}$ (df = 8; 92) & 12.07$^{***}$ (df = 8; 92) \\ 
\hline 
\hline \\[-1.8ex] 
\textit{Note:}  & \multicolumn{6}{r}{$^{*}$p$<$0.1; $^{**}$p$<$0.05; $^{***}$p$<$0.01} \\ 
\end{tabular} }
\end{table*}

\begin{table*}[!htbp] \centering { \tiny 
  \caption{OLS for predicting female-male gender gap in online literacy estimates. Models include alternative predictors and no interaction term. Values in parentheses are the lower and upper bounds of the 95\% confidence intervals of the estimated effects.} 
  \label{tab:model-gender-3v} 
\begin{tabular}{@{\extracolsep{-10pt}}lcccccc} 
\\[-1.8ex]\hline 
\hline \\[-1.8ex] 
\\[-1.8ex] & \multicolumn{6}{c}{DV: female-male gender gap} \\ 
 & (a) & (b) & (c) & (d) & (e) & (f) \\ 
\hline \\[-1.8ex] 
 women civic & 0.39$^{***}$ & 0.39$^{***}$ & 0.40$^{***}$ & 0.37$^{***}$ &  &  \\ 
  & (0.22, 0.56) & (0.22, 0.57) & (0.20, 0.60) & (0.15, 0.59) &  &  \\ 
  & & & & & & \\ 
 civic (all) &  &  &  &  & 0.21$^{**}$ & 0.21$^{**}$ \\ 
  &  &  &  &  & (0.05, 0.38) & (0.01, 0.41) \\ 
  & & & & & & \\ 
 education (gap) &  & 0.05 &  & 0.02 &  &  \\ 
  &  & ($-$0.12, 0.23) &  & ($-$0.19, 0.24) &  &  \\ 
  & & & & & & \\ 
 education (all) & 0.24$^{***}$ &  & 0.24$^{**}$ &  & 0.27$^{***}$ & 0.24$^{**}$ \\ 
  & (0.07, 0.42) &  & (0.06, 0.43) &  & (0.09, 0.46) & (0.04, 0.43) \\ 
  & & & & & & \\ 
 \% Internet (gap) &  &  & $-$0.03 & 0.10 &  & 0.15 \\ 
  &  &  & ($-$0.25, 0.20) & ($-$0.17, 0.37) &  & ($-$0.06, 0.36) \\ 
  & & & & & & \\ 
 \% Internet (all) & $-$0.05 & 0.04 &  &  & 0.07 &  \\ 
  & ($-$0.24, 0.14) & ($-$0.16, 0.23) &  &  & ($-$0.12, 0.26) &  \\ 
  & & & & & & \\ 
 Constant & $-$0.25$^{***}$ & $-$0.25$^{***}$ & $-$0.26$^{***}$ & $-$0.25$^{***}$ & $-$0.24$^{***}$ & $-$0.25$^{***}$ \\ 
  & ($-$0.40, $-$0.09) & ($-$0.41, $-$0.09) & ($-$0.42, $-$0.09) & ($-$0.42, $-$0.08) & ($-$0.41, $-$0.07) & ($-$0.42, $-$0.07) \\ 
  & & & & & & \\ 
\hline \\[-1.8ex] 
Out-of-sample RMSE & 0.83 & 0.87 & 0.85 & 0.88 & 0.91 & 0.92 \\ 
Out-of-sample R2 & 0.19 & 0.13 & 0.21 & 0.13 & 0.09 & 0.1 \\ 
Observations & 101 & 101 & 98 & 98 & 101 & 98 \\ 
R$^{2}$ & 0.27 & 0.22 & 0.28 & 0.22 & 0.17 & 0.19 \\ 
Adjusted R$^{2}$ & 0.25 & 0.19 & 0.25 & 0.20 & 0.15 & 0.17 \\ 
AIC & 248.88 & 256.07 & 243.76 & 250.52 & 261.40 & 254.21 \\ 
BIC & 261.96 & 269.14 & 256.69 & 263.44 & 274.48 & 267.13 \\ 
Residual Std. Error & 0.81 (df = 97) & 0.83 (df = 97) & 0.81 (df = 94) & 0.84 (df = 94) & 0.86 (df = 97) & 0.86 (df = 94) \\ 
F Statistic & 11.93$^{***}$ (df = 3; 97) & 8.89$^{***}$ (df = 3; 97) & 11.96$^{***}$ (df = 3; 94) & 9.07$^{***}$ (df = 3; 94) & 6.77$^{***}$ (df = 3; 97) & 7.58$^{***}$ (df = 3; 94) \\ 
\hline 
\hline \\[-1.8ex] 
\textit{Note:}  & \multicolumn{6}{r}{$^{*}$p$<$0.1; $^{**}$p$<$0.05; $^{***}$p$<$0.01} \\ 
 & \multicolumn{6}{r}{Models (c,d,f) have fewer observations (and slightly lower prediction error) due to missing data in the new predictor.} \\ 
\end{tabular} }
\end{table*}

\begin{table*}[!htbp] \centering {\tiny 
  \caption{OLS for predicting female-male gender gap in online literacy estimates. Models include alternative predictors and controls for geographical groups. Values in parentheses are the lower and upper bounds of the 95\% confidence intervals of the estimated effects.} 
  \label{tab:model-gender-3v-geo} 
\begin{tabular}{@{\extracolsep{-10pt}}lcccccc} 
\\[-1.8ex]\hline 
\hline \\[-1.8ex] 
\\[-1.8ex] & \multicolumn{6}{c}{DV: female-male gender gap} \\ 
 & (a) & (b) & (c) & (d) & (e) & (f) \\ 
\hline \\[-1.8ex] 
 women civic & 0.41$^{***}$ & 0.43$^{***}$ & 0.35$^{***}$ & 0.37$^{***}$ &  &  \\ 
  & (0.23, 0.58) & (0.26, 0.61) & (0.16, 0.53) & (0.17, 0.56) &  &  \\ 
  & & & & & & \\ 
 civic (all) &  &  &  &  & 0.16$^{*}$ & 0.16 \\ 
  &  &  &  &  & ($-$0.02, 0.34) & ($-$0.04, 0.37) \\ 
  & & & & & & \\ 
 education (gap) &  & 0.13$^{*}$ &  & 0.10 &  &  \\ 
  &  & ($-$0.02, 0.29) &  & ($-$0.08, 0.29) &  &  \\ 
  & & & & & & \\ 
 education (all) & 0.19$^{**}$ &  & 0.19$^{**}$ &  & 0.21$^{**}$ & 0.18$^{**}$ \\ 
  & (0.04, 0.34) &  & (0.02, 0.36) &  & (0.05, 0.38) & (0.01, 0.36) \\ 
  & & & & & & \\ 
 \% Internet (gap) &  &  & $-$0.07 & $-$0.07 &  & 0.08 \\ 
  &  &  & ($-$0.30, 0.15) & ($-$0.33, 0.20) &  & ($-$0.14, 0.30) \\ 
  & & & & & & \\ 
 \% Internet (all) & $-$0.28$^{**}$ & $-$0.30$^{**}$ &  &  & $-$0.11 &  \\ 
  & ($-$0.50, $-$0.06) & ($-$0.54, $-$0.06) &  &  & ($-$0.34, 0.11) &  \\ 
  & & & & & & \\ 
 Central/Southern/Eastern Asia & 1.81$^{***}$ & 1.99$^{***}$ & 1.48$^{***}$ & 1.64$^{***}$ & 1.67$^{***}$ & 1.50$^{***}$ \\ 
  & (1.22, 2.41) & (1.40, 2.59) & (0.92, 2.03) & (1.09, 2.19) & (1.02, 2.32) & (0.91, 2.10) \\ 
  & & & & & & \\ 
 Europe/Oceania/Northern America & 0.84$^{***}$ & 0.97$^{***}$ & 0.46$^{*}$ & 0.58$^{**}$ & 0.83$^{***}$ & 0.53$^{*}$ \\ 
  & (0.30, 1.39) & (0.42, 1.53) & ($-$0.03, 0.96) & (0.07, 1.08) & (0.24, 1.43) & (0.002, 1.05) \\ 
  & & & & & & \\ 
 Latin America \& the Caribbean & 0.37 & 0.44$^{*}$ & 0.18 & 0.25 & 0.39 & 0.17 \\ 
  & ($-$0.07, 0.81) & ($-$0.004, 0.88) & ($-$0.28, 0.65) & ($-$0.22, 0.72) & ($-$0.09, 0.88) & ($-$0.32, 0.67) \\ 
  & & & & & & \\ 
 Northern Africa \& Western Asia & 0.35 & 0.48$^{*}$ & $-$0.16 & $-$0.04 & 0.03 & $-$0.20 \\ 
  & ($-$0.20, 0.90) & ($-$0.07, 1.04) & ($-$0.63, 0.31) & ($-$0.50, 0.42) & ($-$0.58, 0.65) & ($-$0.74, 0.33) \\ 
  & & & & & & \\ 
 Constant & $-$0.73$^{***}$ & $-$0.81$^{***}$ & $-$0.50$^{***}$ & $-$0.58$^{***}$ & $-$0.65$^{***}$ & $-$0.50$^{***}$ \\ 
  & ($-$1.06, $-$0.40) & ($-$1.14, $-$0.48) & ($-$0.80, $-$0.20) & ($-$0.88, $-$0.28) & ($-$1.02, $-$0.28) & ($-$0.82, $-$0.17) \\ 
  & & & & & & \\ 
\hline \\[-1.8ex] 
Out-of-sample RMSE & 0.73 & 0.75 & 0.76 & 0.8 & 0.8 & 0.82 \\ 
Out-of-sample R$^{2}$ & 0.37 & 0.35 & 0.35 & 0.31 & 0.28 & 0.29 \\ 
Observations & 101 & 101 & 98 & 98 & 101 & 98 \\ 
R$^{2}$ & 0.50 & 0.48 & 0.48 & 0.46 & 0.41 & 0.42 \\ 
Adjusted R$^{2}$ & 0.46 & 0.44 & 0.44 & 0.42 & 0.36 & 0.37 \\ 
AIC & 218.59 & 222.02 & 218.82 & 222.81 & 235.76 & 230.10 \\ 
BIC & 242.12 & 245.56 & 242.09 & 246.07 & 259.30 & 253.36 \\ 
Residual Std. Error & 0.68 (df = 93) & 0.69 (df = 93) & 0.70 (df = 90) & 0.72 (df = 90) & 0.74 (df = 93) & 0.75 (df = 90) \\ 
F Statistic & 13.29$^{***}$ (df = 7; 93) & 12.40$^{***}$ (df = 7; 93) & 12.00$^{***}$ (df = 7; 90) & 11.01$^{***}$ (df = 7; 90) & 9.13$^{***}$ (df = 7; 93) & 9.30$^{***}$ (df = 7; 90) \\ 
\hline 
\hline \\[-1.8ex] 
\textit{Note:}  & \multicolumn{6}{r}{$^{*}$p$<$0.1; $^{**}$p$<$0.05; $^{***}$p$<$0.01} \\ 
 & \multicolumn{6}{r}{Models (c,d,f) have fewer observations (and slightly lower prediction error) due to missing data in the new predictor.} \\ 
\end{tabular} }
\end{table*}

\begin{table*}[!htbp] \centering {\tiny 
  \caption{OLS for predicting within-country regional disparity in online literacy estimates. Models (d) corresponds to the figure in the main text. Models (b,c,e,f) include alternative interaction terms. Values in parentheses are the lower and upper bounds of the 95\% confidence intervals of the estimated effects.} 
  \label{tab:model-region-main} 
\begin{tabular}{@{\extracolsep{-10pt}}lcccccc} 
\\[-1.8ex]\hline 
\hline \\[-1.8ex] 
\\[-1.8ex] & \multicolumn{6}{c}{DV: within-country regional disparity} \\ 
 & (a) & (b) & (c) & (d) & (e) & (f) \\ 
\hline \\[-1.8ex] 
 unequal edu & 0.47$^{**}$ & 0.34$^{**}$ & 0.32$^{*}$ & 0.41$^{**}$ & 0.25 & 0.25 \\ 
  & (0.10, 0.84) & (0.02, 0.66) & ($-$0.02, 0.66) & (0.03, 0.80) & ($-$0.10, 0.60) & ($-$0.12, 0.63) \\ 
  & & & & & & \\ 
 (unequal edu):(Gini) & 0.18 &  &  & 0.25$^{*}$ &  &  \\ 
  & ($-$0.07, 0.44) &  &  & ($-$0.02, 0.51) &  &  \\ 
  & & & & & & \\ 
 (unequal edu):(\% Internet) &  & $-$0.09 &  &  & $-$0.15 &  \\ 
  &  & ($-$0.36, 0.18) &  &  & ($-$0.46, 0.17) &  \\ 
  & & & & & & \\ 
 (unequal edu):(education) &  &  & 0.003 &  &  & 0.01 \\ 
  &  &  & ($-$0.37, 0.38) &  &  & ($-$0.38, 0.41) \\ 
  & & & & & & \\ 
 Gini & $-$0.22$^{**}$ & $-$0.22$^{**}$ & $-$0.24$^{**}$ & $-$0.35$^{**}$ & $-$0.32$^{**}$ & $-$0.37$^{**}$ \\ 
  & ($-$0.42, $-$0.02) & ($-$0.43, $-$0.02) & ($-$0.46, $-$0.02) & ($-$0.61, $-$0.09) & ($-$0.61, $-$0.04) & ($-$0.67, $-$0.08) \\ 
  & & & & & & \\ 
 \% Internet (all) & 0.40$^{**}$ & 0.33$^{*}$ & 0.35$^{**}$ & 0.46$^{**}$ & 0.34$^{*}$ & 0.41$^{*}$ \\ 
  & (0.08, 0.72) & (0.0000, 0.66) & (0.01, 0.70) & (0.08, 0.83) & ($-$0.05, 0.74) & ($-$0.02, 0.83) \\ 
  & & & & & & \\ 
 education (all) & $-$0.29$^{**}$ & $-$0.27$^{*}$ & $-$0.31 & $-$0.33$^{**}$ & $-$0.29$^{*}$ & $-$0.36 \\ 
  & ($-$0.55, $-$0.02) & ($-$0.56, 0.02) & ($-$0.83, 0.21) & ($-$0.60, $-$0.06) & ($-$0.59, $-$0.003) & ($-$0.90, 0.19) \\ 
  & & & & & & \\ 
 Central/Southern/Eastern Asia &  &  &  & $-$0.17 & $-$0.05 & $-$0.24 \\ 
  &  &  &  & ($-$1.06, 0.72) & ($-$1.04, 0.94) & ($-$1.16, 0.68) \\ 
  & & & & & & \\ 
 Europe/Oceania/Northern America &  &  &  & $-$0.54 & $-$0.42 & $-$0.47 \\ 
  &  &  &  & ($-$1.51, 0.43) & ($-$1.40, 0.56) & ($-$1.47, 0.54) \\ 
  & & & & & & \\ 
 Latin America \& the Caribbean &  &  &  & 0.26 & 0.27 & 0.19 \\ 
  &  &  &  & ($-$0.36, 0.88) & ($-$0.38, 0.92) & ($-$0.46, 0.84) \\ 
  & & & & & & \\ 
 Northern Africa \& Western Asia &  &  &  & $-$0.09 & $-$0.03 & $-$0.18 \\ 
  &  &  &  & ($-$0.89, 0.72) & ($-$0.90, 0.83) & ($-$1.03, 0.67) \\ 
  & & & & & & \\ 
 Constant & $-$0.04 & $-$0.06 & 0.004 & 0.03 & $-$0.06 & 0.11 \\ 
  & ($-$0.24, 0.16) & ($-$0.33, 0.21) & ($-$0.30, 0.31) & ($-$0.47, 0.52) & ($-$0.66, 0.54) & ($-$0.51, 0.73) \\ 
  & & & & & & \\ 
\hline \\[-1.8ex] 
Out-of-sample RMSE & 0.9 & 0.92 & 0.92 & 0.92 & 0.95 & 0.95 \\ 
Out-of-sample R$^{2}$ & 0.14 & 0.1 & 0.11 & 0.09 & 0.08 & 0.06 \\ 
Observations & 79 & 79 & 79 & 79 & 79 & 79 \\ 
R$^{2}$ & 0.25 & 0.23 & 0.23 & 0.29 & 0.26 & 0.25 \\ 
Adjusted R$^{2}$ & 0.19 & 0.18 & 0.17 & 0.19 & 0.17 & 0.16 \\ 
AIC & 207.37 & 209.02 & 209.48 & 210.91 & 213.57 & 214.53 \\ 
BIC & 223.96 & 225.60 & 226.07 & 236.98 & 239.63 & 240.59 \\ 
Residual Std. Error & 0.86 (df = 73) & 0.86 (df = 73) & 0.87 (df = 73) & 0.86 (df = 69) & 0.87 (df = 69) & 0.88 (df = 69) \\ 
F Statistic & 4.75$^{***}$ (df = 5; 73) & 4.36$^{***}$ (df = 5; 73) & 4.24$^{***}$ (df = 5; 73) & 3.09$^{***}$ (df = 9; 69) & 2.73$^{***}$ (df = 9; 69) & 2.61$^{**}$ (df = 9; 69) \\ 
\hline 
\hline \\[-1.8ex] 
\textit{Note:}  & \multicolumn{6}{r}{$^{*}$p$<$0.1; $^{**}$p$<$0.05; $^{***}$p$<$0.01} \\ 
\end{tabular} }
\end{table*}

\begin{table*}[!htbp] \centering {\tiny 
  \caption{OLS for predicting within-country regional disparity in online literacy estimates. Models include alternative predictors and no interaction term. Values in parentheses are the lower and upper bounds of the 95\% confidence intervals of the estimated effects.} 
  \label{tab:model-region-4v} 
\begin{tabular}{@{\extracolsep{-10pt}}lcccccc} 
\\[-1.8ex]\hline 
\hline \\[-1.8ex] 
\\[-1.8ex] & \multicolumn{6}{c}{DV: within-country regional disparity} \\ 
 & (a) & (b) & (c) & (d) & (e) & (f) \\ 
\hline \\[-1.8ex] 
 unequal edu & 0.33$^{**}$ & 0.15 & 0.18 & 0.26 & 0.13 & 0.15 \\ 
  & (0.01, 0.64) & ($-$0.17, 0.46) & ($-$0.12, 0.47) & ($-$0.10, 0.61) & ($-$0.22, 0.48) & ($-$0.19, 0.50) \\ 
  & & & & & & \\ 
 Gini & $-$0.24$^{**}$ & $-$0.28$^{***}$ & $-$0.28$^{***}$ & $-$0.37$^{***}$ & $-$0.34$^{**}$ & $-$0.33$^{**}$ \\ 
  & ($-$0.44, $-$0.05) & ($-$0.49, $-$0.08) & ($-$0.48, $-$0.08) & ($-$0.63, $-$0.10) & ($-$0.61, $-$0.06) & ($-$0.60, $-$0.06) \\ 
  & & & & & & \\ 
 \% Internet (all) & 0.35$^{**}$ &  &  & 0.40$^{**}$ &  &  \\ 
  & (0.04, 0.67) &  &  & (0.02, 0.77) &  &  \\ 
  & & & & & & \\ 
 education (all) & $-$0.31$^{**}$ & $-$0.20 & $-$0.20 & $-$0.34$^{**}$ & $-$0.24 & $-$0.27$^{*}$ \\ 
  & ($-$0.57, $-$0.05) & ($-$0.49, 0.08) & ($-$0.46, 0.05) & ($-$0.61, $-$0.06) & ($-$0.54, 0.07) & ($-$0.55, $-$0.003) \\ 
  & & & & & & \\ 
 income &  & $-$0.01 &  &  & $-$0.06 &  \\ 
  &  & ($-$0.32, 0.31) &  &  & ($-$0.45, 0.33) &  \\ 
  & & & & & & \\ 
 civic (all) &  &  & 0.08 &  &  & 0.17 \\ 
  &  &  & ($-$0.15, 0.32) &  &  & ($-$0.11, 0.45) \\ 
  & & & & & & \\ 
 Central/Southern/Eastern Asia &  &  &  & $-$0.23 & $-$0.03 & 0.13 \\ 
  &  &  &  & ($-$1.14, 0.67) & ($-$0.96, 0.89) & ($-$0.83, 1.08) \\ 
  & & & & & & \\ 
 Europe/Oceania/Northern America &  &  &  & $-$0.46 & 0.11 & 0.02 \\ 
  &  &  &  & ($-$1.44, 0.52) & ($-$0.94, 1.17) & ($-$0.86, 0.90) \\ 
  & & & & & & \\ 
 Latin America \& the Caribbean &  &  &  & 0.19 & 0.47 & 0.51 \\ 
  &  &  &  & ($-$0.43, 0.82) & ($-$0.16, 1.10) & ($-$0.09, 1.12) \\ 
  & & & & & & \\ 
 Northern Africa \& Western Asia &  &  &  & $-$0.17 & 0.22 & 0.39 \\ 
  &  &  &  & ($-$0.99, 0.64) & ($-$0.60, 1.05) & ($-$0.45, 1.22) \\ 
  & & & & & & \\ 
 Constant & 0.002 & 0.001 & $-$0.02 & 0.10 & $-$0.17 & $-$0.23 \\ 
  & ($-$0.19, 0.19) & ($-$0.20, 0.20) & ($-$0.22, 0.19) & ($-$0.40, 0.59) & ($-$0.68, 0.34) & ($-$0.70, 0.25) \\ 
  & & & & & & \\ 
\hline \\[-1.8ex] 
Out-of-sample RMSE & 0.91 & 0.92 & 0.92 & 0.95 & 0.96 & 0.95 \\ 
Out-of-sample R$^{2}$ & 0.12 & 0.09 & 0.1 & 0.08 & 0.05 & 0.05 \\ 
Observations & 79 & 79 & 79 & 79 & 79 & 79 \\ 
R$^{2}$ & 0.23 & 0.18 & 0.18 & 0.25 & 0.21 & 0.22 \\ 
Adjusted R$^{2}$ & 0.18 & 0.13 & 0.14 & 0.17 & 0.12 & 0.13 \\ 
AIC & 207.48 & 212.39 & 211.87 & 212.54 & 217.18 & 215.68 \\ 
BIC & 221.70 & 226.61 & 226.09 & 236.23 & 240.87 & 239.37 \\ 
Residual Std. Error & 0.86 (df = 74) & 0.89 (df = 74) & 0.89 (df = 74) & 0.87 (df = 70) & 0.90 (df = 70) & 0.89 (df = 70) \\ 
F Statistic & 5.38$^{***}$ (df = 4; 74) & 3.94$^{***}$ (df = 4; 74) & 4.09$^{***}$ (df = 4; 74) & 2.97$^{***}$ (df = 8; 70) & 2.30$^{**}$ (df = 8; 70) & 2.52$^{**}$ (df = 8; 70) \\ 
\hline 
\hline \\[-1.8ex] 
\textit{Note:}  & \multicolumn{6}{r}{$^{*}$p$<$0.1; $^{**}$p$<$0.05; $^{***}$p$<$0.01} \\ 
\end{tabular} }
\end{table*} 

\clearpage


\begin{backmatter}




\end{backmatter}
\end{document}